\documentclass[a4paper,11pt]{article}
\pdfoutput=1

\usepackage{jheppub}

\usepackage{graphicx}
\usepackage{amsmath}
\usepackage{amssymb}
\usepackage{xspace}

\usepackage{slashed}
\usepackage{mdwlist}
\usepackage{xcolor}
\usepackage{soul}

\newcommand{\eq}[1]{eq.~\eqref{eq:#1}}
\newcommand{\eqs}[2]{eqs.~\eqref{eq:#1} and \eqref{eq:#2}}
\renewcommand{\sec}[1]{sec.~\ref{sec:#1}}

\newcommand{\subsec}[1]{sec.~\ref{subsec:#1}}
\newcommand{\subsecs}[2]{secs.~\ref{subsec:#1} and \ref{subsec:#2}}
\newcommand{\app}[1]{app.~\ref{app:#1}}
\newcommand{\fig}[1]{fig.~\ref{fig:#1}}

\newcommand{\abs}[1]{\lvert#1\rvert}

\newcommand{\ord}[1]{\mathcal{O}(#1)}
\newcommand{\ORd}[1]{\mathcal{O}\Bigl(#1\Bigr)}

\newcommand{\MAe}[3]{\Bigl\langle#1\Bigr\rvert#2\Bigr\rvert#3\Bigr\rangle}

\newcommand{\Rzero}{R_0}

\newcommand{\df}{\mathrm{d}}

\newcommand{\MS}{\overline{\rm MS}}

\newcommand{\Li}{\textrm{Li}}

\newcommand{\bfT}{{\bf T}} 

\newcommand{\sdt}{\!\cdot\!}
\newcommand{\tr}{\textrm{tr}}

\newcommand{\al}{\alpha}

\newcommand{\ga}{\gamma}

\newcommand{\de}{\delta}
\newcommand{\eps}{\epsilon}

\newcommand{\si}{\sigma}

\newcommand{\cL}{{\mathcal L}}

\newcommand{\Tau}{{\mathcal T}}

\newcommand{\bn}{\bar{n}}

\newcommand{\nn}{\nonumber}

\newcommand{\lqcd}{\Lambda_\mathrm{QCD}}
\newcommand{\LQCD}{\Lambda_\mathrm{QCD}}

\newcommand{\Rv}{R_{\rm veto}}
\newcommand{\Gcusp}{\Gamma_{\rm cusp}}

\newcommand{\csoft}{csoft\xspace}
\newcommand{\Csoft}{Csoft\xspace}
\newcommand{\Ecm}{E_\mathrm{cm}}

\newcommand{\one}{{(1)}}

\newcommand{\SCETa}{\ensuremath{{\rm SCET}_{\rm I}}\xspace}
\newcommand{\SCETb}{\ensuremath{{\rm SCET}_{\rm II}}\xspace}
\newcommand{\SCETp}{\ensuremath{{\rm SCET}_+}\xspace}

\allowdisplaybreaks[2]

\title{Factorization for Jet Radius Logarithms in Jet Mass Spectra at the LHC}

\author[a]{Daniel W. Kolodrubetz,}
\author[b]{Piotr Pietrulewicz,}
\author[a]{Iain W.~Stewart,}
\author[b]{Frank J.~Tackmann,}
\author[c,d]{and Wouter J.~Waalewijn}

\affiliation[a]{Center for Theoretical Physics, Massachusetts Institute of Technology, Cambridge, MA~02139, U.S.A.}
\affiliation[b]{Theory Group, Deutsches Elektronen-Synchrotron (DESY), D-22607 Hamburg, Germany}
\affiliation[c]{ITFA, University of Amsterdam, Science Park 904, 1018 XE, Amsterdam, The Netherlands}
\affiliation[d]{Nikhef, Theory Group, Science Park 105, 1098 XG, Amsterdam, The Netherlands}

\emailAdd{dkolodru@mit.edu}
\emailAdd{piotr.pietrulewicz@desy.de}
\emailAdd{iains@mit.edu}
\emailAdd{frank.tackmann@desy.de}
\emailAdd{w.j.waalewijn@uva.nl}

\abstract{
To predict the jet mass spectrum at a hadron collider it is crucial to account for the resummation of logarithms between the transverse momentum of the jet and its invariant mass $m_J$. For small jet areas there are additional large logarithms of the jet radius $R$, which affect the convergence of the perturbative series. We present an analytic framework for exclusive jet production at the LHC which gives a complete description of the jet mass spectrum including realistic jet algorithms and jet vetoes. It factorizes the scales associated with $m_J$, $R$, and the jet veto, enabling in addition the systematic resummation of jet radius logarithms in the jet mass spectrum beyond leading logarithmic order. We discuss the factorization formulae for the peak and tail region of the jet mass spectrum and for small and large $R$, and the relations between the different regimes and how to combine them. Regions of experimental interest are classified which do not involve large nonglobal logarithms.  We also present universal results for nonperturbative effects and discuss various jet vetoes.
}

\preprint{\vbox{
\hbox{DESY 16-080}
\hbox{MIT--CTP 4800} 
\hbox{NIKHEF 2016-021}}}

\begin{document}

\maketitle

\section{Introduction}
\label{sec:intro}

The field of jet substructure has continued to expand over the past few years, providing valuable tools to study processes in the challenging environment at the LHC~\cite{Altheimer:2012mn,Altheimer:2013yza,Adams:2015hiv}. This is e.g.~due to the fact that massive resonances (top quarks, $W$ bosons, etc.) which may be part of a new physics signal are often boosted and the discrimination of their collimated decay products from QCD jets critically relies on jet substructure techniques.  This field has flourished due to the excellent performance of the ATLAS and CMS detectors and the development of new substructure techniques. Theoretically one has to predict the dynamics and distribution of radiation inside jets produced by different particles.  Most  theoretical studies still rely strongly on Monte Carlo parton showers, which are limited in their precision. However, there has been a recent push to developing analytic frameworks which provide theoretical uncertainties and put predictions on a firmer footing. Such calculations may also suggest ways to improve observables, see e.g.~refs.~\cite{Dasgupta:2013ihk,Larkoski:2014wba,Larkoski:2014gra}. While the description of jets originating from the decay of highly boosted massive particles (e.g.~for $pp \to Z(\to \ell \bar{\ell})Z(\to 1\,{\rm jet})$) can be carried out to high precision with standard methods (see e.g.~ref.~\cite{Feige:2012vc}), the associated process with the jet originating from color-correlated emissions (e.g.~for $pp \to Z(\to\ell\bar\ell)+ 1$ jet) is much more difficult to handle analytically.

A basic and important benchmark observable for studying the radiation inside a jet is the invariant mass $m_J$ of a jet, given by the square of the total four-momentum of the jet constituents, $m_J^2 = ( \sum_{i\in J} p_i^\mu )^2$. The jet mass spectrum provides key information about the influence of Sudakov double logarithms and soft radiation in a hadronic environment and in particular probes the
dependence on the jet algorithm and jet size $R$, color flow, initial and final state partonic channels, hadronization, and underlying event.  The best sensitivity to these effects comes from studying jets in their primal state, without using jet-grooming techniques to change the nature of the jet constituents. While useful for tagging studies, jet grooming fundamentally changes the nature of the jet mass observable, and is known to reduce its utility as a probe of these physical effects~\cite{Dasgupta:2013ihk,Frye:2016okc,Frye:2016aiz}.

In the past few years, several analytic ungroomed jet mass calculations for hadron colliders have been carried out~\cite{Dasgupta:2012hg,Chien:2012ur,Jouttenus:2013hs,Stewart:2014nna,Liu:2014oog,Hornig:2016ahz}. In ref.~\cite{Dasgupta:2012hg}, the inclusive jet mass spectrum in $pp \to 2$ jets and $Z+ 1$ jet was calculated at next-to-leading-logarithmic (NLL) order.  In ref.~\cite{Chien:2012ur},  next-to-next-to-leading-logarithmic (NNLL) order results were obtained for the $pp \to \ga+1$ jet, by examining the jet mass spectrum while expanding around a threshold limit.  A similar setup was used in Ref.~\cite{Liu:2014oog} to obtain the jet mass spectrum for $pp \to$ dijets.   In ref.~\cite{Jouttenus:2013hs}, the jet mass spectrum was directly calculated for $pp \to H + 1$ jet at NNLL order, where a veto on additional jets was imposed to obtain an exclusive 1-jet sample. The utility of the first moment of the jet mass spectrum as a mechanism to disentangle different sources of soft radiation underlying the hard interaction was discussed in ref.~\cite{Stewart:2014nna}.  Recently, in ref.~\cite{Hornig:2016ahz} the study of jet mass was extended to angularities for $pp\to 2$ jets at NLL$'$, with an exclusive 2-jet sample without a veto beyond a certain rapidity cut.

In this paper, we improve the analytic description of jet mass spectra at the LHC, by systematically taking the effects of realistic jet algorithms into account with factorization formulae. In particular, for small jet sizes the exclusive $N$-jet cross section contains Sudakov double logarithms of the jet radius $R$, in conjunction with logarithms of the jet mass and jet veto, and our results enable their resummation at any perturbative order. This allows in particular for NNLL resummation using known anomalous dimensions and the relations provided here. This factorization in the small-$R$ regime is our main focus. We also consider the tail of the jet mass spectrum where the $R$ dependence is important because of the kinematic bound $m_J \lesssim p_T^J R$,\footnote{For a uniform energy distribution inside the jet the upper bound is $m_J < p_T^J R/\sqrt{2}$ (up to $\ord{R^2}$ corrections). For a jet consisting of two particles this reduces to $m_J < p_T^J R/2$ for clustering algorithms like anti-$k_T$.} where $p_T^J$ is the transverse momentum of the jet.

For definiteness, we consider the jet mass spectrum for $pp \to L+1$ jet, where $L$ is a hard color-singlet state (e.g. $\gamma$, $W$, $Z$, $H$) recoiling against the jet. The jet region is determined by a factorization-friendly jet algorithm like anti-$k_T$ clustering~\cite{Cacciari:2008gp} or the $N$-jettiness partitioning used in XCone~\cite{Stewart:2015waa, Thaler:2015xaa}, with a jet radius parameter $R$ controlling its size.  The hard signal jet of interest is uniquely identified by imposing a veto on additional jets, for which we consider a range of possibilities, including beam thrust~\cite{Stewart:2009yx} and the standard $p_T$ jet veto. Although jet mass measurements typically use $R \approx 1$, see e.g.~refs.~\cite{ATLAS:2012am,Chatrchyan:2013vbb},
we will find that the $\mathcal{O}(\alpha_s)$ corrections for $m_J \ll p_T^J R$ are still well approximated by the small-$R$ result, such that the actual expansion parameter is rather $(R/R_0)^2$ with $R_0 \simeq 2$.
Throughout the paper we will often leave the factors of $R_0$ implicit when indicating that there are power correction of $\mathcal{O}(R^2)$ and logarithms $\ln R$.

To treat the small-$R$ effects, we build on the recent work of ref.~\cite{Chien:2015cka}, which discussed the systematic resummation of jet radius logarithms for $e^+ e^- \to 2$ cone jets with an energy veto on the radiation outside the jets. This process was also studied in ref.~\cite{Becher:2015hka} using a similar SCET framework. It was found that the resummation of jet-radius logarithms requires an extension of Soft-Collinear Effective Theory (SCET)~\cite{Bauer:2000ew, Bauer:2000yr, Bauer:2001ct, Bauer:2001yt}, most often called \SCETp, which contains additional modes that are simultaneously collinear and soft~\cite{Bauer:2011uc, Procura:2014cba, Larkoski:2015zka, Pietrulewicz:2016nwo}. Recently, in~\cite{Hornig:2016ahz} the $\ln R$ resummation of ref.~\cite{Chien:2015cka} was extended to $pp\to $ dijets away from the endpoint of the angularity distribution. Note that the resummation of jet radius logarithms at leading logarithmic order was also developed earlier in ref.~\cite{Dasgupta:2014yra} for several types of jet observables, including the inclusive jet spectrum. However, for these observables the structure of logarithms is different than the jet mass measurements considered here, since no Sudakov double logarithms of the jet radius (of the identified hard jet) arise. For the inclusive jet spectrum the small $R$ expansion also works well for $R\lesssim 1$, as recently discussed in ref.~\cite{Dasgupta:2016bnd}.

\begin{figure}
\centering
\includegraphics[width=.5\textwidth]{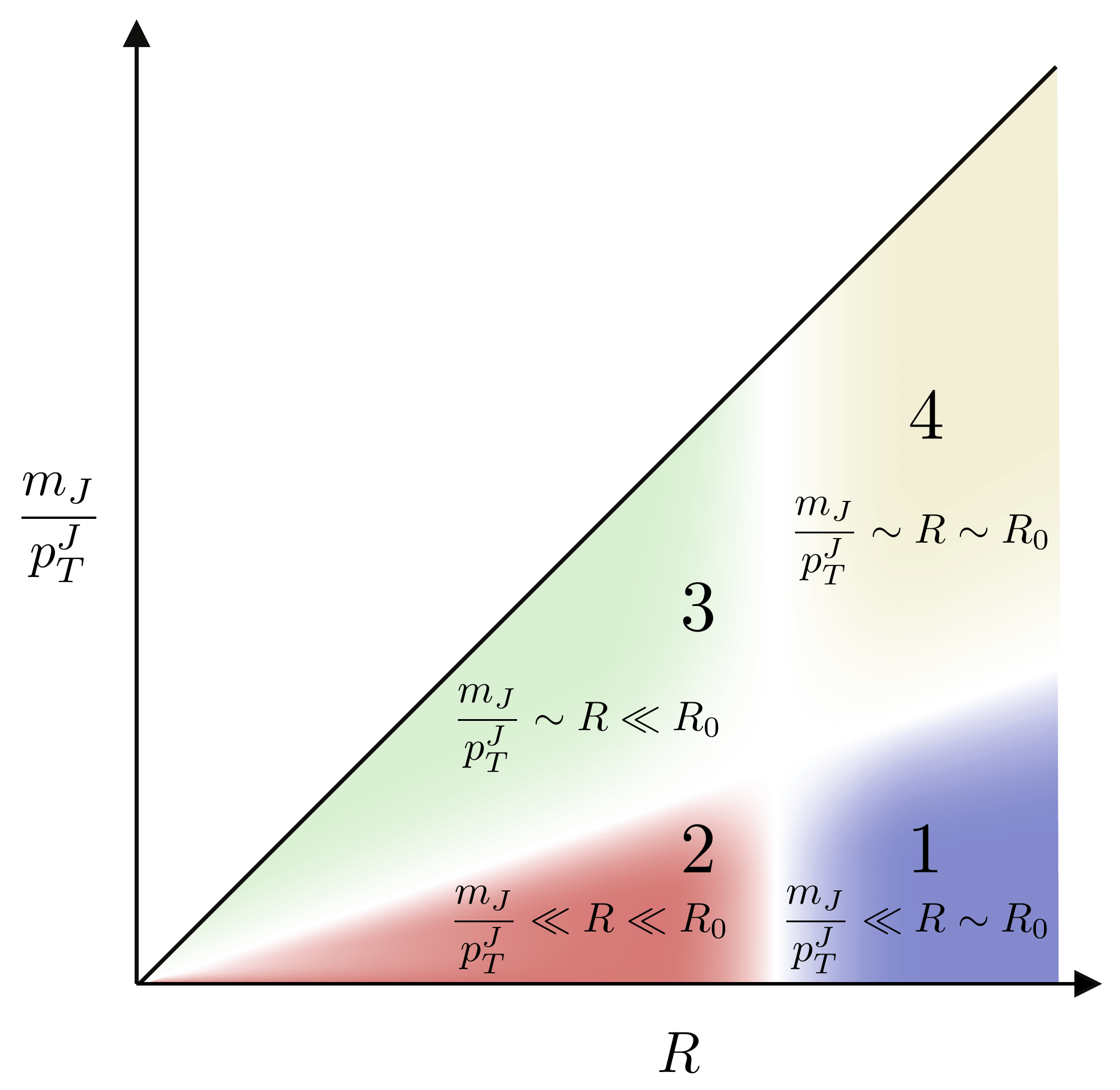}
\caption{Illustration of the various hierarchical regimes for jet mass measurements in the $R$- and $m_J/p_T^J$-plane.
\label{fig:regions_diagram}}
\end{figure}

To organize our discussion, we divide the treatment of jet mass and jet radius into several distinct cases.  As illustrated in \fig{regions_diagram}, one can distinguish four different regimes with different hierarchies between $R$ and $R_0$ and the scales $m_J$ and $p_T^J R$:
\begin{itemize}
 \item regime 1: large-$R$ jets ($R\sim R_0$) for small $m_J$: $m_J \ll p_T^J R \sim p_T^J$
 \item regime 2: small-$R$ jets ($R\ll R_0$) for small $m_J$: $m_J \ll p_T^J R \ll p_T^J$
 \item regime 3: small-$R$ jets ($R\ll R_0$) for large $m_J$: $m_J \sim p_T^J R \ll p_T^J$
 \item regime 4: large-$R$ jets ($R\sim R_0$) for large $m_J$: $m_J \sim p_T^J R \sim p_T^J$
\,.\end{itemize}
All of these require distinct factorization formulae to resum the corresponding large logarithms. Specifically, in regimes 1 and 2 these are logarithms of $m_J/p_T^J$, and in regimes 2 and 3 logarithms of $R/R_0$. We also discuss how to appropriately combine these regimes to obtain a complete description for any value of $m_J$ and $R$.

In carrying out jet mass resummation, an additional complication is that the restrictions on the radiation inside and outside the jet imposed by the measurement lead to nonglobal (NG) structures.  If the kinematic scales related to these constraints are widely separated, the nonglobal contributions can contain parametrically large nonglobal logarithms (NGLs)~\cite{Dasgupta:2001sh}. In ref.~\cite{Dasgupta:2012hg} the NGLs were resummed in the large-$N_c$ approximation and found to be significant in the peak region for the inclusive jet calculation considered there. Although NGLs were not resummed in ref.~\cite{Chien:2012ur}, their estimated size agreed with ref.~\cite{Dasgupta:2012hg}. In contrast, if a veto on additional hard jets is imposed, it changes the structure of the nonglobal terms, providing regions of phase space where NGLs are not large and other regions where they are~\cite{Jouttenus:2013hs}.\footnote{The mitigation of NGLs through  additional measurements was first addressed in ref.~\cite{Berger:2003iw}.}  The NGLs may still have a sizeable relative impact for unnormalized spectra, but in the factorization framework their effects on the small $m_J$-spectrum are tamed by having the same Sudakov suppression as all other terms. For normalized spectra the dependence on the jet veto largely drops out and the effects due to NG structures remain moderate. In particular, in regime 1 with $R\sim R_0$ and a range of jet-veto scales there are no large NGLs over the majority of the jet-mass spectrum~\cite{Jouttenus:2013hs}. We will see that in regime 2 with $R\ll R_0$ large NGLs can similarly be avoided. However, the associated parametric condition on the jet veto cannot be satisfied over the full jet mass spectrum including the far tail of the spectrum corresponding to regime 3. On the other hand, we will demonstrate that in regimes 2 and 3 the leading NGLs are simply those of hemisphere soft functions, which have been studied extensively in the literature, see e.g.~refs.~\cite{Kelley:2011ng, Hornig:2011iu, Schwartz:2014wha}. This has also been seen in the explicit $\mathcal{O}(\alpha_s^2)$ computation for jet shapes in the small $R$ limit in ref.~\cite{Banfi:2010pa}.
Approaches for their resummation beyond the large-$N_c$ leading logarithms~\cite{Banfi:2002hw} have been developed recently, see refs.~\cite{Hatta:2013iba, Larkoski:2015zka, Caron-Huot:2015bja, Neill:2015nya, Becher:2016mmh}, and can be directly applied to our case.

The outline of the paper is as follows: In \sec{fact}, we present the factorized cross sections relevant for regimes 1, 2, and 3, focusing on the case of a global generalized beam thrust jet veto. We also discuss the relations among the regimes and their combination and briefly comment on regime 4. The definitions and one-loop expressions of the relevant ingredients are discussed in \sec{softandjet}, with calculational details relegated to \app{softfct}. In \sec{softandjet}, we also validate the relations between the factorization formulae, discuss the leading nonperturbative effects, and compare the predictions of our factorization framework with earlier jet mass calculations. We discuss the extension to transverse energy/momentum vetoes and jet-based vetoes in \sec{jetveto} including a study on the small $R$ expansion of the fixed-order cross section at $\mathcal{O}(\alpha_s)$, and conclude in \sec{conclusions}. Consistency of RG running is exploited in \app{anomdim} to determine the anomalous dimensions which allow for the NNLL resummation of jet mass, jet radius, and jet veto logarithms.

\section{Factorization for jet mass with jet radius effects}
\label{sec:fact}

\begin{figure}
\centering
\includegraphics[width=.85\textwidth]{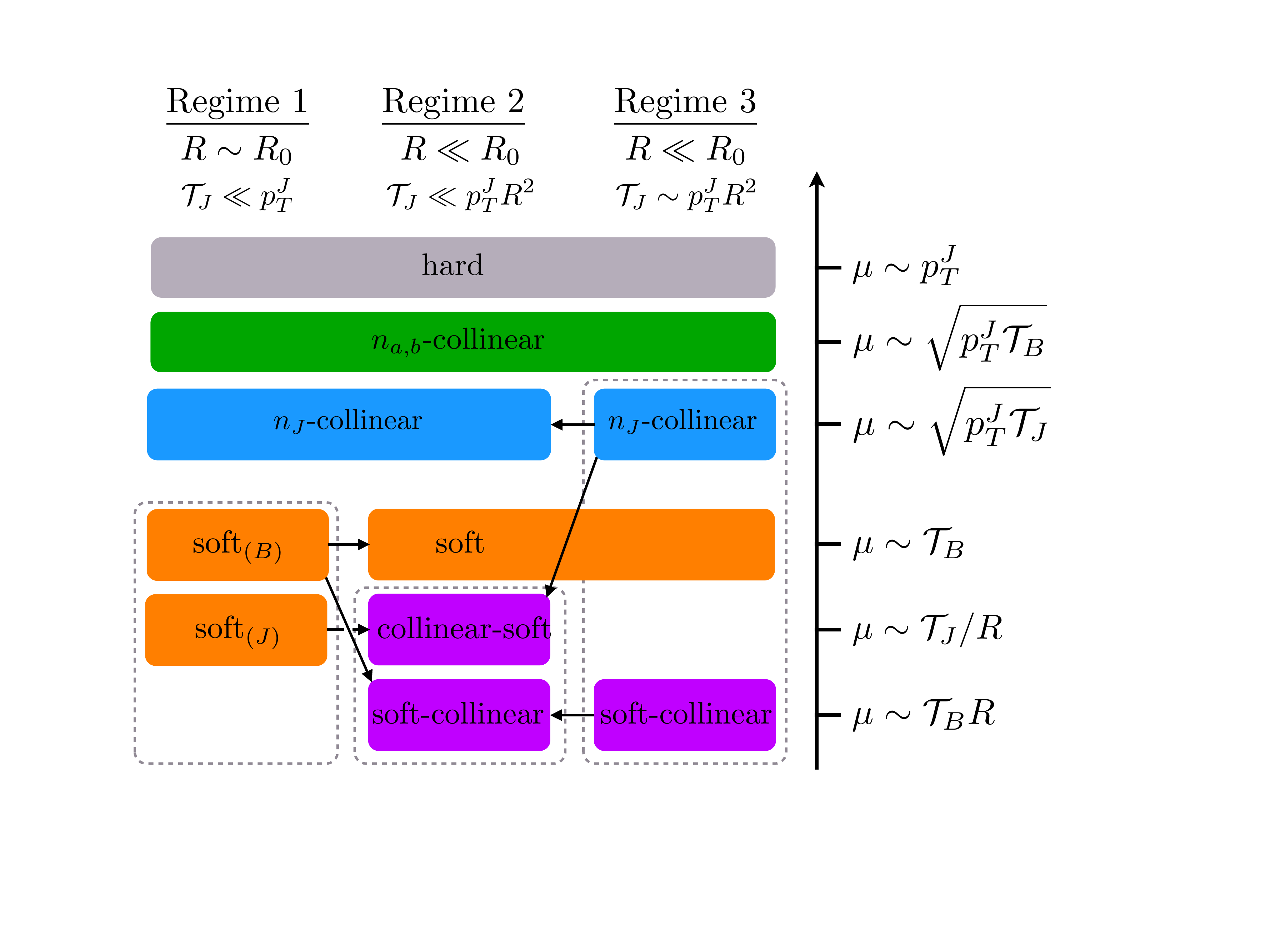}
\caption{Characteristic invariant mass scales of the modes for the regimes 1, 2, and 3. The arrows indicate the relations among them, while the boxes indicate nonglobal correlations. Specifically, as discussed below \eq{csoft_scaling}, in regime 2 for the scaling $\Tau_B R^2 \sim \Tau_J$, the soft-collinear and collinear-soft modes merge into a single $n_J$-csoft mode. Similarly, as discussed below \eq{soft_scaling}, in regime 1 with the scaling $\Tau_B \sim \Tau_J$, the soft$_{(B)}$ and soft$_{(J)}$ modes merge into a single soft mode. In regime 3, the $n_J$-collinear and soft-collinear modes cannot be merged into a single mode by a scaling choice when employing a jet veto $\Tau_B \ll p_T^J$. \label{fig:modes}}
\end{figure}

To study the jet radius dependence in a jet mass spectrum we consider exclusive $pp \to L+1$ jet processes with the hard jet recoiling against a generic color-singlet state $L$. We first summarize the basic setup and kinematics of the process in \subsec{kinematics}. We then discuss the modes of the relevant EFT setup and present the associated factorization formulae for each regime in turn. In \subsec{case1}, we review the jet mass spectrum for $m_J \ll p_T^J R$ and large-$R$ jets~\cite{Jouttenus:2013hs} (regime 1), which can be described with standard SCET. In \subsec{case2}, we discuss regime 2, where $m_J \ll p_T^J R$ but now has narrow jets $R\ll R_0$, which is described using SCET$_+$. The region where the jet mass spectrum turns off, i.e.~$m_J \sim p_T^J R$, is discussed for small-$R$ jets (regime 3) in \subsec{case3}, and briefly for large-$R$ jets (regime 4) in \subsec{case4}. In \subsec{summary}, we show how the theories for these different hierarchies are related to each other, the relations this implies between the ingredients of the factorization formulae, and how to systematically combine the latter including all relevant kinematic power corrections. The modes and corresponding logarithms appearing for regimes 1, 2, and 3 are summarized in table~\ref{tab:regimes}, and their relations and scaling are illustrated in \fig{modes}.

\subsection{Kinematics and measurements}
\label{subsec:kinematics}

The hard (Born) kinematics of the exclusive $pp \to L+1$ jet process is characterized by five independent variables, which we choose to be the jet transverse momentum $p_T^J$, jet pseudorapidity $\eta_J$, azimuthal angle $\phi_J$ of the jet, and the  rapidity $Y_L$ and total invariant mass $q_L^2$ of the recoiling color-singlet state $L$.

The Born-level momentum conservation (corresponding to the label-momentum conservation in the EFT) is given by
\begin{align}
\omega_a \frac{n_a^\mu}{2} + \omega_b \frac{n_b^\mu}{2} &= p_J^\mu + q_L^\mu
\,,\end{align}
with 
\begin{align}
\omega_a = x_a \Ecm \, , \qquad \omega_b = x_b \Ecm \, ,
\end{align}
where $\Ecm$ is the hadronic center-of-mass energy, and the direction of beams $a$ and $b$ are denoted as
\begin{align}
n^\mu_a= (1,0,0,1) = \bn_b^\mu
\,, \qquad
n_b^\mu =  (1,0,0,-1) = \bn_a^\mu
\,.\end{align}
In terms of the hard kinematic variables, the momentum components can be written as
\begin{align}
\omega_a &= m_T e^{Y_L} + p_T^J e^{\eta_J}
\,, \quad
\omega_b = m_T e^{-Y_L} + p_T^J e^{-\eta_J}
\,, \quad m_T = \sqrt{p_T^{J\,2} + q_L^2} 
\,, \nn \\
p_J^\mu &= p_T^J \bigl(\cosh\eta_J, \cos \phi_J, \sin \phi_J, \sinh \eta_J\bigr)
\quad\text{with}\quad
p_J^2 = 0
\,, \nn \\
q_L^\mu &=  (m_T \cosh Y_L, -p_T^J \cos \phi_J, -p_T^J \sin \phi_J, m_T \sinh Y_L)
\,, \nn \\
Q^2 &\equiv \omega_a \omega_b = m_T^2 + p_T^{J\,2} + 2 m_T p_T^J \cosh (Y_L - \eta_J)
\,.\end{align}
Here, $Q$ is the invariant mass of the $L+$jet system and is a derived quantity with our choice of independent variables. Note that in the hard kinematics the jet (label) momentum is represented by a massless four-vector $p_J^\mu$. For future convenience, we introduce the following shorthand for the hard phase space measure
\begin{align} \label{eq:hardkin}
\df\Phi &= \frac{1}{2 \Ecm^2} \,\frac{\df x_a}{x_a} \,\frac{\df x_b}{x_b} \,\frac{\df^4 p_J}{(2 \pi)^4} \,\frac{\df^4 q_L}{(2 \pi)^4} \,2 \pi \delta(p_J^2)\theta(p_J^0) \,(2 \pi)^4 \delta^{(4)} \Bigl(\omega_a \frac{n_a}{2} + \omega_b \frac{n_b}{2} - p_J - q_L\Bigr) \df \Phi_L(q_L)
\nn \\
 &= \frac{p_T^J}{8\pi \Ecm^2 Q^2}\, \df p_T^J\, \df \eta_J\,\df Y_L\,  \df q_L^2\, \frac{ \df\phi_J}{2\pi}\,
   \df \Phi_L(q_L)
\,.\end{align}

In the following, we always assume that the jet is hard and not too forward, i.e.~$p_T^J \sim Q$ and $e^{|\eta_J|} \sim 1$. The factorization in the case $p_T^J \ll Q$ where the jet is soft or close to one of the beams can be performed using SCET$_+$ as in refs.~\cite{Bauer:2011uc, Larkoski:2015zka, Larkoski:2015kga, Pietrulewicz:2016nwo} for large $R$ jets, and could be extended to narrow jets by combining it with the setup discussed in this paper.

We assume that the shape of the jet region is determined by a jet algorithm which clusters collinear radiation first before assigning soft radiation to either the jet or the beam region, with a jet radius parameter $R$ controlling its size. This includes both the anti-$k_T$ algorithm as well as XCone~\cite{Stewart:2015waa,Thaler:2015xaa} based on $N$-jettiness minimization~\cite{Stewart:2010tn,Thaler:2010tr}. For these jet algorithms, narrow jets are all roughly circular, and deviations are power suppressed in $R$. We will present results for a jet radius defined in $(\eta,\phi)$ coordinates.\footnote{For small $R$ this is equivalent to an angular radius in $(\theta,\phi)$ space of $R/\cosh \eta_J$.} The jet mass measurement is encoded by
\begin{align}
\Tau_J = \cosh \eta_J \sum_{i \in \text{jet}} n_J \cdot p_i 
\, ,\end{align}
where
\begin{align}
n^\mu_J \equiv (1,\vec n_J) = \frac{p_J^\mu}{p_T^J\cosh\eta_J}
= \Bigl(1,\frac{\cos \phi_J}{\cosh \eta_J},\frac{\sin \phi_J}{\cosh \eta_J}, \tanh \eta_J\Bigr)
\,, \quad
\bn_J^\mu =  (1,- \vec n_J)
\,,\end{align}
where $\vec{n}$ is the jet (label) direction, which we identify with the jet direction found by the jet algorithm.
We will often write momenta in terms of light-cone coordinates along either the jet or beam directions,
\begin{align} \label{eq:lc}
p^\mu &= n_J \sdt p\,\frac{\bn_J^\mu}{2} + \bar{n}_J \sdt p\,\frac{n_J^\mu}{2} + p_{\perp,J}^\mu \equiv  (n_J \sdt p,\bar{n}_J \sdt p,p_{\perp,J}) \equiv  (p^+,p^-,p_\perp)_J \, ,
\\
&=n_a \sdt p\,\frac{\bn_a^\mu}{2} + \bar{n}_a \sdt p\,\frac{n_a^\mu}{2} + p_{\perp,B}^\mu \equiv  (n_a \sdt p,\bar{n}_a \sdt p,p_{\perp,B}) \equiv  (p^+,p^-,p_\perp)_B \,.
\nn
\end{align}
The relation between the jet mass and $\Tau_J$ (which is more convenient in the following) is in general given by
\begin{align} \label{eq:TauJ}
  m_J^2 &= \Bigl(\sum_{i \in \text{jet}} p_i\Bigr)^2 = 2p_T^J  \Tau_J \biggl[1+ \ORd{\frac{m_J^2}{p_T^{J\,2}}}\biggr]
\,.\end{align}
As long as $\vec n_J$ is chosen along the direction of the total jet momentum the exact relation is $m_J^2 = (E_J + |\vec p_J|) \Tau_J/\cosh\eta_J$, which becomes $m_J^2 =2 p_T^J \Tau_J$ in the singular limit $m_J^2 \ll p_T^{J\,2}$.  Since $m_J \lesssim p_T^J R$ for narrow jets, the corrections in \eq{TauJ} are also power suppressed for regime 3, where $m_J/p_T^J \sim R \ll 1$. As shown below in \eq{cross-section-Jacobian}, this means that the singular part of the differential cross section for $m_J$ and $\Tau_J$ are simply related by a Jacobian.

Additional jets are vetoed with a measurement in the beam region. For simplicity, we discuss in this section first the case of a global jet veto using the generalized beam thrust~\cite{Stewart:2009yx} observable 
\begin{align} \label{eq:TauB}
  \Tau_B = \sum_{i \notin \text{jet}} p_{Ti}  f_B(\eta_i) 
\,,\end{align}
where $\eta_i$ and $p_{Ti}$ are the pseudorapidity and transverse momentum of the $i$-th particle outside the identified jet. We assume that $f_B(\eta) \to e^{-|\eta|}$ for $\eta \to \pm \infty$, which includes beam thrust and the C-parameter measure (discussed e.g.~in ref.~\cite{Gangal:2014qda}), with
\begin{align}\label{eq:vetofct}
f_B^{\tau}(\eta) = e^{-|\eta|}  \quad {\rm and} \quad  f_B^C(\eta) = \frac{1}{2\cosh \eta} 
\,,\end{align}
respectively.
The asymptotic behavior of $f_B(\eta)$ implies that the measurement is described by \SCETa, which contains collinear and soft modes at different invariant mass scales, and that the virtuality-dependent beam functions~\cite{Stewart:2010qs, Berger:2010xi, Gaunt:2014xga, Gaunt:2014cfa} can be used to describe the collinear initial-state radiation. In \sec{jetveto} we will discuss other types of jet vetoes, including a transverse-energy veto where collinear and soft modes are instead separated in rapidity and described by \SCETb, as well as corresponding jet-based vetoes that depend on a jet algorithm.

In the following we write the $1$-jet cross section with additional kinematic constraints $X$ (e.g.~in terms of bins in $p_T^J$ and $\eta_J$, and with cuts on the final color-singlet state $L$) as
\begin{align}\label{eq:dsigma_Nbody}
\frac{\df \sigma (X)}{\df \Tau_B \,\df \Tau_J} = \int\! \df \Phi \, \sum_{\kappa}\frac{\df \sigma (\Phi,\kappa)}{\df \Tau_B \,\df \Tau_J} \, X(\Phi)
\,.\end{align}
The sum over the partonic channels $\kappa = \{\kappa_a,\kappa_b;\kappa_J\}$ runs over all flavors of the colliding partons and the energetic parton initiating the jet.

We can write the full cross section in terms of the resummed leading power (``singular'') cross sections in SCET denoted by $\df \sigma_{1,2,3}$ in regimes 1, 2, and 3, and their respective power-suppressed (``nonsingular'') corrections. In each regime, we will present a factorization formula for the singular part of the cross section and give the parametric size of the associated nonsingular corrections. The singular cross sections can be easily rewritten to be differential in $m_J$ rather than $\Tau_J$ by taking into account a simple Jacobian factor,
\begin{align} \label{eq:cross-section-Jacobian}
\frac{\df \sigma_{1,2,3}(X)}{\df m_J}
= \frac{m_J}{p_T^J} \,\frac{\df \sigma_{1,2,3}(X)}{\df \Tau_J} \biggl|_{\Tau_J=\frac{m_J^2}{2p_T^J}}
\, .\end{align}
The nonsingular corrections are different for $m_J$ and $\Tau_J$ due to the power corrections indicated in \eq{TauJ}.

{\renewcommand{\arraystretch}{1.6}

\begin{table}[t!]
\begin{center}
\scalebox{1}{
\begin{tabular}{l || c | c | c}
\hline \hline
 & regime 1 &  regime 2 & regime 3
\\[-1.5ex]
 modes & $R \sim \Rzero\,,\, \Tau_J \ll p_T^J R^2$ & $R \ll \Rzero\,,\, \Tau_J \ll p_T^J R^2$  &  $R \ll \Rzero\,,\, \Tau_J \sim p_T^J R^2$
\\[0.5ex] \hline\hline
$n_B$-collinear & $\Bigl(\Tau_B,p_T^J,\sqrt{p_T^J \Tau_B}\Bigr)_{\!\!B}$  & $\Bigl(\Tau_B,p_T^J,\sqrt{p_T^J \Tau_B}\Bigr)_{\!\!B}$ & $\Bigl(\Tau_B,p_T^J,\sqrt{p_T^J \Tau_B}\Bigr)_{\!\!B}$
\\[0.5ex] \hline
$n_J$-collinear & $\Bigl(\Tau_J,p_T^J,\sqrt{p_T^J \Tau_J}\Bigr)_{\!\!J}$  & $\Bigl(\Tau_J,p_T^J,\sqrt{p_T^J \Tau_J}\Bigr)_{\!\!J}$  & $p_T^J (R^2,1,R)_J$
\\[0.5ex] \hline
soft & $ \Tau_B(1,1,1)$  & $\Tau_B(1,1,1)$  & $\Tau_B(1,1,1)$
\\[-1ex]
 & $ \Tau_J(1,1,1)$  &  &
\\ \hline
$n_J$-csoft &  & $\dfrac{\Tau_J}{R^2} (R^2,1,R)_J$  &
\\
 &   & $\Tau_B (R^2,1,R)_J$  & $\Tau_B (R^2,1,R)_J$
\\ \hline\hline
resummed logs & $\ln\dfrac{p_T^J}{\Tau_B}$, $\ln\dfrac{p_T^J}{\Tau_J}$ & $\ln\dfrac{p_T^J}{\Tau_B}$, $\ln\dfrac{p_T^J}{\Tau_J}$, $\ln R$ & $\ln\dfrac{p_T^J}{\Tau_B}$, $\ln R$
\\[1ex] \hline
(potential) NGLs & $\alpha_s^2 \ln^2 \dfrac{\Tau_B}{\Tau_J}$ & $\alpha_s^2 \ln^2 \dfrac{\Tau_B R^2}{\Tau_J}$ & $\alpha_s^2 \ln^2 \dfrac{\Tau_B }{p_T^J}$
\\[1ex] \hline \hline
\end{tabular}}
\end{center}
\caption{Summary of the EFT modes setup, the resummed logarithms and the potentially large nonglobal logarithms for the different regimes. For all regimes we take $\Tau_B\ll p_T^J$. By default, we consider the situation where the listed NGLs are not large logarithms in regimes 1 and 2. In a situation where these logarithms become large, the corresponding soft and $n_J$-csoft modes split into multiple modes, as indicated in \fig{modes}.}
\label{tab:regimes}
\end{table}
}

\subsection{Regime 1: Large-$R$ jet with $m_J \ll p_T^J R \sim p_T^J$}
\label{subsec:case1}

This regime, describing the case of a small jet mass for a jet with a wide opening angle $R \sim R_0$, was discussed in detail in Ref.~\cite{Jouttenus:2013hs}. The EFT modes are summarized on the left in table~\ref{tab:regimes} and \fig{modes}.
The collinear radiation carries the large jet momentum and its scaling is fixed by the jet mass measurement. In terms of lightcone coordinates along the jet axis,
\begin{align}
n_J\text{-collinear:}
\qquad p_{n_J}^\mu \sim \Bigl(\frac{m_J^2}{p_T^J},p_T^J,m_J\Bigr)_J
  \sim \Bigl(\Tau_J, p_T^J, \sqrt{p_T^J \Tau_J}\Bigr)_J
\,,\end{align}
Similarly, the scaling of the collinear initial-state radiation is fixed by the hard momentum $Q \sim p_T^J$ it carries and the measurement constraint from $\Tau_B$. In terms of light-cone coordinates along the beam axis,
\begin{align}
n_a \text{-collinear:} 
&\qquad
p_{n_a}^\mu \sim \Bigl(\Tau_B,p_T^J,\sqrt{p_T^J \Tau_B}\Bigr)_B
\,, \nn \\
n_b \text{-collinear:} 
&\qquad
p_{n_b}^\mu \sim \Bigl(p_T^J,\Tau_B,\sqrt{p_T^J \Tau_B}\Bigr)_B
\,.\end{align}
The soft radiation is isotropic and communicates between the collinear radiation along the beams and jet. Its momentum scaling is determined by the fact that it is constrained by either the $\Tau_J$ measurement in the jet region or the jet veto in the beam region,
\begin{align}\label{eq:soft_scaling}
\text{soft:} &\qquad  p_{s}^\mu\sim \Tau_J (1,1,1) 
  \qquad  ( \text{soft$_{(J)}$} )\, ,\nn \\
 &\qquad  p_{s}^\mu \sim \Tau_B (1,1,1) 
  \qquad (\text{soft$_{(B)}$})
\,,\end{align}
written in terms of any lightcone direction. (Sometimes these modes are called ultrasoft in the SCET literature.) In this regime nonglobal structure appears through functions of $\Tau_B/\Tau_J$, and to derive a factorization formula we must assume a power counting for $\Tau_J$ relative to $\Tau_B$. Phenomenologically the most important hierarchy is $\Tau_B\sim \Tau_J$ as it can be applied to a large region of parameter space, and hence we will focus on this case.  In this situation there is a single soft mode and the NGLs are not larger than other nonglobal contributions, all of which are fully captured by the soft function.
Large NGLs appear when $\Tau_B/\Tau_J\gg 1$ or $\Tau_B/\Tau_J\ll 1$, arising from the sensitivity to two parametrically different soft scales (which are conceptually more difficult than the case we treat). 

Going through the usual steps, where the hard scattering interaction is integrated out when matching onto \SCETa and the modes are subsequently decoupled in the Lagrangian, leads to the following factorization formula for the singular part of the cross section~\cite{Stewart:2009yx, Stewart:2010tn, Jouttenus:2013hs}
\begin{align} \label{eq:sigmaTau1}
\frac{\df \sigma_1 (\Phi,\kappa)}
 {\df \Tau_B \, \df\Tau_J}
&=
H_\kappa(\Phi, \mu)
\int\!\df s_a\, B_{\kappa_a}(s_a, x_a, \mu)
\int\!\df s_b\, B_{\kappa_b}(s_b, x_b, \mu)
\int\!\df s_J\, J_{\kappa_J}(s_J, \mu)
\nn \\* &\quad\times
S_\kappa \biggl(\Tau_J - \frac{s_J}{2p_T^J}, \Tau_B - \frac{s_a}{\omega_a}- \frac{s_b}{\omega_b},  \eta_J, R, \mu\biggr) \,,
\\
\frac{\df \sigma (\Phi,\kappa)}
{\df \Tau_B \, \df\Tau_J}
&=
\frac{\df \sigma_1 (\Phi,\kappa)}
{\df \Tau_B \, \df\Tau_J} \biggl[1 +\mathcal{O}\biggl(\frac{\Tau_J}{p_T^J},\frac{\Tau_B}{p_T^J}\biggr)\biggr] \nn
\,.\end{align}
For active-parton scattering this factorization formula does not include contributions from perturbative Glauber gluon exchange that start at ${\cal O}(\alpha_s^4)$~\cite{Gaunt:2014ska,Zeng:2015iba}.  These terms can be simply calculated and included using the Glauber operator framework of Ref.~\cite{Rothstein:2016bsq}, which will modify the structure of the product of beam functions.%
\footnote{For proton initial states this factorization formula also does not account for spectator forward scattering effects, since the Glauber Lagrangian of Ref.~\cite{Rothstein:2016bsq} has been neglected in the derivation of \eq{sigmaTau1}.}
The ${\cal O}(\Tau_{J}/p_T^J,\Tau_{B}/p_T^J)$ terms indicated on the last line are nonsingular corrections, which may be included with fixed-order perturbation theory or by connecting to a factorization formula in regime 4.

The hard function $H_\kappa$ in \eq{sigmaTau1} contains the short-distance matrix element for producing the nonhadronic $L$ plus a jet and depends on the hard kinematic phase space $\Phi$. The beam functions describe the process of extracting a parton out of the proton and the formation of an initial-state jet characterized by the scale $s_{a,b} \sim Q \Tau_B$. The inclusive jet function describes the invariant mass contribution $s_J\sim m_J^2$ of the final-state collinear radiation to the jet mass and is not sensitive to the jet boundary since $m_J \ll p_T^J R$. Finally, the soft function $S_\kappa$ captures the soft radiation effects and depends on the angles between the collinear directions (and thus the pseudorapidity of the jet $\eta_J$),
the jet boundary determined by the jet algorithm and jet radius $R$,
as well as the jet and beam measurements with the jet veto specified by $f_B(\eta)$ in \eq{TauB}. 
 
The factorization formula enables the resummation of the logarithms of $\Tau_J/p_T^J$ and $\Tau_B/p_T^J$ corresponding to the ratios between the hard, beam, and soft scales. Each function only involves a single parametric scale, corresponding to the typical virtuality of that mode. By evaluating each function at its natural scale and evolving them to a common scale $\mu$ using the RG evolution, the logarithms are resummed, i.e.,
\begin{align}\label{eq:evolution1}
H_\kappa(\Phi, \mu) & = U_{H,\kappa} (\Phi, \mu, \mu_H) \, H_\kappa (\Phi, \mu_H) \, ,  \\
B_{\kappa_B}(s,\mu) & = \int\! \df s'  \, U_{B,\kappa_B} (s-s' ,\mu, \mu_B)  \, B_{\kappa_B}(s',\mu_B)  \, ,\nn \\
 J_{\kappa_J}(s,\mu) & = \int\! \df s'   \, U_{J,\kappa_J} (s-s' ,\mu, \mu_J)\,  J_{\kappa_J}(s',\mu_J)  \, ,\nn \\
 S_{\kappa}(\ell_J,\ell_B,\eta_J,R,\mu) & = \int\! \df \ell_J' \, \df \ell_B'   \,U_{S,\kappa} (\ell_J-\ell_J',\ell_B-\ell_B',\eta_J,\mu, \mu_S) \, S_{\kappa}(\ell_J',\ell_B',\eta_J,R,\mu_S)  \, , \nn
\end{align}
where 
\begin{align}\label{eq:scales1}
\mu_H \sim  p_T^J \sim Q \, , \quad \mu_B \sim \sqrt{Q \Tau_B} \, , \quad \mu_J \sim \sqrt{p_T^J \Tau_J}\sim m_J \, , \quad \mu_S \sim \Tau_J \sim \Tau_B \, .
\end{align}
The evolution factors $U$ for the individual functions are the solutions of the renormalization group equations, which read e.g.~for the soft function
\begin{align}\label{eq:U_evolS}
\mu \frac{\df}{\df \mu} \,U_{S,\kappa} (\ell_J,\ell_B,\eta_J,\mu,\mu_S) & = \int \df \ell_J' \, \df\ell_B'  \,\gamma_{S}^{\kappa} (\ell_J-\ell_J',\ell_B-\ell_B',\eta_J,\mu) \, U_{S,\kappa}  (\ell_J',\ell_B',\eta_J,\mu,\mu_S)
\,.
\end{align}
The explicit expressions for the evolution factors and anomalous dimensions can be found in the appendix of Ref.~\cite{Jouttenus:2013hs}. Using \eq{evolution1} with the factorized cross section in \eq{sigmaTau1}, the logarithms $\ln (\mu_H/\mu_B)$, $\ln (\mu_H/\mu_J)$, $\ln (\mu_B/\mu_S)$ and $\ln (\mu_J/\mu_S)$ are resummed and the dependence on the final renormalization scale $\mu$ cancels exactly at any resummed order due to consistency of RG running.

By RGE consistency of the factorization formula the anomalous dimension for the soft function factorizes, as discussed in Refs.~\cite{Fleming:2007xt,Jouttenus:2013hs}. For the case considered here, this consistency gives
\begin{align}\label{eq:anomdim_softBJ1a}
\gamma_{S}^{\kappa} (\ell_J,\ell_B,\eta_J,\mu) = \gamma^{\kappa(J)}_{S} (\ell_J,\eta_J,\mu) \, \delta(\ell_B) +  \gamma^{\kappa(B)}_{S} (\ell_B,\eta_J,\mu) \, \delta(\ell_J)
  + \gamma^{\kappa(\delta)}_{S}(\eta_J,\mu) \, \delta(\ell_J) \delta(\ell_B)
\, .\end{align}
This uniquely assigns the $\ell_B$ and $\ell_J$-dependent cusp terms in the anomalous dimension to the beam and jet.
The remaining $\de(\ell_J) \de(\ell_B)$ noncusp terms can also be factorized, but the precise division requires more care, as discussed below. Together this yields
\begin{align}\label{eq:anomdim_softBJ1b}
\gamma_{S}^{\kappa} (\ell_J,\ell_B,\eta_J,\mu)
= \gamma^{\kappa}_{S^{(J)}} (\ell_J,\eta_J,R,\mu) \, \delta(\ell_B) +  \gamma^{\kappa}_{S^{(B)}} (\ell_B,\eta_J,R,\mu) \, \delta(\ell_J)
\, .\end{align}
Here $\gamma^{\kappa}_{S^{(B)}}$ and $\gamma^{\kappa}_{S^{(J)}}$ each depend on the jet boundary and jet radius $R$, but this dependence cancels in the sum.\footnote{This $R$ dependence becomes even easier to understand when we take $R\ll 1$ in regime 2.} Solving the RGE with these factorized anomalous dimensions allows us to factorize the soft function together with its evolution as
\begin{align} \label{eq:US_refact}
  & \int\!\! \df\ell_J'\, \df\ell_B'\:  
   U_{S,\kappa} (\ell_J\!-\!\ell_J',\ell_B\!-\!\ell_B',\eta_J,\mu,\mu_S) \, S_{\kappa}(\ell_J',\ell_B',\eta_J,R,\mu_S) 
   \\
  & \quad =\int\!\! \df\ell_J'\, \df\ell_B'\: 
  U_{S,\kappa}^{(B)}\big(\ell_B\!-\!\ell_B',\eta_J,R,\mu,\mu_S^{(B)}\big)\,
  U_{S,\kappa}^{(J)}\big(\ell_J\!-\!\ell_J',\eta_J,R,\mu,\mu_S^{(J)}\big) 
  \nn\\
  & \quad \qquad 
  \times  S_{\kappa}\big(\ell_J',\ell_B',\eta_J,R,\mu_S^{(J)},\mu_S^{(B)}\big) 
  \,. \nn
\end{align}
Here we have decomposed the full soft function as
\begin{align} \label{eq:S_refact}
S_\kappa (\ell_J, \ell_B,  \eta_J, R, \mu_S^{(J)}, \mu_S^{(B)})
 &= \int\!\! \df\ell_J'  \, \df\ell_B'\:
 S^{(J)}_{\kappa} (\ell_J -\ell_J', \eta_J, R, \mu_S^{(J)} \big) 
\,S^{(B)}_{\kappa} (\ell_B-\ell_B', \eta_J, R, \mu_S^{(B)} \big) 
 \nonumber \\
& \quad \times \Big[\delta(\ell_J')\,\delta(\ell_B')  + 
  S^{(\rm NG)}_{\kappa} (\ell_J', \ell_B',  \eta_J, R) \Big]
\, .\end{align}
Equations (\ref{eq:US_refact}) and (\ref{eq:S_refact}) factorize the $\mu$-dependence associated with the beam and jet region for the evolution  and the associated low-scale boundary conditions to all orders in perturbation theory, thus allowing distinct scale choices to be made for $\mu_S^{(J)}$ and $\mu_S^{(B)}$.  At one loop, the terms $S^{(J)}_{\kappa}$ and $S^{(B)}_{\kappa}$ describe a single emission inside and outside the jet region at the scale $\mu^{(J)}_S \sim \Tau_J$ and $\mu_S^{(B)} \sim \Tau_B$, respectively, with $S^{(J)}_{\kappa}$ being analogous to the regional soft function in ref.~\cite{Chien:2012ur} (where it was applied to cone jets). This fixes the ambiguity in splitting the noncusp one-loop anomalous dimension $\gamma_{S}^{\kappa(\delta)}$ into distinct contributions to $\gamma^{\kappa}_{S^{(J)}}$ and $\gamma^{\kappa}_{S^{(B)}}$ in \eq{anomdim_softBJ1b}. Here $\gamma^{\kappa}_{S^{(J)}}$ and $\gamma^{\kappa}_{S^{(B)}}$ can be given in terms of $R$-dependent  integrals for generic jet algorithms.\footnote{For cone jets an analytic expression for $\gamma^{\kappa}_{S^{(J)}}$ at one loop was found in ref.~\cite{Chien:2012ur}. For anti-$k_T$ jets, $\gamma^{\kappa}_{S^{(J)}}$ can be evaluated analytically in an expansion in terms of $R$, which has been done at one-loop up to $\mathcal{O}(R^2)$ in ref.~\cite{Liu:2014oog} for $pp \to $ dijets.}

Starting at two loops, there are correlated real emissions into both the jet and beam region, which are thus constrained by both the jet and beam measurements, that lead to nonglobal structures.  In \eq{S_refact} these are absorbed into the $\mu$-independent factor $S^{(\rm NG)}_{\kappa}$. At this order, the decomposition in \eq{S_refact} becomes ambiguous without additional input, since correlated emissions must be considered simultaneously with single region emissions  when defining $S^{(J)}_{\kappa}$ and $S^{(B)}_{\kappa}$, which is known for the double hemisphere case~\cite{Kelley:2011ng,Hornig:2011iu}. In regime 2 for $R \ll 1$, we can use symmetry arguments to constrain the small-$R$ terms of $S^{(J)}_{\kappa}$ and $S^{(B)}_{\kappa}$, see \eqs{SR_hemi}{cons12b}, which allows us to fix most of this ambiguity.  

Some of the corrections in $S^{(\rm NG)}_{\kappa}$ would become large nonglobal logarithms $\ln(\Tau_B/\Tau_J)$ if we were in the alternative situations where $\Tau_B \gg \Tau_J$ or $\Tau_J \gg \Tau_B$, and the resummation for these cases requires techniques other than the renormalization group evolution described above. The refactorization in \eqs{US_refact}{S_refact} is essential to avoid introducing ``fake" NGLs $\sim\alpha_s^n \ln^{2n}(\mu_S^{(B)}/\mu_S^{(J)})$ at leading logarithmic order~\cite{Jouttenus:2013hs}.
After this refactorization, the canonical relationships between the scales in regime 1 are given by
\begin{align} \label{eq:regime-1-canonical}
\mu_H \, \mu_{S}^{(B)} \simeq \mu_B^2\,,\qquad\qquad
\mu_H \, \mu_{S}^{(J)}  \simeq \mu_J^2\,.
\end{align}
These relations together with the scaling relations $\mu_H\simeq p_T^J$, $\mu_S^{(B)}\simeq \Tau_B$ and $\mu_S^{(J)}\simeq \Tau_J$, determine the full canonical scaling which allows all large logarithms to be summed in regime 1, at any desired order in perturbation theory.

The factorization in \eq{sigmaTau1} is limited to large jet radii $R \sim 1$, such that $R$ does not introduce additional scales or modes. In many LHC measurements smaller values of $R$ are employed, leading to a hierarchy of scales within the soft sector and associated large double logarithms of $R$ in the soft function $S_\kappa$. We will discuss how to treat these next.

\subsection{Regime 2: Small-$R$ jet in the region $m_J \ll p_T^J R \ll p_T^J$}
\label{subsec:case2}

For narrow jets, the jet radius introduces an additional hierarchy $R \ll R_0$. The mode setup for the associated EFT, which is a version of \SCETp, is shown in the middle in table~\ref{tab:regimes} and \fig{modes}. It is closely related to the one discussed in ref.~\cite{Chien:2015cka}, which considers it for cone jets at $e^+ e^-$ colliders.

For $\Tau_J \ll p_T^J R^2$ the $n_J$-collinear radiation has a resolution angle $|\vec{p}_{\perp}|/\bar{n}_J \sdt {p} \sim m_J/p_T^J \sim (\Tau_J/p_T^J)^{1/2} \ll R$ and is thus still collimated enough to be insensitive to jet boundary effects. The collinear radiation along the beam directions is still determined by the measurement of $\Tau_B$. Hence, the collinear modes are the same as for regime 1,
\begin{align}
n_J\text{-collinear:}
&\qquad p_{n_J}^\mu \sim \Bigl(\frac{m_J^2}{p_T^J},p_T^J,m_J\Bigr)_J
  \sim \Bigl(\Tau_J, p_T^J, \sqrt{p_T^J \Tau_J}\Bigr)_J
\,,\nn \\
n_a \text{-collinear:}
&\qquad
p_{n_a}^\mu \sim \Bigl(\Tau_B,p_T^J,\sqrt{p_T^J \Tau_B}\Bigr)_B
\,, \nn \\
n_b \text{-collinear:}
&\qquad
p_{n_b}^\mu \sim \Bigl(p_T^J,\Tau_B,\sqrt{p_T^J \Tau_B}\Bigr)_B
\,.\end{align}
Wide-angle soft radiation is now only constrained by the $\Tau_B$ measurement,
\begin{equation}\label{eq:soft_2}
\text{soft:}
\qquad  p_{s}^\mu \sim \Tau_B (1,1,1) 
\,.\end{equation}
It cannot resolve the narrow jet and is thus not constrained by the jet measurement. Therefore, to have a complete description of the infrared structure of QCD for this regime, additional modes are required which have the relative scaling $\sim (R^2,1,R)_J$. The scaling of these modes is uniquely fixed by the requirement that they are restricted by the jet or beam measurement, respectively,
\begin{align}\label{eq:csoft_scaling}
& n_J\text{-collinear-soft:}
\qquad
p_{cs}^\mu \sim \frac{\Tau_J}{R^2} (R^2,1,R)_J \sim \Bigl(\Tau_J, \frac{\Tau_J}{R^2}, \frac{\Tau_J}{R}\Bigr)_J  
\, ,\\
& n_J\text{-soft-collinear:}
\qquad p_{sc}^\mu \sim 
\Tau_B (R^2,1,R)_J \label{eq:scoll_scaling}
\,.\end{align}
This nomenclature for the modes follows refs.~\cite{Chien:2015cka}.
To derive a factorization formula we must choose their parametric relation to be either $\Tau_B \gg \Tau_J/R^2$, $\Tau_B \sim \Tau_J/R^2$, or $\Tau_B \ll \Tau_J/R^2$.  We take $\Tau_B \sim \Tau_J/R^2$, in which case the scalings in \eq{csoft_scaling} become degenerate, so there is only a single mode describing these momenta. We will refer to this common intermediate mode as \csoft%
\footnotemark
\footnotetext{We denote the associated theory here \SCETp. Its close connection to the original \SCETp setup for nearby jets (``ninja'') in ref.~\cite{Bauer:2011uc} becomes obvious by boosting to the frame where the jet region becomes a full hemisphere. In this frame, the soft mode in \eq{soft_2} becomes the ninja csoft mode and the csoft mode in \eq{csoft_scaling2} becomes the overall soft mode.}
\begin{align}\label{eq:csoft_scaling2}
 n_J\text{-csoft:}
 \qquad
 p^\mu \sim \frac{\Tau_J}{R^2} (R^2,1,R)_J 
 \sim  \Tau_B (R^2,1,R)_J  
 \,.
\end{align}
If on the other hand their energies differ parametrically, large NGLs of the ratio $\Tau_B R^2/\Tau_J$ arise, in analogy to the situation for the ratio $\Tau_B/\Tau_J$ for soft radiation in \subsec{case1}.

We remark that different hierarchies between the (wide-angle) soft scale $\Tau_B$ and the jet scale $(p_T^J \Tau_J)^{1/2}$ are possible. In the following no specific relation between these scales needs to be assumed to obtain the factorization formula. In particular, the jet axis is determined only from the recoil-free measurement inside the jet region, which avoids nontrivial convolutions between the perpendicular momentum components of the $n_J$-collinear and soft modes~\cite{Larkoski:2014uqa} (which appear e.g.~when measuring jet broadening with the thrust axis~\cite{Dokshitzer:1998kz}). 

Going through the factorization analysis in \SCETp leads to 
\begin{align} \label{eq:sigmaTau2}
\frac{\df \sigma_2 (\Phi,\kappa)}
 {\df \Tau_B \, \df\Tau_J}
&=
H_\kappa(\Phi, \mu)
\int\!\df s_a\, B_{\kappa_a}(s_a, x_a, \mu)
\int\!\df s_b\, B_{\kappa_b}(s_b, x_b, \mu)
\int\!\df s_J\, J_{\kappa_J}(s_J, \mu)
\nn \\ &\quad\times
\int \df k_J \,\df k_B\, S_{R,\kappa_J}(k_J,k_B,\mu)\,
\delta\Bigl(\Tau_J - \frac{s_J}{2p_T^J }- \frac{R\,k_J}{2 } \Bigr)
\nn \\
&\quad\times  S_{B,\kappa} \biggl(\Tau_B - \frac{s_a}{\omega_a}- \frac{s_b}{\omega_b}- \frac{f_B(\eta_J)k_B}{R}, \eta_J, \mu\biggr)
\, ,\\
\frac{\df \sigma (\Phi,\kappa)}
{\df \Tau_B \, \df\Tau_J}
&= \frac{\df \sigma_2 (\Phi,\kappa)}
{\df \Tau_B \, \df\Tau_J}
\biggl[1+\mathcal{O}\biggl(\frac{\Tau_B}{p_T^J},\frac{\Tau_J}{p_T^J R^2},R^2\biggr)\biggr] \,. \nn
\end{align}
The ${\cal O}(\Tau_{B}/p_T^J,\Tau_{J}/(p_T^JR^2),R^2)$ terms indicated on the last line are nonsingular corrections, which can be included with fixed-order perturbation theory or by connecting to the factorization formula in regimes 1 or 3. Once again we neglect Glauber interactions here.

Deriving the factorization in \eq{sigmaTau2} involves a matching onto \SCETp and the decoupling of modes in the Lagrangian. The structure of the relevant operators in \SCETp can be obtained by applying the BPS decoupling~\cite{Bauer:2001yt,Bauer:2002aj}, either by matching onto \SCETp in two steps as was done in ref.~\cite{Bauer:2011uc},%
\footnote{It is convenient to perform these decoupling steps in the boosted ninja frame where the jet region becomes a hemisphere, see footnote 6. Following ref.~\cite{Bauer:2011uc} one then has to first decouple the soft modes in the ninja frame (corresponding to the csoft modes in the lab frame) from the collinear modes before decoupling the csoft modes in the ninja frame (corresponding to the soft modes in the lab frame) from the collinear ones. This also makes it clear which zero-bin subtractions~\cite{Manohar:2006nz} arise between these modes.} or alternatively by matching in one step and using collinear, csoft, and soft gauge invariance and tree-level calculations as in ref.~\cite{Procura:2014cba}.

In addition, \eq{sigmaTau2} requires the factorization of the measurement into contributions from the individual modes,
\begin{align}\label{eq:measurement_fact}
\Tau_J &= \Tau_J^{(n_J)} + \Tau_J^{(cs)} =
\cosh \eta_J
\bigl(n_J\sdt p^{(n_J)}  + n_J\sdt p^{(cs)}_{\rm in}\bigr)= \frac{s_J}{2p_T^J } + \frac{R\,k_J}{2 }
\nn \,, \\
  \Tau_B &= \Tau_B^{(n_a)} +  \Tau_B^{(n_b)} +  \Tau_B^{(cs)} +  \Tau_B^{({s})} 
  = n_{a}\sdt p^{(n_a)} + n_b\sdt p^{(n_b)}
    + \frac{f_B(\eta_J)}{2\cosh \eta_J}\, \bn \sdt p^{(cs)}_{\rm out}
   +  \Tau_B^{({s})}  \nn \\
   & = \frac{s_a}{\omega_a} + \frac{s_b}{\omega_b} + \frac{f_B(\eta_J)k_B}{R} +  \Tau_B^{({s})} 
\,.\end{align}
Here, $p^{(n_a)}$, $p^{(n_b)}$, $p^{(n_J)}$ denote the momentum of the collinear radiation in the $n_a$, $n_b$, and $n_J$ directions, $p^{(cs)}_{\rm in}$, $p^{(cs)}_{\rm out}$ denote the csoft momentum inside or outside the jet, and $\Tau_B^{(s)}$ is the contribution of soft radiation to the jet veto $\Tau_B$. For the csoft modes we used that $f_B(\eta) = f_B (\eta_J) + \mathcal{O}(R)$ and thus $\Tau_B^{(cs)}=f_B(\eta_J) \, p_T^{(cs)} = f_B(\eta_J)\bar{n} \sdt p^{(cs)}/(2 \cosh \eta_J)  $.\footnote{For anti-k$_T$ yielding a circle in the $\eta$-$\phi$ plane centered around $\eta_J$ the corrections from the expansion of $f_B(\eta)$ give a vanishing contribution at $\mathcal{O}(R)$. For general jet algorithms the relative deviation from the circular shape is of $\mathcal{O}(R)$ so that the associated corrections from the expansion of $f_B(\eta)$ give also only $\mathcal{O}(R^2)$ suppressed terms in the cross section.} 

Compared to \eq{sigmaTau1}, the same hard, beam, and jet functions appear in \eq{sigmaTau2}, while the soft function has now been factorized into two functions $S_{B,\kappa}$ and $S_{R,\kappa_J}$. The soft function $S_{B,\kappa}$ encodes the interactions of the wide-angle soft modes. It contains three soft Wilson lines corresponding to the partons participating in the hard collision, but only contributes to the measurement of $\Tau_B$ as the associated soft modes no longer resolve the jet. The \csoft function $S_{R,\kappa_J}$ consists of two back-to-back \csoft Wilson lines in the representation of the parton that initiates the jet, and contributes to both the $\Tau_B$ and $\Tau_J$ measurements as \csoft modes resolve the jet boundary. For convenience, we have chosen the arguments $k_B$ and $k_J$ of the \csoft function as
\begin{align}
k_J = \frac{2}{R} \Tau_J^{(cs)} = \sum_{i \in {\rm jet}}  \frac{2\cosh \eta_J}{R} \,n_J \sdt k_i
\, , \qquad
k_B = \frac{R}{f_B(\eta_J)} \Tau_B^{(cs)} = \sum_{i \notin {\rm jet}} \frac{R}{2 \cosh \eta_J}   \,\bar{n}_J \sdt k_i
\end{align}
to scale out the dependence on the size of the jets, which allows us to identify the \csoft function with the well-known double hemisphere soft function, see \eq{equiv}. This will be discussed more extensively in \sec{softandjet}, where we also give the precise definitions and the one-loop expressions of the soft functions.

The \csoft and soft function are RG evolved via
\begin{align}\label{eq:evolution2}
 S_{R,\kappa_J}(k_J,k_B,\mu) & = \int \df k_J' \, \df k_B'   \,U_{S_R,\kappa_J} (k_J-k_J',k_B-k_B',\mu, \mu_{S_R}) \, S_{R,\kappa_J}(k_J',k_B',\mu_{S_R})  \, , \nn \\
 S_{B,\kappa}(\ell_B,\eta_J,\mu) & = \int  \df \ell_B'   \,U_{S_B,\kappa} (\ell_B-\ell_B',\eta_J,\mu, \mu_{S_B}) \, S_{B,\kappa}(\ell_B',\eta_J,\mu_{S_B})  \, ,
\end{align}
from their natural scales 
\begin{align} \label{eq:regime-2-natural}
\mu_{S_B} \sim \Tau_B \, , \quad  \mu_{S_R} \sim \Tau_B R \sim \frac{\Tau_J}{R}\, .
\end{align}
We give the anomalous dimensions for $S_{B,\kappa}$ derived from RG consistency in \app{anomdim1}. The solution for the evolution factors are in direct analogy to the well-known ones appearing in \eq{evolution1}. Compared to \eq{scales1} for $R\sim R_0$ there is in total one additional evolution factor allowing for the resummation of $\ln (\mu_{S_R}/\mu_{S_B}) \sim \ln R$.

As in \eq{anomdim_softBJ1b} for regime 1, it is convenient to refactorize the \csoft function to avoid spurious nonglobal Sudakov logarithms involving $\ln(k_B/k_J)\sim\ln(\Tau_B R^2/\Tau_J)$. This is achieved by factorizing the anomalous dimension for this hemisphere \csoft function
\begin{align}\label{eq:anomdim_softBJ2}
\gamma_{S_R}^{\kappa_J} (k_J,k_B,\mu) = \gamma^{\kappa_J}_{S_R^{(J)}} (k_J,\mu) \, \delta(k_B) +  \gamma^{\kappa_J}_{S_R^{(B)}} (k_B,\mu) \, \delta(k_J)
\, \end{align}
with $\gamma^{\kappa_J}_{S_R^{(J)}} (k,\mu)=\gamma^{\kappa_J}_{S_R^{(B)}} (k,\mu) \equiv \gamma^{\kappa_J}_{\rm hemi} (k,\mu)$. This allows us to factorize  its evolution as
\begin{align} \label{eq:Urefact}
  & \int\! \df k_J'\, \df k_B'\:  
   U_{S_R,\kappa_J} ( k_J\!-\! k_J', k_B\!-\! k_B',\mu,\mu_{S_R})\,
   S_{R,\kappa_J}( k_J', k_B',\mu_{S_R}) 
   \\
  & \quad =\int\! \df k_J'\, \df k_B'\: 
  U_{S_R,\kappa_J}^{(B)}\big( k_B\!-\! k_B',\mu,\mu_{S_R}^{(B)}\big)\,
  U_{S_R,\kappa_J}^{(J)}\big( k_J\!-\! k_J',\mu,\mu_{S_R}^{(J)}\big) 
\,S_{R, \kappa_J}\big( k_J', k_B',\mu_{S_R}^{(J)},\mu_{S_R}^{(B)}\big) 
  \,. \nn
\end{align}
Here the \csoft function $S_{R,\kappa_J}$ contains two scales and can be written as
\begin{align} \label{eq:refact}
S_{R,\kappa_J}(k_J,k_B,\mu_{S_R}^{(J)},\mu_{S_R}^{(B)})
 &= \int\! \df k_J' \,\df k_B'\:
 S^{(J)}_{R,\kappa_J} (k_J -k_J',\mu_{S_R}^{(J)}) \,
 S^{(B)}_{R,\kappa_J} (k_B-k_B',\mu_{S_R}^{(B)}) 
 \nonumber \\
& \quad \times \Big[\delta(k_J')\,\delta(k_B')  + 
  S^{(\rm NG)}_{R,\kappa_J} (k_J', k_B') \Big]
\,.\end{align}
Equations (\ref{eq:Urefact}) and (\ref{eq:refact}) allow us to choose two different \csoft scales $\mu_{S_R}^{(B)}$ and $\mu_{S_R}^{(J)}$ for the contributions inside the beam and jet region, respectively.
Here $S_{R,\kappa_J}^{(J)}$ mainly describes the collinear-soft radiation at the scale $\mu_{S_R}^{(J)}\sim k_J$, and $S_{R,\kappa_J}^{(B)}$ mainly describes the soft-collinear radiation at the scale $\mu_{S_R}^{(B)} \sim k_B$.
Due to the symmetric nature of the double hemisphere \csoft function $S_{R,\kappa_J}$ it is natural to define these factors to be equal, 
\begin{align}\label{eq:SR_hemi}
  S^{(B)}_{R,\kappa_J}(k,\mu)=S^{(J)}_{R,\kappa_J}(k,\mu)
\, .\end{align}
The contribution $S^{(\text{NG})}_{R,\kappa_J}$ in \eq{refact}  captures nonglobal correlations, and starts at two loops where it contains double and single logarithms as well as nonlogarithmic terms, computed in refs.~\cite{Kelley:2011ng,Hornig:2011iu}. Starting at two loops, the function in \eq{SR_hemi} is a priori not well defined and depends on which $\mu$-independent terms are kept in $S^{(\text{NG})}_{R,\kappa_J}$. One proposal for the decomposition of the double hemisphere soft function to all orders in perturbation theory leading to \eq{refact}  was discussed in ref.~\cite{Hornig:2011iu}.

Some of the corrections in $S^{(\text{NG})}_{R,\kappa_J}$ would become large nonglobal logarithms $\ln(k_B/k_J)\sim \ln(\Tau_B R^2/ \Tau_J)$  if we were in the alternate scenarios where  $k_B \gg k_J$ or $k_J \gg k_B$. Just as for \eq{S_refact}, the factorization of scales in \eq{refact} is essential to avoid introducing ``fake" NGLs at leading logarithmic order. After this refactorization, the canonical relationships between the scales in this region are given by
\begin{align} \label{eq:regime-2-canonical}
\mu_H \, \mu_{S_B} \simeq \mu_B^2\,,\qquad
\mu_H \, \mu_{S_R}^{(B)} \frac{f_B(\eta_J)}{R} \simeq \mu_B^2\,,\qquad
\mu_H \, \mu_{S_R}^{(J)} \frac{R}{2} \simeq \mu_J^2
\,,\end{align}
which together with the scale choices
\begin{equation}
\mu_H \simeq p_T^J
\,,\qquad
\mu_{S_B} \simeq \Tau_B
\,,\qquad
\mu_{S_R}^{(J)} \simeq \frac{2}{R}\,\Tau_J
\,,\end{equation}
determine the full canonical scaling, implying e.g.~$\mu_{S_R}^{(B)} \simeq R\,\Tau_B/f_B(\eta_J)$. This allows for the resummation of all large logarithms in regime 2.

The NGLs become unavoidable in the region where $\Tau_B R^2 \ll \Tau_J$. This is the hierarchy explicitly discussed in Ref.~\cite{Chien:2015cka}, which also does not attempt to resum NGLs. The NGLs arise because the soft-collinear radiation resolves each individual collinear-soft emission, obstructing a simple factorization approach. In particular, each real collinear-soft emission requires an additional soft-collinear Wilson line to describe its interactions with the soft-collinear radiation. The NGLs in the double hemisphere soft function are well-studied and various new approaches systematically capturing their dominant effects have been recently explored~\cite{Hatta:2013iba, Larkoski:2015zka, Caron-Huot:2015bja, Neill:2015nya, Becher:2016mmh}, which can directly be applied to our context due to the equivalence between our \csoft function and the double hemisphere soft function.

\subsection{Regime 3: Small-$R$ jet in the region $m_J \sim p^J_T R \ll p_T^J$}
\label{subsec:case3}

Next we discuss the jet mass spectrum of a narrow jet for $\Tau_J \sim p^J_T R^2$, corresponding to the far tail of the jet mass spectrum. The relevant mode setup in \SCETp is shown on the right in table~\ref{tab:regimes} and \fig{modes}.
The beam-collinear and wide-angle soft modes are as in regime 2 only constrained by the $\Tau_B$ measurement,
\begin{align}
n_a \text{-collinear:}
&\qquad
p_{n_a}^\mu \sim \Bigl(\Tau_B,p_T^J,\sqrt{p_T^J \Tau_B}\Bigr)_B
\,, \nn \\
n_b \text{-collinear:}
&\qquad
p_{n_b}^\mu \sim \Bigl(p_T^J,\Tau_B,\sqrt{p_T^J \Tau_B}\Bigr)_B
\,, \nn \\
\text{soft:}
&\qquad  p_{s}^\mu \sim \Tau_B (1,1,1) 
\,.\end{align}
The collinear radiation in the jet now resolves the jet boundary, since its momentum scales as
\begin{align}
n_J\text{-collinear:}
\qquad
p^\mu_{n_J}\sim \Bigl(\Tau_J,p_T^J,\sqrt{p_T^J \Tau_J}\Bigr)_J \sim p_T^J (R^2,1,R)_J \sim \frac{\Tau_J}{R^2}(R^2,1,R)_J
\,,\end{align}
implying that the collinear-soft mode in \eq{csoft_scaling} cannot be distinguished from the collinear mode anymore, and the two become degenerate. As in \subsec{case2}, the wide-angle soft radiation does not resolve the narrow jet, such that a soft-collinear mode related to the beam measurement with the scaling in \eq{scoll_scaling} is still present,
\begin{align}
& n_J\text{-soft-collinear:}
\qquad p_{sc}^\mu \sim
\Tau_B (R^2,1,R)_J
\,.\end{align}
Assuming a jet veto with $\Tau_B \ll p_T^J \sim Q$ this mode has a parametrically different energy compared to the $n_J$-collinear mode but the same angular resolution, which makes the appearance of large NGLs of $\Tau_B/p_T^J$ unavoidable.\footnote{Removing the jet veto, i.e.~$\Tau_B \sim p_T^J$, large NGLs do not appear in this regime, but this would give rise to large NGLs for $\Tau_J \ll p^J_T R^2$. For the production of a massive boson with a soft jet the relation $\Tau_B \sim p_T^J \ll Q$ could be satisfied, but this regime is of limited relevance for jet mass measurements and presents challenges of its own.}  Completely disentangling the mode fluctuations at the different scales thus requires one to marginalize over all configurations of $n_J$-collinear emissions (which can individually be resolved by a proper low-energy measurement~\cite{Larkoski:2015zka}) each leading to soft-collinear matrix elements involving individually a different number of Wilson lines with different directions, see for example~\cite{Balitsky:1995ub,Weigert:2003mm,Hatta:2013iba,Caron-Huot:2015bja,Larkoski:2015zka,Becher:2015hka,Neill:2015nya}.

Here we do not attempt to entirely carry out this procedure, but instead only disentangle the corrections between the hard, beam-collinear, wide-angle soft, and a global $n_J$-collinear sector (which is not fully factorized). For this case the cross section can be written in a factorized form as
\begin{align} \label{eq:sigmaTau3}
\frac{\df \sigma_3 (\Phi,\kappa)}
 {\df \Tau_B \, \df\Tau_J}
&=  H_\kappa(\Phi, \mu) 
\int\!\df s_a\, B_{\kappa_a}(s_a, x_a, \mu)
\int\!\df s_b\, B_{\kappa_b}(s_b, x_b, \mu) \int \df s_J \, \delta\Bigl(\Tau_J - \frac{s_J}{2p_T^J}\Bigr)
\nn \\ &
\quad \times  \int\df k_B \, \mathcal{J}_{\kappa_J}(s_J, p_T^J R, k_B ,\mu) \, S_{B,\kappa} \biggl(\Tau_B - \frac{s_a}{\omega_a}- \frac{s_b}{\omega_b}- \frac{f_B(\eta_J)k_B}{R}, \eta_J, \mu\biggr) \, ,\nn \\
\frac{\df \sigma (\Phi,\kappa)}
{\df \Tau_B \, \df\Tau_J}
&=\frac{\df \sigma_3 (\Phi,\kappa)}
{\df \Tau_B \, \df\Tau_J}
\biggl[1+\mathcal{O}\biggl(\frac{\Tau_B}{p_T^J},R^2\biggr)\biggr]
\, .\end{align}
The ${\cal O}(\Tau_{B}/p_T^J,R^2)$ terms indicated on the last line are nonsingular corrections, which can be included with fixed-order perturbation theory. The hard function $H_\kappa$, beam functions $B_{\kappa_{a,b}}$, and soft function $S_{B,\kappa}$ are the same as in \eq{sigmaTau2} for regime 2. The collinear function $\mathcal{J}_{\kappa_J}$ encodes the interactions of both soft-collinear and collinear modes.
It depends both on the jet invariant mass $s_J$ and the scale $p_T^J R$, which reflects the sensitivity to the jet boundary, and also contributes to the measurement of $\Tau_B$.

Without any additional refactorization, the collinear function $\mathcal{J}_{\kappa_J}$ contains large unresummed Sudakov double logarithms $\sim \alpha_s^n \ln^{2n}(p_T^J R/k_B)\sim \alpha_s^n \ln^{2n}(p_T^J/\Tau_B)$. To resum the leading double logarithms, we can decompose it as
\begin{align}\label{eq:refactJ}
\mathcal{J}_{\kappa_J}(s_J, p_T^J R, k_B,\mu) 
 &= \sum_n J_R^{(n)} \otimes S_R^{(B,n)}
 \nn\\
 &= \int \df s_J' \,\df k_B' \, J_{R,\kappa_J}(s_J-s_J', p_T^J R, \mu) \,S^{(B)}_{R,\kappa_J}(k_B-k_B',\mu) \nn \\
 & \quad \times  \Bigl[\delta(k_B') \,\delta(s_J')+ \mathcal{J}^{\rm (NG)}_{\kappa_J}(s_J', p_T^J R, k_B')\Bigr]
\, .\end{align}
The sum over $n$ in the first equality indicates a dressed parton expansion like in ref.~\cite{Larkoski:2015zka} (with different soft-collinear matrix elements $S_R^{(B,n)}$ for a different number of resolved collinear emissions and associated directions) and the factor $J_{R,\kappa_J} S_{R,\kappa_J}^{(B)}$ contains the $n=0$ term in this expansion. 
The jet function $J_{R,\kappa_J}$ mainly describes corrections from the energetic $n_J$-collinear modes and depends on the details of the jet algorithm. These types of jet functions were introduced and calculated at one loop for cone jets and the $k_T$ family of jet algorithms in ref.~\cite{Ellis:2010rwa}. We give the one-loop results for the latter explicitly in \subsec{jet}.   The function $S^{(B)}_R$ can be taken to be the same function as in \eq{refact}, and mainly describes corrections from soft-collinear modes. The $\mu$-dependence factorizes between $J_{R,\kappa_J}$  and $S^{(B)}_R $ allowing for a separate evolution of these functions,
\begin{align}\label{eq:evolution3}
 S^{(B)}_{R,\kappa_J}(k_B,\mu) & = \int \df k_B'   \,U^{(B)}_{S_R,\kappa_J} (k_B-k_B',\mu, \mu_{S_R}^{(B)}) \, S^{(B)}_{R,\kappa_J}(k_B',\mu_{S_R}^{(B)})  \, , \nn \\
  J_{R,\kappa_J}(s_J, p_T^J R, \mu) & = U_{J_R,\kappa_R} (p_T^J R,\mu, \mu_{J_R}) \, J_{R,\kappa_J}(s_J, p_T^J R, \mu_{J_R})  \,.
\end{align}
We derive the form of the anomalous dimensions from RG consistency in \app{anomdim1}. The canonical scales are given by
\begin{align}
\mu_{J_R}  \sim m_J \sim (p_T^J \Tau_J)^{1/2} \sim p_T^J R \,,\quad\qquad \mu_{S_R}^{(B)} \sim \Tau_B R 
\,.\end{align}

Note that the evolution of the jet function $J_R$ is local, i.e.~does not involve a convolution, and is identical to the one for the ``unmeasured" jet function~\cite{Chien:2015cka}. Compared to a single evolution of $\mathcal{J}$ the two separate evolutions in \eq{evolution3} resum logarithms $\ln(\mu_{J_R}/\mu_{S_R}^{(B)})\sim \ln (p_T^J/\Tau_B)$ arising from collinear and soft-collinear emissions which are uncorrelated between these two, including in particular the Sudakov double logarithms. However, starting at $\mathcal{O}(\alpha_s^2)$ there are also NGLs of the form $\alpha_s^n \ln^n(p_T^J/\Tau_B)$ in the nonglobal correction $\mathcal{J}^{\rm (NG)}_{\kappa_J}$. Depending on the desired accuracy they may be treated as fixed-order corrections (multiplying the overall evolution factors) as indicated in \eq{refactJ} or (partially) summed using more steps in a dressed parton expansion in close analogy to ref.~\cite{Larkoski:2015zka}. In fact, the leading NGLs relevant for NLL$'$ accuracy arise from a strongly ordered limit (of consecutively less energetic emissions) and can be expected to be the same as for the hemisphere soft function discussed in \subsec{case2}. This has been seen explicitly at $\mathcal{O}(\alpha_s^2)$ for the related case of jet shapes in $e^+e^-$-collisions for small $R$ in ref.~\cite{Banfi:2010pa}.
As mentioned at the end of \subsec{case2} recent approaches for a resummation of NGLs have been applied to this prototypical case.

The canonical relationships between the different scales in regime 3 are then
\begin{align} \label{eq:regime-3-canonical}
\mu_H \mu_{S_B} \simeq \mu_B^2\,,\qquad
\mu_H \mu_{S_R}^{(B)} \frac{f_B(\eta_J)}{R} \simeq \mu_B^2\,,\qquad
\mu_H \frac{R}{2} \simeq \mu_{J_R}
\,.\end{align}
Together with the choices $\mu_H \simeq p_T^J$ and $\mu_{S_B} \simeq \Tau_B$ they determine the full canonical scaling required to resum all logarithms $\ln R$ and a subset of logarithms $\ln(p_T^J/\Tau_B)$ as discussed above.

\subsection{Regime 4: Large-$R$ jets with $m_J \sim p^J_T R \sim p_T^J$}
\label{subsec:case4}

 The situation for large-$R$ jets in the far tail of the spectrum, corresponding to $\Tau_J \sim m_J \sim p_T^J \sim Q$, is also an interesting conceptual hierarchy to consider. In this regime there are no resolved final-state collinear modes and the jet consists only of hard wide-angle emissions. As in regime 3, parametrically large NGLs of $\Tau_B/p_T^J$ appear, due to the fact that soft wide-angle radiation resolves the number of the hard wide-angle emissions in the jet region. One can expect that the additional corrections with respect to the narrow jet case $R \ll R_0$ for typically applied jet radii are quite small at the far tail, so that for phenomenological applications it is most likely sufficient to include them in fixed-order QCD, unless one is interested in the precise behavior at the endpoint of the spectrum. In analogy to \eq{refactJ} for the global collinear function $\mathcal{J}$ in regime 3 one can also resum Sudakov logarithms $\ln(\Tau_B/p_T^J)$ in regime 4 by refactorizing the associated global hard function $\mathcal{H}$ into jet radius and algorithm dependent hard and soft functions.

\subsection{Relations between the different hierarchies}
\label{subsec:summary}

We have investigated the mode setup and factorization for large and small $R$ jets across the jet mass spectrum. The main features are summarized in table~\ref{tab:regimes}, including the logarithms the factorization formula resums. When $\Tau_J \sim \Tau_B R^2$ the nonglobal correlations do not result in large NGLs, but this condition cannot be satisfied for $\Tau_J \sim p_T^J R^2$ (regime 3) without also removing the jet veto. We now discuss in more detail how the different EFTs are related to each other, as illustrated in \fig{modes}, and how the associated factorized cross sections can be combined.
 
The factorized cross section in \eq{sigmaTau2} for regime 2, describing the hierarchy $\Tau_J \ll p_T^J R$ for narrow jets, can be obtained from the result in \eq{sigmaTau1} for regime 1, describing broad jets, by taking the limit $R\ll R_0$ and carrying out an associated factorization of the soft sector. This enables the resummation of logarithms of $R$, and goes hand in hand with the following expansion of the corrections in $R$
\begin{align} \label{eq:cons12}
S_\kappa (\ell_J,  \ell_B, \eta_J, R, \mu) 
&= \frac{2}{R}  \int\! \df k_B \, S_{R,\kappa_J} \biggl(\frac{2 \ell_J}{R}, k_B,\mu \biggr) 
S_{B,\kappa}\biggl(\ell_B - \frac{f_B(\eta_J)k_B}{R}, \eta_J, \mu\biggr) 
\nn \\ & \quad \times 
\bigl[1 + \ord{R^2}\bigr] 
\, ,\end{align}
and using \eqs{S_refact}{refact}, the individual pieces of the soft functions are related by
\begin{align} \label{eq:cons12b}
S^{(B)}_\kappa ( \ell_B, \eta_J, R, \mu) 
&=\int\! \df k_B \, S^{(B)}_{R,\kappa_J} (k_B,\mu ) \,
S_{B,\kappa}\biggl(\ell_B - \frac{f_B(\eta_J)k_B}{R}, \eta_J, \mu\biggr)\, \bigl[1 + \ord{R^2}\bigr] \, ,
\nn \\ 
S^{(J)}_\kappa ( \ell_J, \eta_J, R, \mu) 
&= \frac{2}{R}  \, S^{(J)}_{R,\kappa_J} \biggl(\frac{2 \ell_J}{R},\mu \biggr) \,
\bigl[1 + \ord{R^2}\bigr]
\,, \nn \\
S^{(\rm NG)}_{\kappa} (\ell_J, \ell_B,  \eta_J, R) & = \frac{2}{f_B(\eta_J)}\, S^{(\rm NG)}_{R,\kappa_J} \biggl(\frac{2 \ell_J}{R}, \frac{R \,\ell_B} {f_B(\eta_J)}\biggr)\, \bigl[1 + \ord{R^2}\bigr]
\, .\end{align}

To obtain a combined description valid for regimes 1 and 2, the $\ord{R^2}$ corrections in \eq{cons12} need to be included and combined with the resummation of jet radius logarithms in regime 2. By including the fixed-order matching corrections for the soft functions (or in general for all functions appearing in the factorized cross section) to the same order as the noncusp terms in the anomalous dimension, corresponding to the often utilized N$^k$LL$'$ order counting, this can be conveniently obtained by turning off the resummation in the relevant scale hierarchy. Thus, the cross section for $\Tau_J \ll p_T^J R^2$ with $\ln (m_J/p_T^J)$ and $\ln R$ resummation and including nonsingular corrections with the full $R$ dependence can be written as
\begin{align} \label{eq:sigma12}
&\frac{\df \sigma_{1+2}(\Phi,\kappa)}
  {\df \Tau_B \, \df\Tau_J}
  = \frac{\df \sigma_2 (\Phi,\kappa)}
  {\df \Tau_B \, \df\Tau_J}
  + \biggl(\frac{\df \sigma_1 (\Phi,\kappa)}
  { \df \Tau_B \, \df\Tau_J} - 
  \frac{\df \sigma_2(\Phi,\kappa)}
  {\df \Tau_B \, \df\Tau_J}\Big|_{\mu^{(B)}_{S_R} = \mu_{S_B} = \mu^{(B)}_S \!, \, \mu^{(J)}_{S_R}=\mu^{(J)}_{S}}
  \biggr)
\,.\end{align}
The scale choices in the third term indicate that the jet radius logarithms are included at fixed order only to cancel the corresponding terms in $\df\sigma_1$. Therefore for $R\ll R_0$ the cross section $\df\sigma_{1+2}$ corresponds to the singular resummed cross section from regime 2 plus nonsingular power corrections starting at ${\cal O}(R^2)$ that are determined by the terms in parentheses. At the same time, the scales $\mu_{S_R}^{(B)}$, $\mu_{S_R}^{(J)}$, and $\mu_{S_B}$ in the first term are chosen using suitable profile scales~\cite{Ligeti:2008ac, Abbate:2010xh} such that in the regime 1 limit $R\sim R_0$ the $\ln R$ resummation is turned off and the two terms involving $\df\sigma_2$ in \eq{sigma12} exactly cancel, leaving just the resummed result from regime 1.

Similarly, regime 2 is obtained from regime 3 in the limit $\Tau_J\ll p_T^J R^2$ with an associated factorization of the collinear sector as
\begin{align} \label{eq:cons23}
\mathcal{J}_{\kappa_J}(s_J,p_T^J R,k_B,\mu) = \int \df k_J \, J_{\kappa_J}\bigl(s_J- p_T^J R\,  k_J, \mu\bigr)\, S_{R,\kappa_J}(k_J,k_B,\mu) \biggl[1+\mathcal{O}\biggl(\frac{s_J}{ (p_T^{J} R)^2}\biggr)\biggr] 
\,,
\end{align}
and using \eqs{refactJ}{refact}, the individual pieces are related by
\begin{align}\label{eq:cons23b}
J_{R,\kappa_J}(s_J,p_T^J R,\mu) & = \int \df k_J \, J_{\kappa_J}\bigl(s_J- p_T^J R\,  k_J, \mu\bigr)\, S^{(J)}_{R,\kappa_J}(k_J,\mu) \biggl[1+\mathcal{O}\biggl(\frac{s_J}{(p_T^{J} R)^2}\biggr)\biggr]
\,, \nn \\
\mathcal{J}^{(\rm NG)}_{\kappa_J}(s_J,p_T^J R,k_B) & = \frac{1}{p_T^J R} \,S^{(\rm NG)}_{R,\kappa_J}\biggl(\frac{s_J}{p_T^J R}, k_B\biggr) \biggl[1+\mathcal{O}\biggl(\frac{s_J}{(p_T^{J} R)^2}\biggr)\biggr]
\,.\end{align}
In ref.~\cite{Chien:2015cka}, the first relation has been explicitly demonstrated at one loop and exploited to obtain two-loop corrections to the ``unmeasured'' jet function.

Therefore, one can combine regimes 2 and 3 to obtain a description of the cross section for small-$R$ jets over the whole spectrum with $\ln (m_J/p_T^J)$ and $\ln R$ resummation and including all nonsingular corrections in $m_J/(p_T^JR)$
as follows
\begin{align} \label{eq:comb23}
&\frac{\df \sigma_{2+3} (\Phi,\kappa)}
  {\df \Tau_B \, \df\Tau_J}
  = \frac{\df \sigma_2 (\Phi,\kappa)}
  {\df \Tau_B \, \df\Tau_J}
  + \biggl(\frac{\df \sigma_3 (\Phi,\kappa)}
  {\df \Tau_B \, \df\Tau_J} - 
  \frac{\df \sigma_2 (\Phi,\kappa)}
  {\df \Tau_B \, \df\Tau_J}\Big|_{\mu^{(J)}_{S_R} = \mu_J=\mu_{J_R} }
  \biggr)
\,.\end{align}
As in \eq{sigma12}, this requires to use primed counting for $\df\sigma_2$ and $\mu^{(J)}_{S_R}$ to be chosen as a suitable profile scale that smoothly merges with $\mu_J$ as the endpoint $m_J \sim p_T^J R$ is approached. In the last term of \eq{comb23} the resummation of logarithms of $m_J/(p_T^J R)$ is turned off.

Finally, the full cross section including all fixed-order nonsingular corrections is given by,
\begin{align}
\frac{\df \sigma (\Phi,\kappa)}
  {\df \Tau_B \, \df\Tau_J}
  = \frac{\df \sigma_{1+2+3} (\Phi,\kappa)}
  { \df \Tau_B \, \df\Tau_J}
  + \biggl(\frac{\df \sigma_{\rm FO} (\Phi,\kappa)}
  {\df \Tau_B \, \df\Tau_J} - 
  \frac{\df \sigma_{1+2+3} (\Phi,\kappa)}
  {\df \Tau_B \, \df\Tau_J}\Big|_{\mu_{i} = \mu_{\rm FO}}
  \biggr)
\,,\end{align}
where $\df \sigma_{\rm FO}$ denotes the fixed-order cross section computed in full QCD at the scale $\mu=\mu_{\rm FO}$,  and the terms from the singular regions are combined via
\begin{align}
\frac{\df \sigma_{1+2+3} (\Phi,\kappa)}
  {\df \Tau_B \, \df\Tau_J} \equiv \frac{\df \sigma_{1+2} (\Phi,\kappa)}
    {\df \Tau_B \, \df\Tau_J} + \frac{\df \sigma_{2+3} (\Phi,\kappa)}
      {\df \Tau_B \, \df\Tau_J} - \frac{\df \sigma_2 (\Phi,\kappa)}
        {\df \Tau_B \, \df\Tau_J}
\, .\end{align}

\subsection{Comparison to earlier calculations}
\label{subsec:comparison}

We conclude this section by identifying which jet radius logarithms were accounted for in earlier jet mass calculations. 

In the jet mass calculation of Ref.~\cite{Chien:2012ur} for $pp \to \ga$ + jet, with an expansion around the kinematic threshold, the soft function was refactorized in order to resum Sudakov logarithms between the soft scales. As discussed below \eq{S_refact} their regional soft function corresponds to $S_\kappa^{(J)}$ for a cone jet. Due to \eq{cons12b} this could encode the correct small-$R$ dependence, and they obtain the correct one-loop anomalous dimension $\gamma^{\kappa}_{S^{(J)}}$.
However, their regional soft function does not contain the required $\al_s \ln^2 R$ term, and it is not clear whether the scale they obtain from a numerical minimization procedure satisfies $\mu_S^{(J)} \sim \Tau_J/R$ for $R\ll 1$, as required for $\ln R$ resummation at LL accuracy.\footnote{Since there are three physical low scales to be accounted for in the small-$R$ limit, namely $\Tau_J/R$, $[m_X^2/p_T^J-\Tau_J]$ and $[m_X^2/p_T^J-\Tau_J]/R$ (where $m_X$ denotes the total partonic invariant mass in the final state), but only two soft renormalization scales $\mu_S$ are used, it cannot be expected that $\mu_S^{(J)}$ comes out to have the correct parametric scaling. This is also indicated by the ratio of their scales for $R=0.3$ and $R=0.5$, $\mu_S^{(J)}(R=0.3)/\mu_S^{(J)}(R=0.5) \approx 3$, which differs from the value of $5/3$ that is required for a correct scaling with $R$.}
In ref.~\cite{Liu:2014oog} a similar approach was taken for $pp \to$ 2 jets. They do obtain the correct one-loop expressions for $S_\kappa^{(J)}$ in the small-$R$ limit, but it is again unclear whether they obtain the correct scale from their numerical minimization. Since the jet radius logarithms that multiply the two-loop cusp anomalous dimension are not included, they can at best achieve LL accuracy.

In Ref.~\cite{Jouttenus:2013hs}, the refactorization of the soft function was based on the structure of the anomalous dimension and identifying the correct scale choice $\mu_S^{(J)} \sim \Tau_J/R$. This accounts for the LL resummation of the jet radius logarithms in the normalized spectrum. But this choice alone is not sufficient beyond LL.

Ref.~\cite{Dasgupta:2012hg} considers the inclusive jet mass spectrum without a jet veto, only probing radiation in the jet. This allows for a resummation of $\ln R$ at LL in the normalized spectrum, and even NLL once the $R$ dependence of the NGLs $\sim \ln(p_T^J R^2/\Tau_J)$ are taken into account.  Their final expression resums only logarithms of the ratio $m_J^2/(p_T^J R)^2$, implying that a hard scale of $p_T^J R$ rather than $p_T^J$ was used. They employ a framework tailored to obtain the NLL result, making it difficult to directly compare the functions from our factorization theorem with results from their calculation.

None of the above approaches accounted for the jet radius logarithms in the \emph{normalization} of the cross section for each individual partonic channel, which requires an additional factorization for the soft out-of-jet corrections (corresponding to the first line in \eq{cons12b}). This is crucial for determining the relative contribution of the different partonic channels. Thus, when summing over different partonic channels to obtain the final physical spectrum, the $\ln R$ resummation is not accounted for systematically even at LL. Our factorization theorem presented in regime 2 allows for $\ln R$ resummation in the jet mass cross section at any order in resummed perturbation theory for which the corresponding anomalous dimensions are known.  

While this work was being prepared ref.~\cite{Hornig:2016ahz} appeared, which also builds on ref.~\cite{Chien:2015cka} and discusses dijet angularities for $pp \to$ dijets at small $R$, addressing the nontrivial color space. They achieve NLL precision for a resummation of logarithms associated with both $R$ and the measurement of angularities, one of which is the jet mass. They use a jet-based transverse momentum veto within a certain rapidity range $\abs{\eta} < \eta^{\rm cut}$ and no restrictions beyond. For phenomenologically relevant values of $\eta^{\rm cut}$, their setup does not seem to properly account for the resummation of rapidity logarithms $\ln(p_T^{\rm cut}/p_T^J)$ because it does not take into account the effect of the jet veto on the beam-collinear radiation. Their study focuses on the equivalent of our regime 2, and therefore does not include nonsingular corrections from the regime $\Tau_J \sim p_T^J R$ or perturbative power corrections of $\mathcal{O}(R^2)$. The latter points can be addressed in a straightforward manner by combining their results with the framework presented here.

\section{Jet and soft functions}
\label{sec:softandjet}

In this section we give the definitions and relevant one-loop expressions for the various jet and soft functions that enter the factorization formulae in \sec{fact}. In \subsecs{soft1}{softB} we discuss the wide-angle soft functions for large- and small-$R$ jets appearing in regime 1 ($S_\kappa$) and regimes 2 and 3 ($S_{B,\kappa}$), respectively. The results for $S_{B,\kappa}$ are new. The \csoft function $S_{R,\kappa_J}$ (together with its refactorized form) is given in \subsec{csoft}. In \subsec{jet} we collect the results for the known jet functions. 
The RG consistency of the factorization formulae allows us extract the remaining anomalous dimensions needed for NNLL resummation of the logarithms, as discussed in \app{anomdim}. We verify the relations between the different EFTs given by \eqs{cons12}{cons23} in \subsec{consistency} and discuss nonperturbative effects in \subsec{nonpert}.

\subsection{Wide-angle soft function for large-$R$ jets $S_\kappa$ (regime 1)}
\label{subsec:soft1}

For the large-$R$ jets in regime 1 there is a single soft function $S_\kappa$ that describes the contribution of soft radiation to the jet mass and jet veto. For example, for the partonic channel $\kappa = \{q, \bar q; g\} \equiv \{q\bar q\to g\}$ the matrix element is defined as
\begin{align} \label{eq:S_def}
S_\kappa  (\ell_J, \ell_B, \eta_J)
&= \frac{1}{N_c C_F} \MAe{0}{\tr \Bigl\{\bar{T} \bigl[Y_{n_a}^\dagger T^c \mathcal{Y}_{n_J}^{ce} Y_{n_b}\bigr] \de(\ell_J- \hat \ell_J) \,\de(\ell_B - \hat \ell_B)\,
T \bigl[Y_{n_b}^\dagger T^d \mathcal{Y}_{n_J}^{de} Y_{n_a} \bigr]\Bigr\}}{0} 
\, ,\end{align}
where $Y_{n_a}$ and $Y_{n_b}$ are soft Wilson lines in the fundamental representation along the lightlike directions $n_a$ and $n_b$, and $\mathcal{Y}_{n_J}$ is a Wilson line in the adjoint representation along $n_J$. The trace runs over color, $T$ ($\bar T$) denotes (anti)time ordering, and $\hat \ell_J$ and $\hat \ell_B$ encode the measurements in the jet and beam regions, i.e.,
\begin{align}\label{eq:ellhat}
\hat\ell_J \, |X_s\rangle = \cosh \eta_J \sum_{i \in \text{jet}} n_J \sdt p_i \,  |X_s\rangle
\,, \qquad
\hat\ell_B  \, |X_s\rangle = \sum_{i \notin \text{jet}} f_B(\eta_i)  p_{Ti} \, |X_s\rangle
\,.\end{align}
Here, $\eta_i$ and $p_{Ti}$ are the rapidity and transverse momentum of particle $i$ with respect to the \emph{beam} axis. The representation of the Wilson lines and the overall normalization needs to be appropriately modified for other channels.

The one-loop result of the soft function for $N$-jettiness jets has been computed in Ref.~\cite{Jouttenus:2011wh} (and for $N$-jettiness with generic angularities in ref.~\cite{Kasemets:2015uus}). This procedure can be extended to generic jet algorithms, jet vetoes, and jet measurements at hadron colliders, which will be discussed in detail in a forthcoming paper~\cite{Bertolini:forthcoming}. In general, the soft function up to one-loop order can be written as
\begin{align} \label{eq:S_kappa}
S_\kappa (\ell_J,\ell_B,  & \eta_J, R, \mu) =  \de(\ell_J)\,\de(\ell_B) +
\frac{\alpha_s(\mu)}{4\pi}\biggl\{
\bfT_a \cdot \bfT_b  \biggl[\frac{8}{\mu} \,\cL_1\Bigl(\frac{\ell_B}{\mu}\Bigr) \,\de(\ell_J) 
\nn \\ & \quad
+ s_{ab,1}(R) \Bigl(\frac{1}{\mu}\cL_0\Bigl(\frac{\ell_B}{\mu}\Bigr) \delta(\ell_J) - \frac{1}{\mu}\cL_0\Bigl(\frac{\ell_J}{\mu}\Bigr)\delta(\ell_B)  \Bigr) +  s_{ab,\delta}(R,\eta_J)\,\delta(\ell_B) \,\de(\ell_J)\biggr] 
\nn \\ & \quad
+ \bfT_a \cdot \bfT_J
\biggl[
   \frac{8}{\mu}\,\cL_1\Bigl(\frac{\ell_B}{\mu}\Bigr) \,\delta(\ell_J) 
   + \frac{8}{\mu}\,\cL_1\Bigl(\frac{\ell_J}{\mu}\Bigr) \,\delta(\ell_B)
   \nn \\ & \quad
    + s_{aJ,B}(R,\eta_J) \,\frac{1}{\mu}\, \cL_0\Bigl(\frac{\ell_B}{\mu}\Bigr) \,\delta(\ell_J)    
   + s_{aJ,J}(R,\eta_J)\, \frac{1}{\mu}\, \cL_0\Bigl(\frac{\ell_J}{\mu}\Bigr) \,\delta(\ell_B)
   \nn \\ & \quad     
   + s_{aJ,\delta}(R,\eta_J) \, \delta(\ell_J) \, \delta(\ell_B)  \biggr]
   \biggr\} 
    + \Bigl\{ (a,\eta_J) \leftrightarrow (b,-\eta_J) \Bigr\}
    +  \ord{\al_s^2} \, ,
\end{align}
where $s_{ab,1}(R) =  2/\pi \times \pi R^2 = 2 R^2$~\cite{Stewart:2014nna} is proportional the jet area in the $\eta$-$\phi$ plane.\footnote{For XCone jets with parameter $R$ (and arbitrary values for the parameters $\beta>0$ and $\gamma$) the jet area deviates from $\pi R^2$ by small $\mathcal{O}(R^6)$ terms.} The
$s_{ab,\delta}$ and $s_{aJ,\delta}$ depend on the algorithm determining the jet region and the beam measurement. 
We give the analytic results for the coefficients $s_{ab,\delta}(R,\eta_J)$, $s_{aJ,B}(R,\eta_J)$, $s_{aJ,J}(R,\eta_J)$ and $s_{aJ,\delta}(R,\eta_J)$ in the small $R$ limit in \subsec{consistency}, and compare them to the full numerical results for anti-k${}_T$ jets as a function of $R$. In \eq{S_kappa} $\bfT_a, \bfT_b, \bfT_J$ denote the color charges of the respective hard partons entering the hard interaction.

\subsection{Wide-angle soft function for small-$R$ jets $S_{B}$ (regimes 2 \& 3)}
\label{subsec:softB}

In \eqs{sigmaTau2}{sigmaTau3} the soft function $S_B$ describes the interactions of the wide-angle soft modes, which do not resolve the jet. For the partonic channel $\kappa = \{q,\bar q; g\}$ this matrix element is defined as
\begin{align} \label{eq:SB_def}
S_{B,\kappa}  (\ell_B, \eta_J)
&= \frac{1}{N_c C_F} \MAe{0}{\tr \Bigl\{\bar{T} \bigl[Y_a^\dagger T^c \mathcal{Y}_J^{ce} Y_b\bigr] \de(\ell_B - \hat \ell_B)
T \bigl[Y_b^\dagger T^d \mathcal{Y}_J^{de} Y_a \bigr]\Bigr\}}{0} 
\, ,\end{align} 
with
\begin{align}
\hat\ell_B  \, |X_s\rangle = \sum_{i} f_B(\eta_i)  p_{Ti} \, |X_s\rangle
\,.\end{align}
In contrast to \eq{ellhat}, the sum on $i$ now runs over all particles, since the momentum scaling of particles present in the soft state $|X_s\rangle$ implies that this real radiation cannot resolve the jet area. $S_B$ depends on the choice of jet veto and thus on the function $f_B(\eta)$, for which we consider the two choices in \eq{vetofct}. The one-loop computation can be carried out in close correspondence to the calculation for an energy veto~\cite{Ellis:2010rwa} and is discussed in \app{softfct}. The result for the C-parameter veto reads
\begin{align} \label{eq:S_BC}
   S^C_{B,\kappa} (\ell_B, \eta_J,\mu) &= \delta(\ell_B) +\frac{\alpha_s(\mu)} {4 \pi}  \biggl\{\bfT_a \cdot \bfT_b  \biggl[\frac{8}{\mu} \,\cL_1\Bigl(\frac{\ell_B}{\mu}\Bigr)- \frac{\pi^2}{2}\,\delta(\ell_B)\biggr] + \bfT_a \cdot \bfT_J \biggl[ 8 \ln\Bigl(\frac{1+\tanh \eta_J}{2}\Bigr)
 \nn \\ & \quad \times
 \frac{1}{\mu} \,\cL_0\Bigl(\frac{\ell_B}{\mu}\Bigr)
 + \biggl(4\,\Li_2\Bigl(\frac{1+\tanh \eta_J}{2}\Bigr) +2 \ln^2\Bigl(\frac{1-\tanh \eta_J}{2}\Bigr)
 \nn \\ & \quad
  -8 \ln^2(2\cosh\eta_J)-\frac{2\pi^2}{3}\biggr) \delta(\ell_B) \biggr]
 \biggr\}
 + \Bigl\{ (a,\eta_J) \leftrightarrow (b,-\eta_J) \Bigr\}  
\,.\end{align}
For the beam thrust veto we find
\begin{align} \label{eq:S_BB}
   S^{\tau}_{B,\kappa} (\ell_B, \eta_J,\mu) &= \delta(\ell_B) +\frac{\alpha_s(\mu)} {4 \pi}  \biggl\{\bfT_a \cdot \bfT_b  \biggl[\frac{8}{\mu} \,\cL_1\Bigl(\frac{\ell_B}{\mu}\Bigr)- \frac{\pi^2}{6}\,\delta(\ell_B)\biggr] 
   \nn \\
 & \quad + \bfT_a \cdot \bfT_J \biggl[ 16 \eta_J \,\theta(-\eta_J) 
 \frac{1}{\mu} \,\cL_0\Bigl(\frac{\ell_B}{\mu}\Bigr)
 + \Bigl(-4\,\Li_2\bigl(e^{-2|\eta_J|}\Bigr)-8 \eta_J^2 \, \theta(-\eta_J) \biggr) \delta(\ell_B) \biggr] 
  \biggr\}  \nn \\
& \quad   
 + \Bigl\{ (a,\eta_J) \leftrightarrow (b,-\eta_J) \Bigr\}
 \,.\end{align}
The anomalous dimension of $S_B$ can be obtained at higher orders by exploiting RG consistency in \eq{cons12}, see \app{anomdim}.

\subsection{\Csoft function $S_R$ (regime 2)}
\label{subsec:csoft}

Next, we discuss the \csoft function $S_{R}$ in \eq{sigmaTau2} describing the interactions of the \csoft modes that are a combination of collinear-soft and soft-collinear modes. For a quark jet (i.e.~$\kappa_J = q$) it is defined as 
\begin{align}
  S_{R,q}(k_J,k_B) &=
\frac{1}{N_c} \Bigl\langle 0 \Big| \tr\Bigl\{ \bar T \bigl[X_{n_J}^\dagger(0) X_{\bar{n}_J}(0)\bigr]\, \de\Bigl[k_J - \frac{2 \cosh \eta_J}{R}\, n_J \sdt \hat k_\text{in}\Bigr]
\nn\\ & \qquad\times \de\Bigl[k_B - \frac{ R}{2 \cosh \eta_J} \, \bar{n}_J \sdt \hat k_\text{out}\Bigr] \,
T \bigl[X_{\bar{n}_J}^\dagger(0) X_{n_J}(0)\bigr] \Bigr\} \Big| 0 \Big\rangle
\,.\end{align}
The Wilson lines $X_{n_J}$ and $X_{\bar{n}_J}$ are the \csoft (i.e.~boosted soft) analogs of the (u)soft Wilson line $Y_{n_J}$ and $Y_{\bar n_J}$ and the momentum operators $\hat k_\text{in}$ and $\hat k_\text{out}$ pick out the momentum inside and outside the jet. For a gluon jet the Wilson lines are in the adjoint representation and the overall factor changes from $1/N_c$ to $1/(N_c^2-1)$.

Since the jet is defined through the beam coordinates $\eta$, $\phi$, the angular size of the jet region is $R/\!\cosh \eta_J$. A boost along the jet axis by $\ln[R/(2\cosh \eta_J)]$ turns the jet region into a hemisphere (ignoring $\ord{R^2}$ corrections) while leaving these Wilson lines invariant. This is most easily seen by using reparametrization invariance (RPI-III)~\cite{Manohar:2002fd} to rescale the jet directions via $n_J \to n_J' = n_J \beta$, $\bar{n}_J \to \bar{n}_J' = n_J'/\beta$ with $\beta = R/(2 \cosh \eta_J)$. This boost invariance of the two-direction soft function has been exploited before in Refs.~\cite{Lee:2006nr,Feige:2012vc,Kang:2013nha}.
From this transformation we see that $S_{R}$ is just the hemisphere soft function, and with our choice of variables, is independent of $R$,
\begin{align} \label{eq:equiv}
  S_{R,q}(k_J,k_B) &=
\frac{1}{N_c} \Bigl\langle 0 \Big| \tr\Bigl\{ \bar T \bigl[X_{n'_J}^\dagger(0) X_{\bn'_J}(0)\bigr]\, \de(k_J - n_J'\sdt \hat k_R) \,
\nn\\
 &\qquad\qquad \times 
  \de(k_B - \bar{n}_J'\sdt \hat k_L)\,
T \bigl[X_{\bn_J'}^\dagger(0) X_{n_J'}(0)\bigr] \Bigr\} \Big| 0 \Big\rangle
\,.\end{align}
Here $\hat k_R$ ($\hat k_L$) picks out the momentum going into the right (left) hemisphere with respect to the jet direction, i.e.~for $n_J' \cdot k < \bar{n}_J' \cdot k$ ($n_J' \cdot k > \bar{n}_J' \cdot k$).
Thus up to one-loop order~\cite{Fleming:2007xt,Schwartz:2007ib}
\begin{align}\label{eq:S_J}
   S_{R,\kappa_J}(k_J,k_B,\mu)&=\delta(k_J)\,\de(k_B) +\frac{\alpha_s(\mu) \bfT_J^2}{4\pi} \biggl[- \frac{8}{\mu}\, \mathcal{L}_1\Bigl(\frac{k_J}{\mu}\Bigr) \de(k_B) - \frac{8}{\mu}\, \mathcal{L}_1\Bigl(\frac{k_B}{\mu}\Bigr) \de(k_J)
   \nn \\ & \quad
    +\frac{\pi^2}{3} \,\delta(k_J)\,\de(k_B) \biggr] + \ord{\al_s^2}
\,,\end{align}
where the color charge $\bfT_J^2$ is equal to $C_F$ for quark jets and $C_A$ for gluon jets.
The refactorization in \eq{refact} is trivial at one-loop order, since only one parton contributes to either the beam or jet region. As these regions correspond to hemispheres after the boost the collinear-soft and soft-collinear function are thus given by the same one-loop function
\begin{align}\label{eq:S_hemi}
   S^{(J)}_{R,\kappa_J}(k,\mu) &= S^{(B)}_{R,\kappa_J}(k,\mu) =\delta(k) +\frac{\alpha_s(\mu) \bfT_J^2}{4\pi} \biggl[- \frac{8}{\mu}\, \mathcal{L}_1\Bigl(\frac{k}{\mu}\Bigr) +\frac{\pi^2}{6} \,\delta(k)\biggr] + \ord{\al_s^2}
\,.\end{align}

\subsection{Jet functions (regimes 1, 2 \& 3)}
\label{subsec:jet}

The inclusive jet functions in \eqs{sigmaTau1}{sigmaTau2} measuring the invariant mass of the collinear radiation are well known and given by a vacuum correlator of two jet fields. Up to one-loop order they are given by~\cite{Bauer:2003pi, Fleming:2003gt, Becher:2009th}
\begin{align}\label{eq:jet0}
 J_q(s,\mu^2) &=\delta(s)+\frac{\alpha_s(\mu)C_F}{4\pi}\biggl\{\frac{4}{\mu^2}\, \mathcal{L}_1\Bigl(\frac{s}{\mu^2}\Bigr)-\frac{3}{\mu^2}\, \mathcal{L}_0\Bigl(\frac{s}{\mu^2}\Bigr)+(7-\pi^2)\,\delta(s)\biggr\}
   \, ,\\
    J_g(s,\mu^2) &=\delta(s)+\frac{\alpha_s(\mu)}{4\pi}\biggl\{\frac{4C_A}{\mu^2}\, \mathcal{L}_1\Bigl(\frac{s}{\mu^2}\Bigr)-\frac{\beta_0}{\mu^2}\, \mathcal{L}_0\Bigl(\frac{s}{\mu^2}\Bigr)+\Bigl[\Bigl(\frac{4}{3}-\pi^2\Bigr)C_A+\frac{5}{3}\beta_0\Bigr]\delta(s)\biggr\}
\,.\nn \end{align}

The jet function $J_{R}$, obtained from the collinear function $\mathcal{J}$ in \eq{sigmaTau3} after the decomposition in \eq{refactJ}, encodes the fact that the energetic $n_J$-collinear radiation is constrained to lie within the jet region and explicitly depends on the jet algorithm, as discussed in refs.~\cite{Jouttenus:2009ns, Ellis:2010rwa}.
Following \eq{cons23b}, we write $J_R$ as
\begin{align}\label{eq:J_alg}
J_{R,\kappa_J}(s,p_T^J R,\mu) = \int \df k \, J_{\kappa_J}( s-p_T^J R \, k,\mu)\,  S^{(J)}_{R,\kappa_J}( k,\mu) + \Delta  J^{\rm alg}_{\kappa_J}(s,p_T^J R,\mu)  \, ,
\end{align}
where the term $\Delta  J^{\rm alg}_{\kappa_J}(s,p_T^J R,\mu)$ contains the algorithm dependent terms, which are power suppressed in regime 2 where $\Tau_J \ll p_T ^J R^2$. $\Delta  J^{\rm alg}$ has been computed at one loop for different jet algorithms in $e^+ e^-$-colliders in refs.~\cite{Jouttenus:2009ns, Ellis:2010rwa}. Adapting their expressions to the hadron collider case, the one-loop result for k$_T$-type clustering algorithms reads
\begin{align} \label{eq:Delta_J}
\Delta J_q^{{\rm k}_T} (s,p_T^J R,\mu) &= \frac{\alpha_s(\mu) C_F}{4\pi} \frac{1}{s} \biggl\{ \theta\biggl(\frac{(p_T^J R)^2}{4} - s\biggr) \Bigl[ 4\ln\Bigl(\frac{1-x_1}{x_1}\Bigr)+ 6x_1 - 3\Bigr] \nn \\
& \quad + 4 \ln\Bigl(\frac{s }{(p_T^J R)^2}\Bigr) + 3 \biggr\} 
\,, \nn \\
\Delta J_g^{{\rm k}_T} (s,p_T^J R,\mu) &= \frac{\alpha_s(\mu)}{4\pi} \frac{1}{s}  \biggl\{ \theta\biggl(\frac{(p_T^J R)^2}{4} - s\biggr) \biggl[C_A \biggl(4\ln\Bigl(\frac{1-x_1}{x_1}\Bigr)
- 6x_1^3 + 9 x_1^2 - 3 x_1 \biggr)
\nn \\ & \quad
 + \beta_0 (2x_1^3 - 3x_1^2 + 3x_1 - 1) \biggr] 
 +  4C_A\ln\Bigl(\frac{s}{ (p_T^J R)^2}\Bigr) + \beta_0 \biggr\} 
\,,\end{align}
with
\begin{align}
x_1 &=\frac{1}{2}\biggl(1-\sqrt{1-\frac{4 s}{ (p_T^J R)^2}}\biggr)
\,.
\end{align}
The anomalous dimension of $J_R$ can be obtained at higher orders by exploiting RG consistency of \eq{J_alg}, as discussed in \app{anomdim}. 

The jet function $J_R$ is related to the algorithm-dependent jet function $J^{\rm alg.}_{\kappa_J}$ in refs.~\cite{Ellis:2010rwa,Chien:2015cka} via
\begin{align}\label{eq:J_alg2}
J_{R,\kappa_J}(s,p_T^J R,\mu) & = \int \df k \, J^{\rm alg.}_{\kappa_J}( s-p_T^J R \, k,\mu)\,  S^{(J)}_{R,\kappa_J}( k,\mu)
\, , \nn \\
J^{\rm alg.}_{\kappa_J}( s,p_T^J R,\mu) & = J_{\kappa_J}( s,\mu)+ \Delta  J^{\rm alg. \scriptsize\text{\cite{Chien:2015cka}}}_{\kappa_J}(s,p_T^J R,\mu) \, .
\end{align}
Thus, refs.~\cite{Ellis:2010rwa,Chien:2015cka} effectively combine the algorithm-dependent fixed-order corrections in regime 3 ($m_J \sim p_T^J R$) with the inclusive jet function, thereby including nonsingular correction in the regime-2 limit $m_J \ll p_T^J R$ in a definite way. In our description of regime 3 in \eqs{sigmaTau3}{refactJ}, the single function $J_{R,\kappa_J}$ encodes the contributions of the energetic collinear radiation to the jet measurement (corresponding to the fact that collinear and collinear-soft modes present in regime 2 become degenerate in regime 3).%
\footnote{The direct computation of $J^{\rm alg.}_{\kappa_J}$ in \cite{Jouttenus:2009ns, Ellis:2010rwa} required nontrivial (collinear-)soft zero bin subtractions on the $n_J$-collinear modes. In our mode setup for regime 3 with a single energetic $n_J$-collinear mode these subtractions do not appear. Thus our $J_{R,\kappa_J}$ differs from $ J^{\rm alg.}_{\kappa_J}$ by these zero-bin subtractions, which correspond exactly to our collinear-soft function $S_R^{(J)}$. This was also observed in Ref.~\cite{Chay:2015ila} in a related context.}

\subsection{Verification of the relation between different regimes} 
\label{subsec:consistency}

Using the perturbative results in secs.~\ref{subsec:soft1} -- \ref{subsec:jet} we can explicitly verify that the relations between the different EFTs hold at the one-loop level. First, \eqs{cons23b}{J_alg} imply that the algorithm dependent correction $\Delta J^{\rm alg}_{\kappa_J}$ needs to vanish when $m_J \ll  p_T^J R$, i.e.~by taking $x_1\to 0$,
\begin{align}
s \,\Delta J^{\rm alg}_{\kappa_J}(s,p_T^J R,\mu) = \mathcal{O}\biggl(\frac{s}{(p_T^{J} R)^2}\biggr)\, ,
\end{align}
which can be verified directly at one loop using \eq{Delta_J}.

Next, the relation in \eq{cons12} between the small-$R$ and large-$R$ jets for  $m_J \ll  p_T^J R$ implies that at one-loop order\footnote{The leading power corrections in this relation are only $\mathcal{O}(R^2)$ for a smooth jet veto. For the beam thrust veto at $\eta_J=0$ the power corrections are in fact $\mathcal{O}(R)$. Consequently the small $R$ limit is not as good an approximation to the full result for $|\eta_J|<R$.}
\begin{align} 
S_\kappa^\one (\ell_J,  \ell_B, \eta_J, R, \mu) 
&= \biggl[S_{B,\kappa}^\one(\ell_B, \eta_J, \mu)\, \de(\ell_J) +
\frac{2}{f_B(\eta_J)}\, S^\one_{R,\kappa_J} \biggl(\frac{2 \ell_J}{R}, \frac{R \,\ell_B}{f_B(\eta_J)},\mu \biggr)  \biggr] 
\nn \\ 
& \quad \times \bigl[1 + \ord{R^2}\bigr]
\,.\end{align}
Exploiting color conservation,
\begin{align}
  \bfT_J^2 = -\bfT_a \sdt \bfT_J - \bfT_b \sdt \bfT_J
\,,\end{align}
this requires the coefficients of the wide-angle soft function $S_\kappa$ in \eq{S_kappa} to satisfy
\begin{align}\label{eq:S_kappa_smallR}
s_{aJ,B}(R,\eta_J) & = 8 (\eta_J+ \ln R) +\mathcal{O}(R^2)  
\, , \nn \\
s_{aJ,J}(R,\eta_J) & = -8\ln\frac{R}{2}+\mathcal{O}(R^2) 
\, ,  \nn  \\
s^C_{ab,\delta}(R,\eta_J) & = -\frac{\pi^2}{2}+\mathcal{O}(R^2) 
\, , \nn \\
s^{\tau}_{ab,\delta}(R,\eta_J) & = -\frac{\pi^2}{6} +\mathcal{O}(R^2) 
\, , \nn \\
s^C_{aJ,\delta}(R,\eta_J) & = 4 \,\Li_2\Bigl(\frac{1\!+\!\tanh \eta_J}{2}\Bigr) - 2 \ln^2\Bigl(\frac{1\!+\!\tanh \eta_J}{2}\Bigr) +4\eta_J^2 + 8 \ln^2 R + 8 \ln R \, \ln \cosh \eta_J  \nn  \\
& \quad +4\ln^2 2 - \pi^2 +\mathcal{O}(R^2) 
\, ,  \nn  \\
s^{\tau}_{aJ,\delta}(R,\eta_J) & = -4\,\Li_2\bigl(e^{-2|\eta_J|}\bigr)+4 \eta_J^2 [\theta(\eta_J) - \theta(-\eta_J)] +8  \ln^2 R  + 8 \ln R \bigl[|\eta_J|-\ln 2\bigr]
   \nn \\ & \quad 
+4\ln^2 2- \frac{\pi^2}{3} +\mathcal{O}(R^2) 
\,,
\end{align}
in the small-$R$ limit.
Here we encounter logarithms (in particular also Sudakov double logarithms) of the jet radius which are not resummed without the factorization of the soft function in regime 2.
Furthermore, we remark that consistency of the anomalous dimensions implies that any choice of \SCETa-type veto only alters the coefficient of the local terms in momentum space proportional to $\delta(\ell_J) \, \delta(\ell_B)$ and thus gives the same results for $s_{aJ,B}$ and $s_{aJ,J}$.\footnote{Additional terms in the combination $1/\mu \,\mathcal{L}_0 (\ell_B/\mu) \, \delta(\ell_J) -1/\mu \, \mathcal{L}_0 (\ell_J/\mu) \,\delta(\ell_B)$ do not affect this consistency and in fact appear in general for large $R$ jets. However, these are only related to algorithm dependent deviations of the jet region (and not to the employed beam measurement) which are power suppressed in the small $R$ limit.}

\begin{figure}
  \centering
     \includegraphics[width=0.5\textwidth]{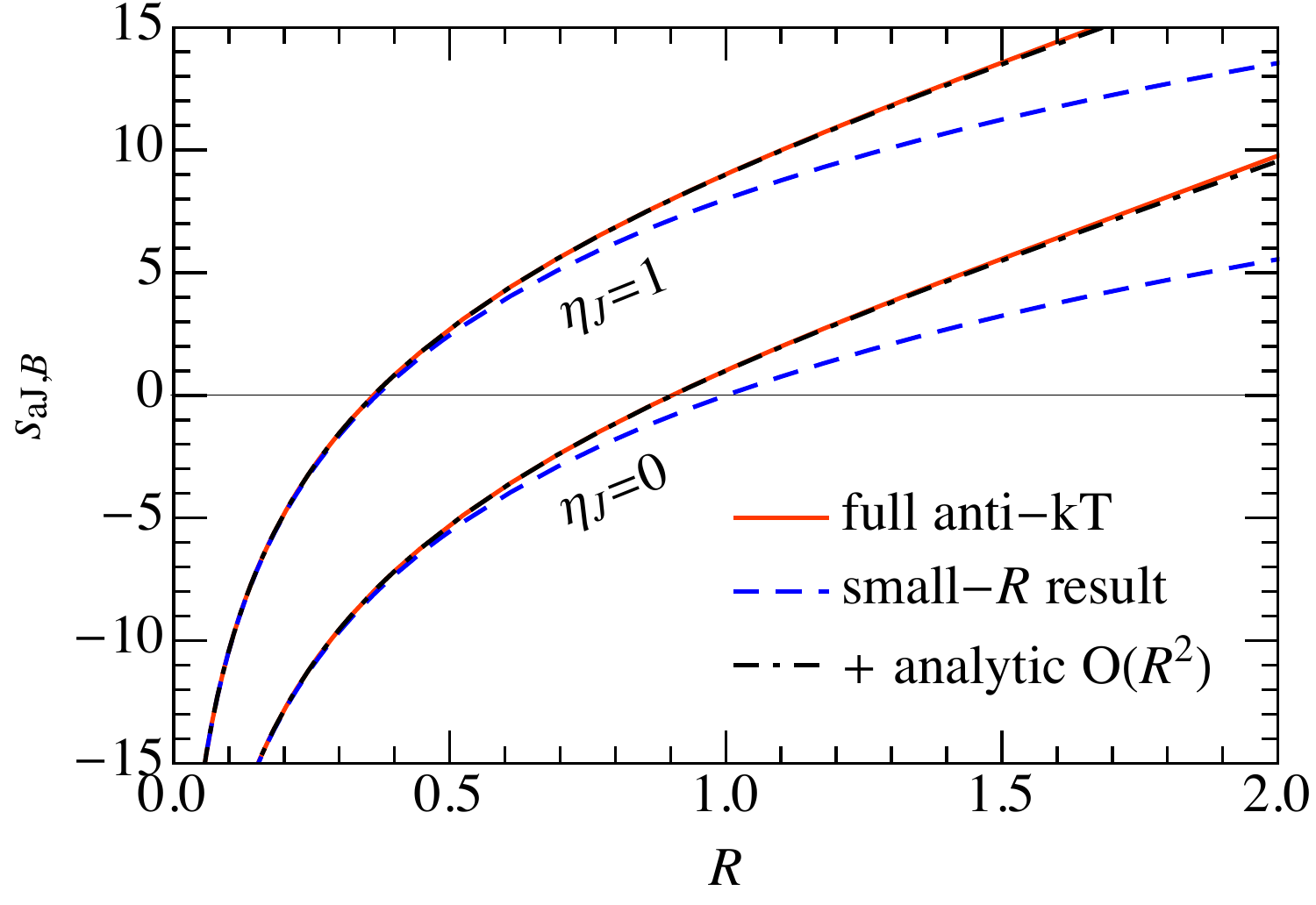}\hfill%
    \includegraphics[width=0.5\textwidth]{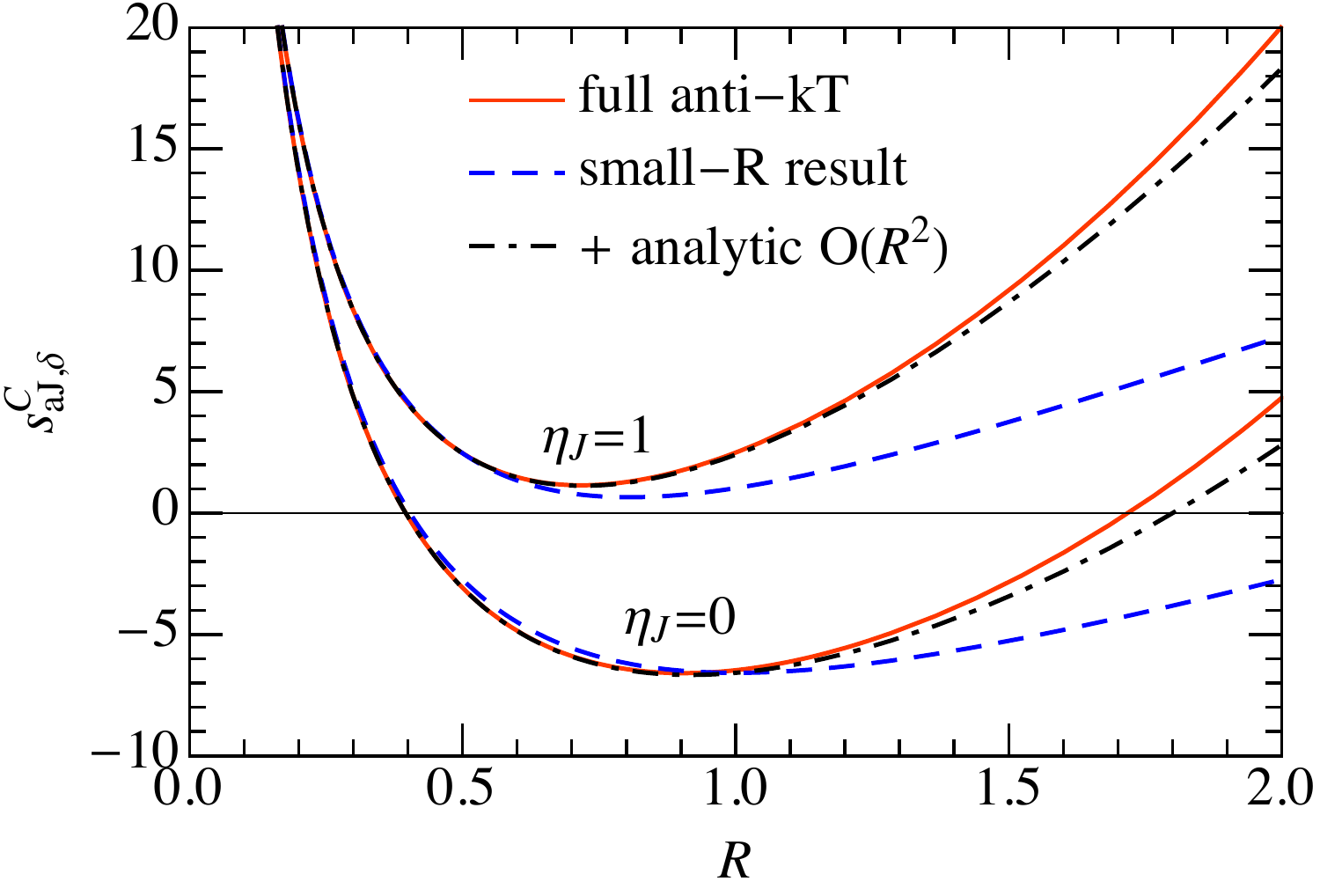}%
    \caption{Coefficients of the soft function $S_\kappa$ for the C-parameter jet veto and for anti-k${}_T$ jets. Shown are the exact results (solid red) together with the corresponding results in the small-$R$ limit (dashed blue) and including the first $\ord{R^2}$ corrections (dot-dashed black) for two values $\eta_J=0$ and $\eta_J=1$. 
\label{fig:deltaS1C}} 
\end{figure}

By performing appropriate expansions of the integral expressions for the coefficients of $S_\kappa$, one can confirm analytically that these relations are indeed satisfied~\cite{Bertolini:forthcoming}. In \fig{deltaS1C}, we show the full numerical results for the coefficients together with the small $R$ result in \eq{S_kappa_smallR} for the C-parameter veto. We also display the coefficients when including corrections at $\mathcal{O}(R^2)$ in a small-$R$ expansion, which can be calculated analytically and will be given explicitly in ref.~\cite{Bertolini:forthcoming}.
One can see that the small-$R$ results approximate the full coefficients very well for $R \ll R_0$. We have verified that this holds also for the beam thrust veto and an arbitrary jet rapidity. Including $\mathcal{O}(R^2)$ corrections one obtains an excellent approximation of the full result even for $R\gtrsim 1$. This suggests that the small-$R$ limit (including terms at $\mathcal{O}(R^2)$) is a good approximation for phenomenological jet mass studies at the LHC.\footnote{For the beam-beam dipole the $\mathcal{O}(R^2)$ corrections are typically larger and can be quite sizable also for smaller values of the jet radius $R \sim 0.5$.}
Such an expansion has been applied in \cite{Dasgupta:2012hg,Liu:2014oog} for the inclusive jet mass spectrum with the result that $\mathcal{O}(R^4)$ corrections have a negligible impact for phenomenologically relevant values of $R$.
We see from \fig{deltaS1C} that the expansions are valid up to jet radii $R \sim 2$ implying that $R_0\gtrsim 2$ is a more appropriate radius of convergence than $R_0 \simeq 1$.\footnote{For central jets with a cone radius $R^{\rm cone}_0=\pi/2\approx 1.6$ the jet region becomes a full hemisphere, which is a naive estimate for the radius of convergence. Using a radius in the $\eta-\phi$ plane instead implies a significantly smaller jet area and a wider range of convergence, so that a value $R_0\gtrsim 2$ is plausible.}

\subsection{Leading nonperturbative effects}
\label{subsec:nonpert}

We conclude this section by discussing the leading nonperturbative effects on the jet mass spectrum.
The leading nonperturbative effects are in particular relevant in the peak and tail region where $p_T^J R^2 \gg \Tau_J \gtrsim \LQCD$ and thus affect the factorization formulae in \subsecs{case1}{case2}.
Nonperturbative corrections to the jet veto are ignored, since their effect is negligible for normalized spectra, which are measured experimentally.
We start by briefly summarizing the findings of ref.~\cite{Stewart:2014nna} for large-$R$ jets, before moving on to small-$R$ jets. 

The wide-angle soft function can be decomposed into a perturbative component $S^{\rm pert}_\kappa$ and a nonperturbative function $F_\kappa$~\cite{Korchemsky:1999kt, Hoang:2007vb, Ligeti:2008ac},
\begin{align} \label{eq:Sfact}
S_\kappa(\ell_J, \ell_B,\eta_J, R,\mu)
&= \int\! \df k\, S_\kappa^{\rm pert}(\ell_J - k, \ell_B, \eta_J, R,\mu)
F_\kappa(k, \eta_J, R)\, \Bigl[1 + {\cal O}\Bigl(\frac{\lqcd}{\ell_B} \Bigr) \Bigr]
 .\end{align}
Expanding in $\lqcd \ll \ell_J$, one obtains
\begin{align} \label{eq:SOPE}
S_\kappa(\ell_J, \ell_B, \eta_J, R,\mu)
= S_\kappa^\mathrm{pert}\Bigl(\ell_J- \Omega_\kappa(R),\ell_B, \eta_J, R,\mu \Bigr)
 \biggl[1+ \mathcal{O}\biggl(\frac{\lqcd^2}{\ell_J^2}, \frac{\alpha_s\lqcd}{\ell_J},\frac{\lqcd}{\ell_B} \biggr)\biggr]
\,.\end{align}
Thus the leading nonperturbative effect leads to a shift in the jet mass,
\begin{align} \label{eq:mshift}
  m_J^2 = (m_J^2)^{\rm pert} + 2p_T^J \Omega_\kappa(R)
  \,, \qquad
  \Omega_\kappa(R) = \int\!\df k\, k\, F_\kappa(k, \eta_J, R)
 \, .\end{align}
In ref.~\cite{Stewart:2014nna} it was shown that $\Omega_\kappa$ depends only on the jet radius $R$ and channel $\kappa$ but not on the jet rapidity $\eta_J$, and that for small jet radii
\begin{align} \label{eq:omega_exp}
\Omega_\kappa(R) = \frac{R}{2}\,\Omega_{\kappa_J} \bigl[1+ \ord{R^2}\bigr]
\,,\end{align}
where as indicated, the $R$-independent nonperturbative parameter $\Omega_{\kappa_J}$ depends only on whether the jet is initiated by a quark or a gluon. Here $\Omega_q$ is the nonperturbative correction for thrust in deep-inelastic scattering (DIS)~\cite{Kang:2013nha}, and $\Omega_g$ is its analog for gluons. Technically, once hadron mass effects are accounted for the function $F_\kappa$ and parameter $\Omega_\kappa$ also have renormalization group evolution between the hadronic and soft scales, and there is another matching coefficient at the soft scale~\cite{Mateu:2012nk}. This does not change the universality discussion above, and hence this complication is suppressed for simplicity.

We now show that the same conclusion follows directly from the factorization formula for small $R$ in \eq{sigmaTau1}. The leading nonperturbative effects come from the \csoft function $S_{R}$, which is identical to the (DIS) double hemisphere soft function, as argued in \subsec{csoft}. The leading nonperturbative correction is therefore
\begin{align}
S_{R,\kappa_J}(k_J, k_B, \mu)
= S_{R,\kappa_J}^\mathrm{pert}\bigl(k_J-\Omega_{\kappa_J}, k_B, \mu \bigr)
 \biggl[1+ \mathcal{O}\biggl(\frac{\lqcd^2}{k_J^2}, \frac{\alpha_s\lqcd}{k_J},\frac{\lqcd}{k_B} \biggr)\biggr]
\,.\end{align}
This correspond to a shift in the perturbative jet mass spectrum given by
\begin{align}
 m_J^2 & = 2 p_T^J \, \Bigl(\Tau^{\rm pert}_J + \frac{R}{2}\, \Omega_{\kappa_J}\Bigr) = (m_J^2)^{\rm pert} + p_T^J R\, \Omega_{\kappa_J}
\,,\end{align}
in agreement with \eqs{mshift}{omega_exp}.

In addition to the above nonperturbative effects, which are associated with hadronization, the jet mass
spectrum is also affected by underlying-event contributions associated with multiple partonic interactions,
which has perturbative and nonperturbative components. These effects scale like $R^4$~\cite{Dasgupta:2007wa}
and are thus not very relevant at small $R$. Note that contributions from primary soft radiation, which
share some underlying-event characteristics in that they also scale as $R^4$, are fully captured by the
soft function(s)~\cite{Stewart:2014nna}.

\section{Application to \SCETb and jet-based vetoes}
\label{sec:jetveto}

In this section, we consider other classes of jet vetoes, focussing our attention on regime 2 in \subsec{case2}, which has the largest number of hierarchies, $\Tau_J \ll p_T^J R^2$ and $R\ll R_0$. Specifically, we discuss the transverse energy veto as an example of a \SCETb-type beam measurement, as well as jet-based vetoes.

\subsection{Transverse energy veto}
\label{subsec:pt}

Here we discuss the mode setup and factorization formula for a veto on the transverse energy outside the jet, 
\begin{align} \label{eq:ETdef}
E_T 
\equiv \Tau_B^{(f_B=1)} 
= \sum_{i \notin \text{jet}} p_{Ti}
\,,\end{align}
i.e.~with $f_B(\eta) = 1$ in \eq{TauB}. This combines features of \SCETa for the jet mass measurement, \SCETb for the $E_T$ jet veto~\cite{Tackmann:2012bt}, and \SCETp for the inclusion of jet radius effects. Compared to the case of a generalized beam thrust veto, the scaling of the modes changes for a transverse energy veto. The $n_J$-collinear radiation has the same parametric scaling as before,
\begin{align}
n_J\text{-collinear:}
\qquad p_{n_J}^\mu \sim \Bigl(\frac{m_J^2}{p_T^J},p_T^J,m_J\Bigr)_J \sim \Bigl(\Tau_J, p_T^J, \sqrt{p_T^J \Tau_J}\Bigr)_J
\,,\end{align}
because it is fixed by the large jet momentum and jet mass measurement. Since we consider the hierarchy $\Tau_J \ll p_T^J R^2$ these modes are too collimated to resolve the jet boundary. The collinear initial-state radiation still has an energy $Q \sim p_T^J$ and the scaling is fixed by the measurement constraint through $E_T$,
\begin{align}
n_a \text{-collinear:} 
&\qquad
p_{n_a}^\mu \sim \Bigl(\frac{E_T^2}{p_T^J},p_T^J,E_T\Bigr)_B
\,, \nn \\
n_b \text{-collinear:} 
&\qquad
p_{n_b}^\mu \sim \Bigl(p_T^J, \frac{E_T^2}{p_T^J},E_T\Bigr)_B
\,.\end{align}
The scaling of the wide-angle soft radiation for narrow jets follows from the transverse energy measurement in the beam region,
\begin{align}
\text{soft:}\qquad  p_{s}^\mu\sim E_T (1,1,1)
\,.\end{align}
The initial-state collinear and soft modes are now only separated in rapidity leading to the emergence of rapidity divergences for the associated individual bare corrections, requiring additional regularization. The additional collinear-soft and soft-collinear modes that probe the jet boundary and are defined by the restrictions due to the measured jet mass and imposed jet veto are
\begin{align}
& n_J\text{-collinear-soft:}
\qquad
p_{cs}^\mu \sim \frac{\Tau_J}{R^2} (R^2,1,R)_J \sim \Bigl(\Tau_J, \frac{\Tau_J}{R^2}, \frac{\Tau_J}{R}\Bigr)_J  
\, ,\\
& n_J\text{-soft-collinear:}
\qquad p_{sc}^\mu \sim 
E_T (R^2,1,R)_J 
\,.\end{align}
We assume that $\Tau_J \sim E_T R^2$, such that the modes are degenerate,
\begin{align}
& n_J\text{-csoft:}
\qquad
p_{cs}^\mu \sim \frac{\Tau_J}{R^2} (R^2,1,R)_J  \sim E_T (R^2,1,R)_J 
\,,\end{align}
and large NGLs are avoided as in \subsec{case2}. This leads to the following factorized cross section
\begin{align} \label{eq:sigmaET}
\frac{\df \sigma_2 (\Phi,\kappa)}
{\df E_T \, \df\Tau_J}
&=
H_\kappa(\Phi, \mu)
\int\!\df E_{Ta}\, B_{\kappa_a}\Bigl(E_{Ta}, x_a, \mu,\frac{\nu}{\omega_a}\Bigr)
\int\!\df E_{Tb}\, B_{\kappa_b}\Bigl(E_{Tb}, x_b, \mu,\frac{\nu}{\omega_b}\Bigr)
\nn \\ &\quad\times
\int\!\df s_J\, J_{\kappa_J}(s_J, \mu)
\int \df k_J \,\df k_B\, S_{R,\kappa_J}(k_J,k_B,\mu)\,
\delta\Bigl(\Tau_J - \frac{s_J}{2p_T^J}- \frac{R\,k_J}{2} \Bigr)
\nn \\
&\quad\times  S_{B,\kappa} \Bigl(E_T - E_{Ta} - E_{Tb} - \frac{k_B}{R},\eta_J, \mu, \nu\Bigr) \, ,\nn \\
\frac{\df \sigma (\Phi,\kappa)}
{\df E_T \, \df\Tau_J}
&=\frac{\df \sigma_2 (\Phi,\kappa)}
{\df E_T \, \df\Tau_J}
\biggl[1+\mathcal{O}\biggl(\frac{E_T}{p_T^J},\frac{\Tau_J}{p_T^J R^2},R^2\biggr)\biggr]
\, .\end{align}
Once again, the indicated $\mathcal{O}({E_T}/{p_T^J},{\Tau_J}/{(p_T^J R^2)},R^2)$ nonsingular corrections can be obtained by considering the correspondence with other regimes or fixed-order calculations.
Compared to \eq{sigmaTau2} the same hard, jet, and collinear-soft functions appear, while the beam functions and soft functions are different and depend also on an additional rapidity renormalization scale $\nu$~\cite{Chiu:2011qc,Chiu:2012ir}. The natural scales for the beam functions are $\mu_B \sim E_T$ and $\nu_B \sim \omega_{a,b} \sim p_T^J$. The natural scales for the soft function $S_B$ are $\mu_S \sim \nu_S \sim E_T$. Since the rapidity regulator breaks boost invariance, $S_B$ still depends on $\eta_J$. 

At one loop, the matching coefficients in the beam functions encode only up to one real emission and therefore correspond to the transverse-momentum dependent beam functions in refs.~\cite{Becher:2010tm, Chiu:2012ir, Ritzmann:2014mka, Luebbert:2016itl}. We calculate the one-loop correction for the soft function $S_B$ in \app{softfct} using the $\eta$-regulator in refs.~\cite{Chiu:2011qc, Chiu:2012ir}. The result reads
\begin{align}\label{eq:S_BT}
S^{E_T}_{B,\kappa}(\ell_B, \eta_J,\mu,\nu) &= \delta(\ell_B) + \frac{\alpha_s(\mu)}{4\pi} \biggl\{\bfT_a \cdot \bfT_b  \biggl[\frac{8}{\mu} \,\cL_1\Bigl(\frac{\ell_B}{\mu}\Bigr)-\frac{8}{\mu} \,\cL_0\Bigl(\frac{\ell_B}{\mu}\Bigr) \ln \Bigl(\frac{\nu}{\mu}\Bigr)+ \frac{\pi^2}{6}\,\delta(\ell_B)\biggr]
\nn \\
& \quad + \bfT_a \cdot \bfT_J \, \biggl[ -\frac{8}{\mu}\,\mathcal{L}_0\Bigl(\frac{\ell_B}{\mu}\Bigr) \ln \Bigl(\frac{\nu \, e^{-\eta_J}}{\mu}\Bigr)+ \frac{\pi^2}{3}\,\delta(\ell_B)
\biggr] \biggr\}
\nn\\ & \quad
+  \biggr\{ (a,\eta_J) \leftrightarrow (b,-\eta_J) \biggl\}
\, .\end{align}
We verify in \app{anomdim} that this result is in agreement with the RG consistency of the factorized cross section.

\subsection{Jet-based vetoes}
\label{subsec:taujet}

In this section we consider the corresponding jet-based versions of the global \SCETa and \SCETb jet vetoes, as discussed e.g.~in refs.~\cite{Tackmann:2012bt, Gangal:2014qda}. These local jet-veto variables are based on identifying additional jets $j(\Rv)$ using a jet algorithm with radius $\Rv$ in the beam region and considering the largest contribution from a single jet. (The jet algorithms and radii for the identification of the hard signal jet and for the vetoing of additional jets can in principle be different.)
We consider the jet vetoes $\Tau_B^{\rm cut}$ and $p_{T}^{\mathrm{cut}}$ defined through
\begin{align} \label{eq:TauBjetdef}
\max_{j(\Rv)} \bigl\{ |\vec{p}_{T j} | f_B(\eta_{j})\bigr\} &\leq \Tau_B^{\rm cut} 
\,  ,\nn \\
\max_{j(\Rv)} |\vec{p}_{Tj}| &\leq   p_{T}^{\mathrm{cut}}
\,.\end{align}
The clustering effects due to the jet veto affect both collinear initial-state radiation as well as soft and csoft radiation (outside the identified jet), introducing a dependence on $\Rv$ in the beam and soft functions. For a small value of $\Rv$, the jet clustering of collinear and soft radiation is power-suppressed by $\mathcal{O}(\Rv^2)$~\cite{Tackmann:2012bt, Liu:2012sz} so that the veto on additional jets is separately imposed on the collinear initial-state radiation and soft radiation. One can also argue that the clustering of soft and csoft modes is predominantly performed within each sector for $\Rv \ll 1$, such that the measurement also factorizes between these sectors. The price to pay for this factorization is the appearance of clustering logarithms $\ln \Rv$ (closely related to NGLs) starting at $\mathcal{O}(\alpha_s^2)$, whose systematic resummation is beyond the scope of this paper. In the following we consider only the resummation of the jet radius logarithms $\ln R$ related to the observed jet. 

The EFT mode setup for the jet-based vetoes is identical to that for the corresponding global veto. For the $\Tau_B^{\rm cut}$ veto the modes are as for the generalized beam thrust veto in \sec{fact} and summarized in table \ref{tab:regimes}, with the identification $\Tau_B \to \Tau_B^{\rm cut}$, leading to the factorized cross section 
\begin{align}
\frac{\df \sigma_2}
{ \df\Tau_J} (\Tau_B^\text{cut}, \Phi,\kappa)
&=
H_\kappa(\Phi, \mu) \,
B_{\kappa_a}(\omega_a \Tau_B^\text{cut}, x_a,\Rv, \mu) \,
B_{\kappa_b}(\omega_b \Tau_B^\text{cut}, x_b,\Rv, \mu)
\nn \\ &\quad\times
\int\!\df s_J\, J_{\kappa_J}(s_J, \mu)\int \df k_J \, S_{R,\kappa_J}\Big(k_J,\frac{\Tau_B^\text{cut} R}{f_B(\eta_J)},\Rv, \mu\Big) \delta\Bigl(\Tau_J - \frac{s_J}{2p_T^J}- \frac{R\,k_J}{2} \Bigr)
\nn \\ &\quad\times  S_{B,\kappa} (\Tau_B^\text{cut}, \eta_J,\Rv, \mu) \, ,\nn \\
\frac{\df \sigma}
{ \df\Tau_J} (\Tau_B^\text{cut}, \Phi,\kappa)
& = \frac{\df \sigma_2}
{ \df\Tau_J} (\Tau_B^\text{cut}, \Phi,\kappa) \biggl[1+\mathcal{O}\biggl(\frac{\Tau_B^{\text{cut}}}{p_T^J},\frac{\Tau_J}{p_T^J R^2},R^2,\Rv^2\biggr)\biggr]\, .
\end{align}
As in previous cases, the nonsingular corrections indicated in the last line can be obtained by using the correspondence with other regimes and fixed-order calculations.
At one loop, the beam functions $B$, the soft function $S_B$ and the collinear soft function $S_R$ describe a single emission, such that the clustering algorithm in the beam region does not play any role and their expressions are the cumulant of the matrix elements in \subsec{case2}. We emphasize that the structure of the renormalization differs between the global and local jet-based vetoes.
Starting at two loops, the analytic structure of the expressions changes, accounting now for the jet clustering as indicated by the additional dependence on $\Rv$. The renormalization is multiplicative in the arguments associated with the jet veto, as required by the structure of the factorization theorem~\cite{Tackmann:2012bt, Gangal:2014qda}. For example, the renormalization of the csoft function $S_R$ is multiplicative in $k_B^\mathrm{cut}$ but involves a convolution in $k_J$ as can be seen from the associated RG equation
\begin{align}
  \mu\frac{\df}{\df\mu}\, S_{R,\kappa_J}(k_J,k_B^\text{cut},\Rv, \mu) = \int \df k_J'\, \ga_{S_R,\kappa_J}(k_J -k_J', k_B^\text{cut},\Rv,\mu)\, S_{R,\kappa_J}(k_J',k_B^\text{cut},\Rv, \mu)
\, .\end{align}

Next, we consider the jet-based transverse momentum veto, $p_T^{\rm cut}$, which is the standard choice used by the experiments. This combines the features discussed above with the mode setup for the \SCETb veto in \subsec{pt} (with $E_T\to p_T^{\rm cut}$) and leads to the factorization formula
\begin{align}\label{eq:fact_pTjet}
\frac{\df \sigma_2}
{\df\Tau_J} (p_T^{\rm cut},\Phi,\kappa)
&=
H_\kappa(\Phi, \mu) \,
B_{\kappa_a}\Bigl(p_T^{\rm cut}, x_a,\Rv, \mu,\frac{\nu}{\omega_a}\Bigr) \,
B_{\kappa_b}\Bigl(p_T^{\rm cut}, x_b,\Rv, \mu,\frac{\nu}{\omega_b}\Bigr)
\nn \\
& \quad \times	\int\!\df s_J\, J_{\kappa_J}(s_J, \mu) \int\df k_J \, S_{R,\kappa_J}(k_J,p_T^{\rm cut} R,\Rv, \mu)\, \delta\Bigl(\Tau_J - \frac{s_J}{2p_T^J}- \frac{R\,k_J}{2} \Bigr) \, ,
\nn \\ &\quad\times  \,
S_{B,\kappa} (p_T^{\rm cut}, \eta_J,\Rv, \mu,\nu)\, \nn \\
\frac{\df \sigma}
{\df\Tau_J} (p_T^{\rm cut},\Phi,\kappa)
&=\frac{\df \sigma_2}
{\df\Tau_J} (p_T^{\rm cut},\Phi,\kappa)
\biggl[1+\mathcal{O}\biggl(\frac{p_T^{\rm cut}}{p_T^J},\frac{\Tau_J}{p_T^J R^2},R^2,\Rv^2\biggr)\biggr]
\, .\end{align}
The one-loop correction to the wide-angle soft function $S_{B}$  reads in direct analogy to \eq{S_BT}
\begin{align}\label{eq:S_pT}
S^{(1)}_{B,\kappa}(p_T^{\rm cut}, \eta_J,\mu,\nu) &=  \frac{\alpha_s(\mu)}{4\pi} \biggl\{\bfT_a \cdot \bfT_b  \biggl[4\ln^2\biggl(\frac{p_T^{\rm cut}}{\mu}\biggr)-8 \ln\biggl(\frac{p_T^{\rm cut}}{\mu}\biggr)\ln \Bigl(\frac{\nu}{\mu}\Bigr)+ \frac{\pi^2}{6}\biggr]
\\
& \quad + \bfT_a \cdot \bfT_J \, \biggl[ -8 \ln\biggl(\frac{p_T^{\rm cut}}{\mu}\biggr)\ln \Bigl(\frac{\nu e^{-\eta_J}}{\mu}\Bigr)+ \frac{\pi^2}{3}
\biggr] \biggr\}  + \biggl\{(a,\eta_J) \leftrightarrow (b,-\eta_J) \biggr\} \nn
\, .\end{align}
To demonstrate explicitly which logarithms are resummed by \eq{fact_pTjet} at higher orders, we give the jet radius and jet mass dependent logarithmic terms predicted by it at NNLO in \app{FO}. 

With the analogous relation to \eq{cons12} the results in \eqs{S_J}{S_pT} allow us also to write the one-loop expression for the associated unfactorized soft function $S_\kappa$ (encoding the contributions from all soft modes) as (with $L_{p_T} \equiv \ln(p_T^{\rm cut}/\mu)$)
\begin{align}
S^{(1)}_{\kappa}(\ell_J,p_T^{\rm cut}, \eta_J,\mu,\nu) &= \frac{\alpha_s(\mu)}{4\pi} \biggl\{\bfT_a \cdot \bfT_b  \biggl[4L^2_{p_T} \delta(\ell_J)-8L_{p_T} \ln \Bigl(\frac{\nu}{\mu}\Bigr)\delta(\ell_J) \nn \\
& \quad  + s_{ab,1}(R) \Bigl(L_{p_T} \delta(\ell_J)  - \frac{1}{\mu}\cL_0\Bigl(\frac{\ell_J}{\mu}\Bigr) \Bigr) +\Bigl(\frac{\pi^2}{6} + \Delta s^{p_T}_{ab,\delta}(R,\eta_J)\Bigr) \delta(\ell_J)\biggr]
\nn \\ & \quad
 + \bfT_a \cdot \bfT_J \, \biggl[\frac{8}{\mu}\,\cL_1\Bigl(\frac{\ell_J}{\mu}\Bigr)+4L^2_{p_T} \delta(\ell_J) -8 L_{p_T}  \ln \Bigl(\frac{\nu e^{-\eta_J}}{\mu \, R}\Bigr)\delta(\ell_J) 
\nn \\ & \quad
+\frac{1}{\mu}\,\mathcal{L}_0\Bigl(\frac{\ell_J}{\mu}\Bigr) \biggl(-8\ln\frac{R}{2} + \Delta s_{aJ}(R,\eta_J) \biggr) -\Delta s_{aJ}(R,\eta_J) L_{p_T} \delta(\ell_J) \nn \\
& \quad + \biggl(4 \ln^2 R +4 \ln^2\!\frac{R}{2} +\Delta  s^{p_T}_{aJ,\delta}(R,\eta_J) \biggr)\delta(\ell_J) \biggr]
 \biggr\}
\nn\\ & \quad
+  \biggl\{(a,\eta_J) \leftrightarrow (b,-\eta_J)\biggr\}
\,,\end{align}
where $s_{ab,1}(R)=2R^2$, $\Delta s^{p_T}_{ab,\delta}(R,\eta_J)$, $\Delta s_{aJ}(R,\eta_J) $ and $\Delta  s^{p_T}_{aJ,\delta}(R,\eta_J)$ will be corrections that start at $\mathcal{O}(R^2)$ and are given in ref.~\cite{Bertolini:forthcoming}.

A related soft function has been also computed for small $R$ in ref.~\cite{Liu:2013hba} in the context of an exclusive $H + 1$ jet analysis without an explicit measurement of the jet mass. The associated result corresponds to the combination of the one-loop soft and soft-collinear corrections (where the latter are encoded in the $S^{(B)}_{R,\kappa_J}$ component of the csoft function) to the jet veto measurement
\begin{align}\label{eq:S_out}
S^{(1)}_{\rm out}(p_T^{\rm cut}, R, \eta_J, \mu, \nu) & \equiv S^{(B,1)}_{R,\kappa_J}(p_T^{\rm cut} R, \mu) + S^{(1)}_{B,\kappa}(p_T^{\rm cut}, \eta_J, \mu, \nu) \\
& = \frac{\alpha_s(\mu)}{4\pi} \biggl\{(\bfT_a^2 +\bfT_b^2)\biggr[-4\ln^2\biggl(\frac{p_T^{\rm cut}}{\mu}\biggr)+8\ln\biggl(\frac{p_T^{\rm cut}}{\mu}\biggr)\ln\biggl(\frac{\nu}{\mu}\biggr)-\frac{\pi^2}{6}\biggl]  \nn \\
& \quad + 8\eta_J(\bfT_b^2-\bfT_a^2)\ln\biggl(\frac{p_T^{\rm cut}}{\mu}\biggr) -4 \bfT_J^2 \ln R\biggl[\ln R +2 \ln\biggl(\frac{p_T^{\rm cut}}{\mu}\biggr)  \biggr] \biggr\}
\,,\nn \end{align}
using \eqs{S_hemi}{S_pT}. This result agrees with the computation in ref.~\cite{Liu:2013hba} (see eq.~(20) therein). Their result is expressed in terms of two-dimensional integrals, which numerically agree with our analytic expression in \eq{S_out} up to ${\cal O}(R^2)$ terms.

\subsection{Fixed-order cross section for small-$R$ jets}
\label{subsec:nlosing}

In \subsec{consistency}, we showed numerical results for the one-loop soft function, demonstrating consistency between regimes 1 and 2, and finding that the small-$R$ results provide a good approximation even up to rather large values of $R$. To lend more credence to this conclusion, we show numerical results for the cross section in this section, comparing the results of regime 2 with regime 1. The comparison is performed at NLO and thus only tests the validity of the small $R$ expansion at fixed order and not the effect due to $\ln R$ resummation. A detailed phenomenological study of the effects due to resummation of $\ln R$ terms will be presented in the future.

\begin{figure}
 \centering
 \includegraphics[width=0.5\textwidth]{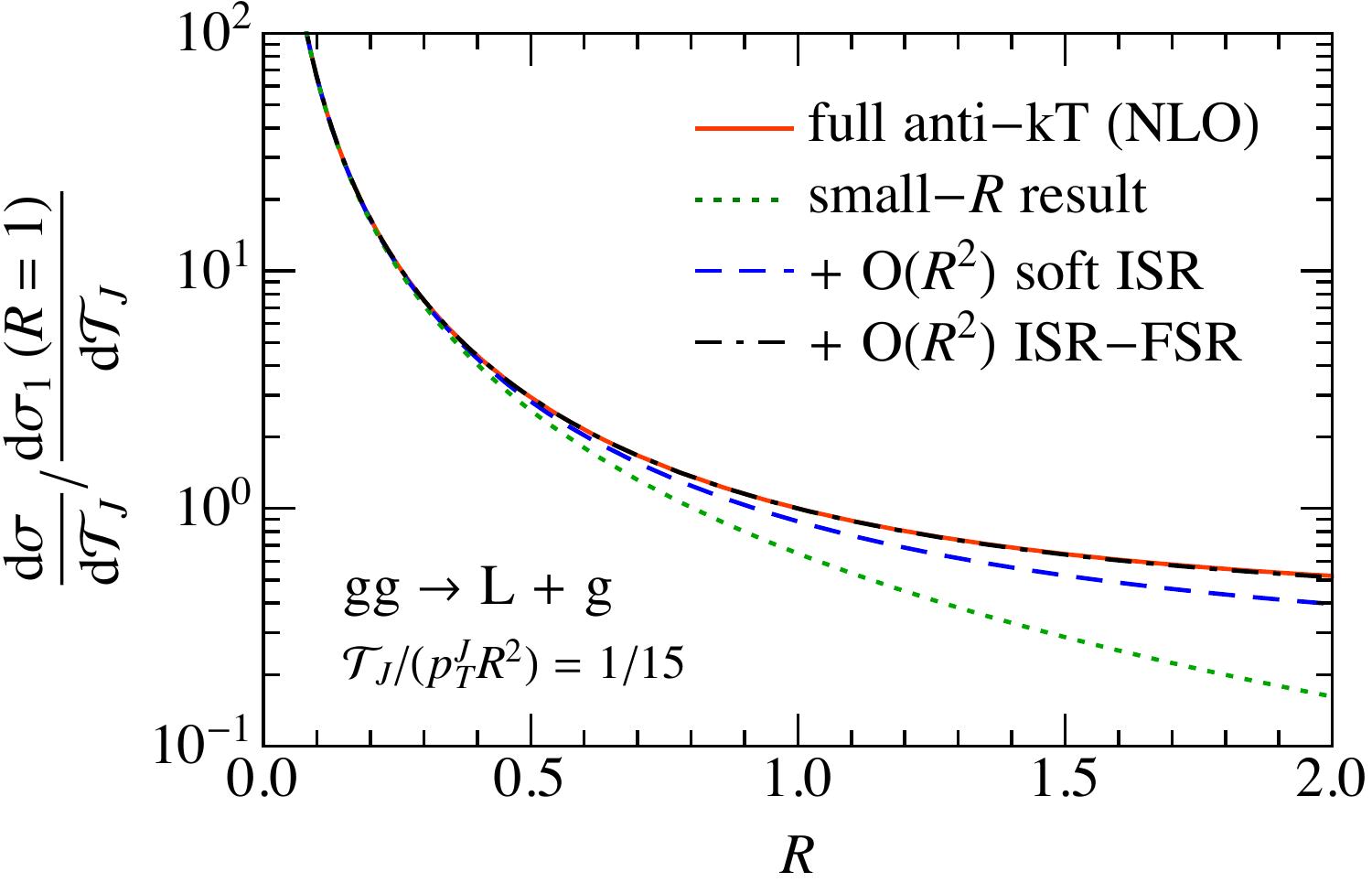}\hfill%
  \includegraphics[width=0.5\textwidth]{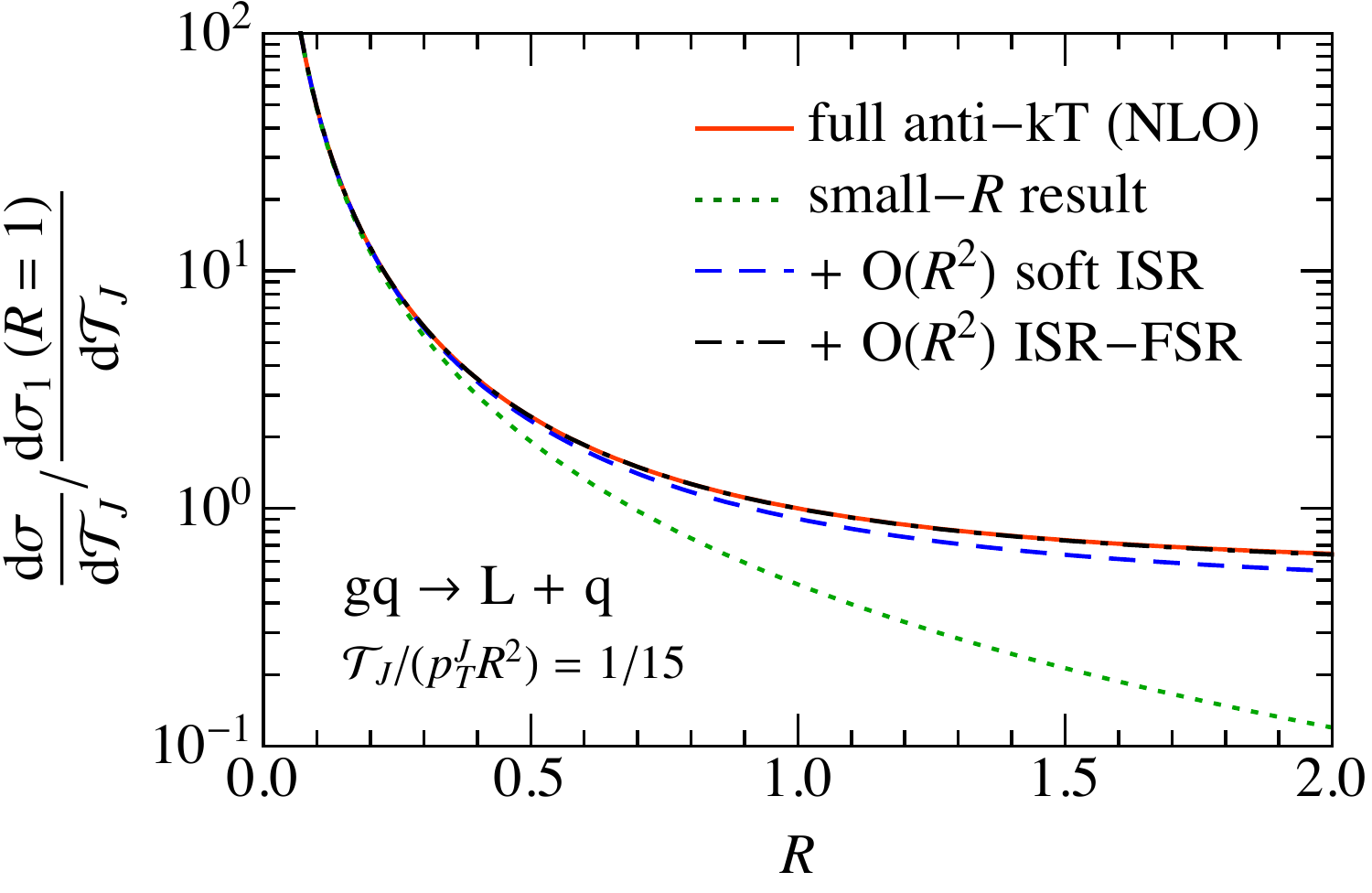}%
  \\
  \includegraphics[width=0.5\textwidth]{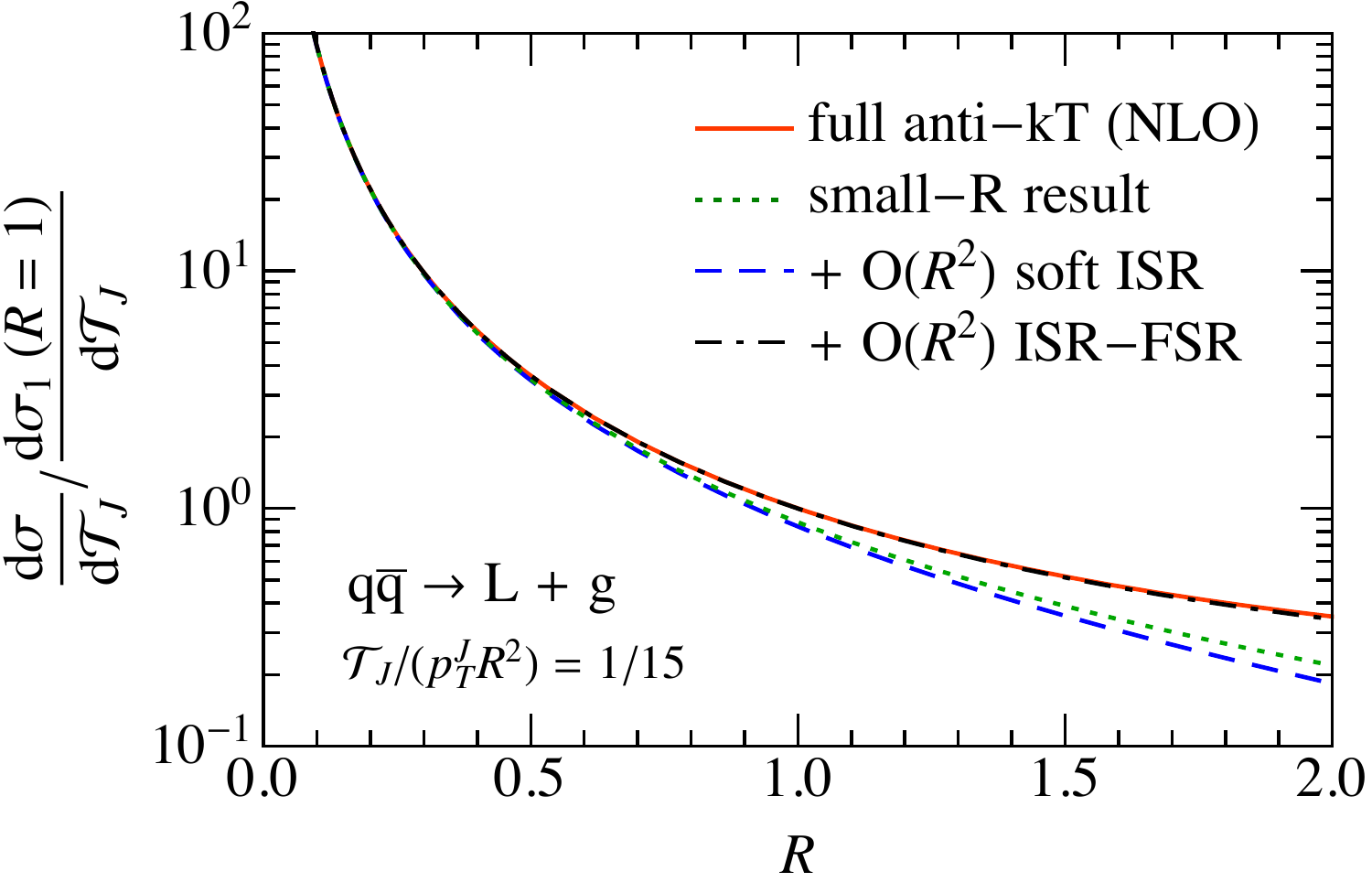}
  \caption{Jet radius dependence of the spectrum at next-to-leading order, as defined in \eq{plot}. Shown are the full anti-$k_T$ result (red solid), the small $R$ result (green dotted), including the $\ord{R^2}$ soft ISR (blue dashed) and including the full set of analytic corrections at $\ord{R^2}$ (black dot-dashed), always normalized to the full anti-$k_T$ result for $R=1$. We take $\Tau_J/(p_T^J R^2) =1/15 \ll 1$, which allows us to restrict ourselves to the singular terms.
\label{fig:spectrumFO} }
\end{figure}

\begin{figure}
 \centering
  \includegraphics[width=0.5\textwidth]{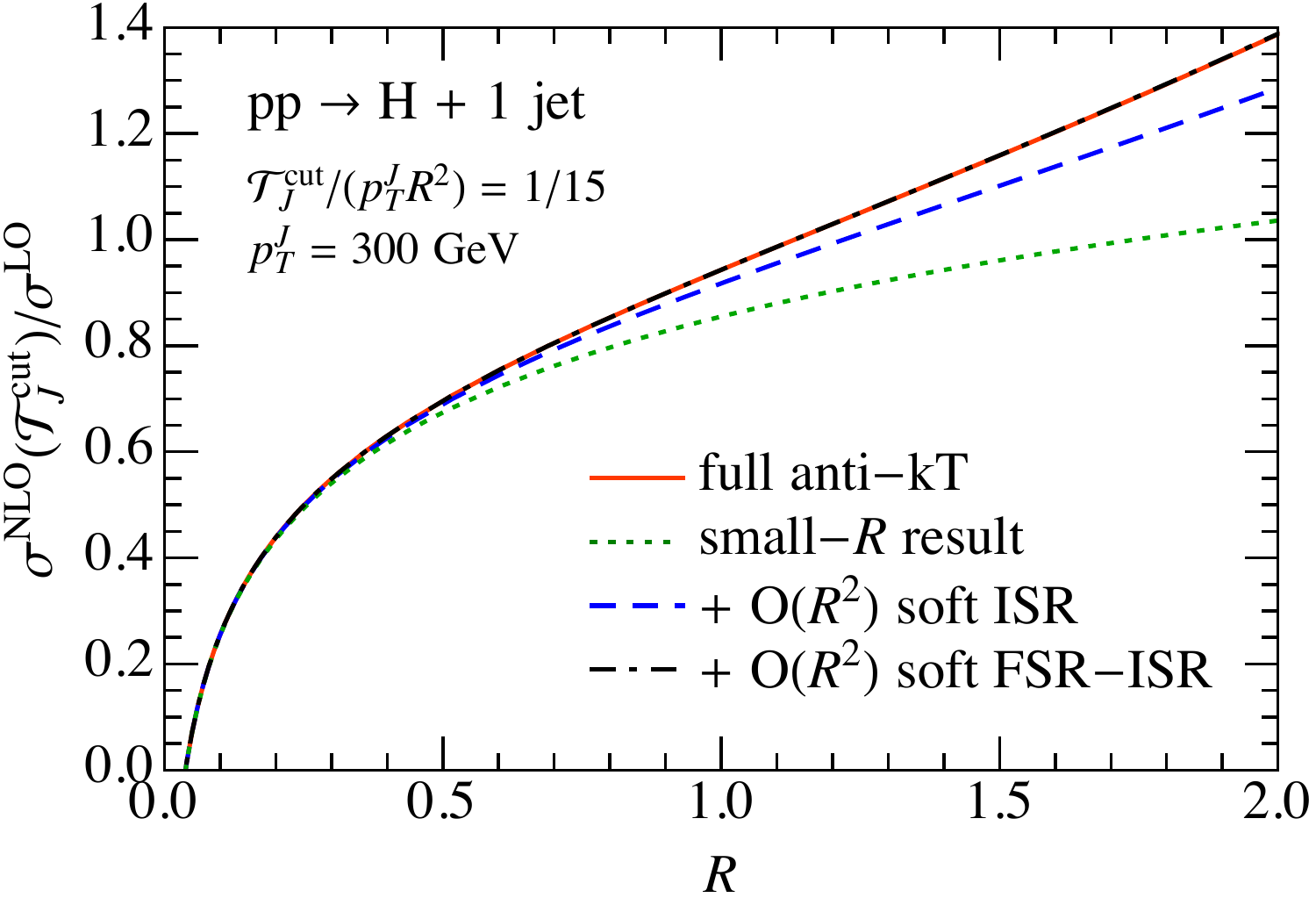}\hfill%
  \includegraphics[width=0.5\textwidth]{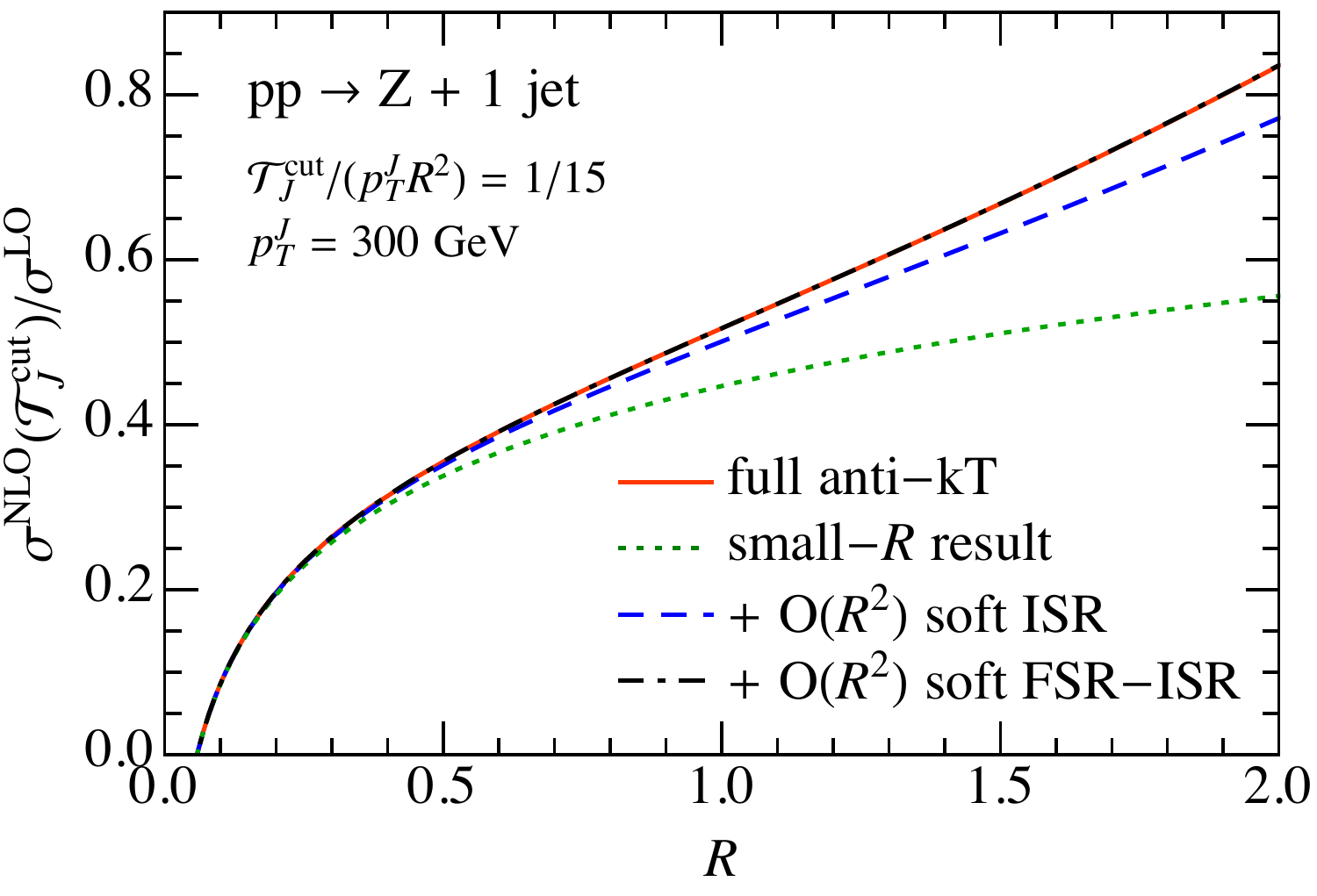}
  \caption{Jet radius dependence of the fixed-order cumulant at $\mathcal{O}(\alpha_s)$, normalized to the tree-level result, for an anti-$k_T$ jet with $p_T^J = 300$ GeV and $\Tau^{\rm cut}_J/(p_T^J R^2)= 1/15 $ for $pp\to H+1$ jet and $pp\to Z+1$ jet.
\label{fig:cumulantFO} }
\end{figure}

To investigate the range where the small $R$ expansion is valid, we show results for the spectrum and its cumulative distribution
\begin{align}\label{eq:plot}
  \text{\fig{spectrumFO}:}&\qquad  \Big(\frac{\df \si_i^\text{NLO}(R)}{\df \Tau_J}\Big)\Big/ \Big(\frac{\df \si_1^\text{NLO}(R=1)}{\df \Tau_J}\Big)
\,, \nn \\ 
  \text{\fig{cumulantFO}:}&\qquad  \Big(\int_0^{\Tau_J^\text{cut}}\!\df\Tau_J\, \frac{\df \si_i^\text{NLO}(R)}{\df \Tau_J}\Big)\Big/ \Big(\int_0^{\Tau_J^\text{cut}}\!\df\Tau_J\, \frac{\df \si^\text{LO}(R)}{\df \Tau_J}\Big)
\,.\end{align}
The (N)LO cross section is obtained by expanding the factorization formula for regime $i=1,2$ to this order and taking all scales equal to $\mu=p^J_T$. In the ratio of jet mass spectra most ingredients drop out, e.g. for $i=2$
\begin{align}\label{eq:ratio_NLO}
\Big(\frac{\df \si_2^\text{NLO}(R)}{\df \Tau_J}\Big)\Big/ \Big(\frac{\df \si_1^\text{NLO}(R=1)}{\df \Tau_J}\Big) &= 
\biggl[ 2 p_T^J\, J_{\kappa_J}^{(1)}(2p_T^J \Tau_J, \mu) +
 \frac{2}{R}\,S_{R,\kappa_J}^{(J,1)}\Big(\frac{2\Tau_J}{R},\mu\Big)\biggr]
 \\ & \quad \times
\biggl[ 2 p_T^J\, J_{\kappa_J}^{(1)}(2p_T^J \Tau_J, \mu) +
S_\kappa^{(J,1)} (\Tau_J,  \eta_J, R=1, \mu)\biggr]^{-1}
,\nn\end{align}
because only for a single real emission radiated into the jet region one does obtain a nonvanishing spectrum at NLO. The ratio in \eq{ratio_NLO} is in particular independent of the jet veto and hard process, and only depends on the partonic channel, the jet radius R, the ratio $\Tau_J/(p_T^J R^2)$, which we take to be $1/15$ for our plots. This value corresponds for example to $\Tau_J=5$ GeV and $p_T^J=300$ GeV for a jet radius $R \sim 0.5$, which would satisfy the requirement $\Tau_J \sim p_T^{\rm cut} R^2/2$ for avoiding large NGLs with a jet veto $p_T^{\rm cut}=30$ GeV.

The results are shown in \fig{spectrumFO} for anti-$k_T$ jets with the full $R$ dependence (red solid) from regime 1 and the leading small-$R$ result (green dotted) from regime 2. Furthermore, we display the small-$R$ result including the $\ord{R^2}$ correction arising from soft initial-state radiation (blue dashed), which corresponds to including the $s_{ab,1}=2 R^2$ term in \eq{S_kappa}, and including all analytic corrections to $\ord{R^2}$ (black dot-dashed), which will be given in~\cite{Bertolini:forthcoming}.
The small-$R$ approximation works quite well for $R \lesssim 0.5$, and its range of validity is considerably extended by including the soft ISR correction. This is not surprising, because the contribution of soft ISR to the jet mass only starts at $\ord{R^2}$, whereas other $\ord{R^2}$ corrections only account for deviations in the shape of the jet region and are comparably small. Including also all remaining corrections at $\mathcal{O}(R^2)$ coming from soft ISR-FSR interference the full result for anti-$k_T$ jets is almost exactly approximated even for a jet radius $R \sim 2$. This confirms the statement that the effective expansion parameter is $R/R_0$ with $R_0\gtrsim 2$. For the $\kappa=\{q,\bar{q};g\}$ channel the soft ISR correction appears with a numerically small color factor $C_F - C_A/2 = -1/6$, compared to $C_A/2 = 3/2$ for the other channels, as pointed out in ref.~\cite{Stewart:2014nna}, so that already the leading result of the small-$R$ expansion gives a good approximation even for large values of the jet radius.

In \fig{cumulantFO}, we show the jet radius dependence of the cumulative distribution for 
$pp \to H +1$ jet (left panel) and $pp \to Z +1$ jet (right panel), using the second line of \eq{plot}. We employ the jet-based transverse momentum veto discussed in \subsec{taujet}, and use $\Tau_J/(p_T^J R^2) =1/15$, $p_T^{\rm cut}=30$ GeV, $p_T^J = 300$ GeV, $\eta_J=0$, $Y_L=0$, $E_{\rm cm} =13$ TeV. For simplicity, we consider the production of on-shell EW bosons without any subsequent decay.
Compared to the differential spectrum, the small-$R$ approximation seems to work over an even larger range.  Once again, including the soft ISR correction greatly extends the range where the small-$R$ approximation works well. (Also for $pp \to Z +$jet the soft ISR correction gives the dominant $\mathcal{O}(R^2)$ effect, since the contribution from the $\{q,\bar{q};g\}$-channel is small compared to the one from the $\{g,q;q\}$-channel, where the soft ISR correction is large.) The fact that the full result is almost exactly reproduced by including the full set of $\ord{R^2}$ corrections is somewhat specific to anti-$k_T$ jets with a $p_T^{\rm cut}$-veto. For different jet algorithms and vetoes there is in general some visible difference toward large $R$ between the full result and the one containing the corrections to $\mathcal{O}(R^2)$, see for example the $R$-dependence for the C-parameter in the right panel of \fig{deltaS1C}.

\section{Conclusions}
\label{sec:conclusions}

We presented a factorization framework to provide a complete description of jet mass spectra in hadronic collisions including realistic jet algorithms and jet vetoes. It allows to systematically treat jet radius effects in the jet mass spectrum, including the resummation of jet radius logarithms, the jet boundary effects that cut off the spectrum at $m_J \lesssim p_T^J R$, and the inclusion of $\ord{R^2}$-suppressed power corrections.
This description is based on \SCETp, which is an extension of standard Soft-Collinear Effective Theory with additional modes that are simultaneously soft and collinear. We utilized this theory for the jet mass measurement in the process $pp \to L + 1$ jet and discussed the factorization formulae and all relevant ingredients allowing for the systematic higher-order resummation of logarithms of the jet mass, jet radius, and jet veto at NNLL for global vetoes and NLL$^\prime$ for jet based vetoes, and beyond once the relevant ingredients become known.

In the phenomenologically important peak and tail region of the jet mass spectrum with $m_J \ll p_T^J R$, and for appropriate jet veto scales determined by a definite power counting, nonglobal structures do not contain large logarithms and can thus be included at fixed order. In the far tail region, $m_J \sim p_T^J R$, recent progress in the resummation of NGLs can be directly applied to incorporate their dominant effect.
Comparing the perturbative soft corrections at one loop, we found that an expansion in terms of small $R$ gives a good approximation in the peak and tail region for typically adopted jet radii $R\lesssim 1$.

A detailed phenomenological study for experimentally measured jet mass spectra at the LHC including the effects due to $\ln R$ resummation and the relevant power corrections as well as the associated uncertainties will be presented in the future.

\begin{acknowledgments}
We thank Ian Moult and Duff Neill for discussions.
This work was supported by the German Science Foundation (DFG) through the Emmy-Noether Grant No.~TA 867/1-1, the Collaborative Research Center (SFB) 676 Particles, Strings and the Early Universe. 
This work was supported by the Office of Nuclear Physics of the U.S. Department of Energy under the Grant No.~DE-SC0011090. I.S.~was also supported by the Simons Foundation through the Investigator grant 327942.
This work is part of the D-ITP consortium, a program of the Netherlands Organization for Scientific Research (NWO) that is funded by the Dutch Ministry of Education, Culture and Science (OCW).  This work was also supported in part by a Global MISTI Collaboration Grant from MIT.
\end{acknowledgments}

\appendix

\section{Relations between anomalous dimensions}
\label{app:anomdim}	

\subsection{Anomalous dimensions for the generalized beam thrust veto} \label{app:anomdim1}	

The anomalous dimensions of the matrix elements for a global \SCETa beam measurement are defined in analogy to \eq{U_evolS}. For the jet and soft functions appearing in the factorization formula of \subsec{case1} the anomalous dimensions do not depend on the jet radius $R$ and have the structure
\begin{align}\label{eq:anom_J}
\gamma_{J}^{\kappa_J} (s, \mu)  &= -2 \,\bfT_J^2\, \Gamma_{\rm cusp}[\al_s(\mu)] \,\frac{1}{\mu^2} \, \mathcal{L}_0\Bigl(\frac{s}{\mu^2}\Bigr) +  \gamma_{J}^{\kappa_J}[\al_s(\mu)]\, \delta(s)
\,, \\
\gamma_{S}^{\kappa} (\ell_J,\ell_B, \mu)  &= -2\, \Gamma_{\rm cusp}[\al_s(\mu)] \biggl[ \bigl(2 \,\bfT_a \cdot \bfT_b-\bfT_J^2\bigr)\, \frac{1}{\mu} \, \mathcal{L}_0\Bigl(\frac{\ell_B}{\mu}\Bigr)\delta(\ell_J) -\bfT_J^2 \,\frac{1}{\mu} \, \mathcal{L}_0\Bigl(\frac{\ell_J}{\mu}\Bigr) \delta(\ell_B)
\nn \\ & \quad 
- \biggl(\bfT_a \cdot \bfT_J \ln \frac{e^{-\eta_J}}{2}+ \bfT_b \cdot \bfT_J \ln \frac{e^{\eta_J}}{2}\biggr)\, \delta(\ell_J)\,\delta(\ell_B)\biggr]
 + \gamma_{S}^\kappa[\al_s(\mu)] \,\delta(\ell_J)\,\delta(\ell_B) \nn
\,,\end{align}
where in the second line we assumed Casimir scaling for the cusp anomalous dimension which holds at least up to three loops. The cusp and the noncusp anomalous dimension are known at least up to three and two loops, respectively. Analytic expressions using the same notation can be found for example in the appendices of refs.~\cite{Stewart:2010qs, Berger:2010xi, Jouttenus:2013hs, Chien:2015cka}. Here we only infer the structure for the anomalous dimensions of the remaining jet and soft functions involved in the factorization formulae in \eqref{eq:sigmaTau2} and \eqref{eq:sigmaTau3}.

The relation in \eq{cons12} implies that 
\begin{align}
\gamma_{S}^{\kappa} (\ell_J,\ell_B,\mu) & = \gamma_{S_B}^{\kappa} (\ell_B,\mu) \,\delta(\ell_J) +  \frac{2}{f_B(\eta)}\, \gamma_{S_{R}}^{\kappa_J} \biggl(\frac{2 \ell_J}{R},\frac{R\, \ell_B}{f_B(\eta_J)},\mu\biggr)
\,.\end{align}
$S_R$ is the double hemisphere soft function, for which the $\mu$-dependence factorizes, i.e.,
\begin{align} \label{eq:SR_app_factorize}
\gamma_{S_R}^{\kappa_J} (k_J,k_B,\mu) = \gamma^{\kappa_J}_{\rm hemi} (k_J,\mu)\, \delta(k_B) + \gamma^{\kappa_J}_{\rm hemi} (k_B,\mu)\, \delta(k_J)
\,,\end{align}
with
\begin{align}\label{eq:anom_Shemi}
\gamma^{\kappa_J}_{\rm hemi} (k,\mu) =
2 \,\bfT_J^2\,\Gamma_{\rm cusp} [\al_s(\mu)]  \,\frac{1}{\mu} \, \mathcal{L}_0\Bigl(\frac{k}{\mu}\Bigr) +  \gamma_{{\rm hemi}}^{\kappa_J}[\al_s(\mu)]  \,\delta(k)
\,,\end{align}
where $\gamma^{\kappa_J}_{\rm{hemi}}(\alpha_s)$ is half of the noncusp anomalous dimension of the standard double hemisphere soft function. Thus, the anomalous dimension for the function $S_{B,\kappa}$ reads
\begin{align}
\gamma_{S_B}^{\kappa} (\ell_B,\mu) & =  2 \,\Gamma_{\rm cusp}[\al_s(\mu)]  \biggl[-2\, \bfT_a \sdt \bfT_b\,  \frac{1}{\mu} \, \mathcal{L}_0\Bigl(\frac{\ell_B}{\mu}\Bigr) +  \bigl(\bfT_J^2 \,\ln f_B(\eta_J) 
+  (\bfT_b -\bfT_a) \cdot \bfT_J \, \eta_J\bigr)  \delta(\ell_B) \biggr]
\nn \\ & \quad
 + \bigl\{\gamma_{S}^{\kappa}[\al_s(\mu)]  -2 \gamma_{{\rm hemi}}^{\kappa_J}[\al_s(\mu)] \bigr\} \delta(\ell_B)
\,.\end{align}
Using the one-loop cusp anomalous dimension, $\Gamma_{\rm cusp}[\al_s(\mu)]=\alpha_s/\pi +\mathcal{O}(\alpha_s^2)$, and the fact that the soft noncusp dimensions vanish at this order, this is in agreement with the $\mu$-dependence of \eqs{S_BC}{S_BB}.

The relation in \eq{cons23} implies that
\begin{align}
\gamma_{J_R}^{\kappa_J} (s, p_T^J R, \mu) = \gamma_{J}^{\kappa_J} (s, \mu)  +\frac{1}{p_T^J R} \,\gamma_{{\rm hemi}}^{\kappa_J} \biggl(\frac{s}{p_T^J R}, \mu\biggr) 
\, ,\end{align}
such that the anomalous dimension of the jet function $J_R$
\begin{align}
\gamma_{J_R}^{\kappa_J} (s, p_T^J R, \mu)
= \biggl\{-\bfT_J^2 \, \Gamma_{\rm cusp}[\al_s(\mu)] \ln\frac{(p_T^J R)^2}{\mu^2} +\gamma_{J}^{\kappa_J}[\al_s(\mu)] +\gamma_{{\rm hemi}}^{\kappa_J}[\al_s(\mu)]  \biggr\} \delta(s)
\, ,\end{align}
in analogy to the result for the ``unmeasured'' jet function in ref.~\cite{Chien:2015cka}.

\subsection{Anomalous dimensions for the transverse energy veto}

We now determine the structure of the anomalous dimension for $S_{B}$ for the transverse energy veto, and check that this is consistent with the one-loop result in \eq{S_BT}. The beam function $\nu$ and $\mu$ anomalous dimensions read~\cite{Tackmann:2012bt}
\begin{align}\label{eq:anom_B}
\gamma_{\nu,B}^{\kappa_B}(\ell_B,\mu) &=-2 \bfT_B^2 \,  \Gamma_{\rm cusp}[\al_s(\mu)]\, \frac{1}{\mu} \, \mathcal{L}_0\Bigl(\frac{\ell_B}{\mu}\Bigr) +\gamma_{\nu}^{\kappa_B}[\al_s(\mu)]\, \delta(\ell_B)
\, , \nn \\
\gamma_{B}^{\kappa_B}\Bigl(\ell_B,\mu,\frac{\nu}{\omega}\Bigr) &=\biggl\{2\bfT_B^2\, \Gamma_{\rm cusp}[\al_s(\mu)]\,\ln\frac{\nu}{\omega} + \gamma_{B}^{\kappa_B}[\al_s(\mu)]\biggr\} \,\delta(\ell_B)
\,.\end{align}
Here $\bfT_B^2 = \bfT^2_{a,b}$ and $\kappa_B = \kappa_{a,b}$ encode the flavor of the colliding parton coming from the respective beam, and $\omega_B = \omega_{a,b}$ its large momentum component.
This directly leads to the $\nu$-anomalous dimension of $S_B$, 
\begin{align}
\gamma_{\nu,S_B}^{\kappa}(\ell_B,\mu) &=- \gamma_{\nu,B}^{\kappa_a}(\ell_B,\mu)- \gamma_{\nu,B}^{\kappa_b}(\ell_B,\mu)
\\\nn
&=2\Gamma_{\rm cusp}[\al_s(\mu)]\,(\bfT_a^2+\bfT_b^2) \,  \frac{1}{\mu} \, \mathcal{L}_0\Bigl(\frac{\ell_B}{\mu}\Bigr) - (\gamma_{\nu}^{\kappa_a}[\al_s(\mu)] + \gamma_{\nu}^{\kappa_b}[\al_s(\mu)])\, \delta(\ell_B)\, .
\end{align}
Using the one-loop cusp anomalous dimension, $\Gamma_{\rm cusp}[\al_s(\mu)]=\alpha_s/\pi +\mathcal{O}(\alpha_s^2)$, and $\gamma_{\nu}^{\kappa}[\al_s(\mu)]=\mathcal{O}(\alpha_s^2)$ this is in agreement with the $\nu$-dependence of \eq{S_BT}. To check the $\mu$-dependence we give also the hard function anomalous dimension, 
\begin{align}\label{eq:anom_H}
\gamma_{H}^{\kappa}(\omega_i,\eta_J,\mu)
&= \Gamma_{\rm cusp}[\al_s(\mu)] \biggl[\bfT_a^2\, \ln\frac{\omega^2_a e^{-2\eta_J}}{\mu^2} + \bfT_b^2 \, \ln\frac{\omega^2_b e^{2\eta_J}}{\mu^2} + \bfT_J^2 \,\ln\frac{\omega_J^2}{(2\cosh \eta_J)^2\mu^2} \biggr]
\nn \\
& \quad +\gamma_{H}^{\kappa}[\al_s(\mu)]
\,,\end{align}
with $\omega_{J} = 2 p_T^J \cosh \eta_J$, where the quoted form again assumes Casimir scaling of the cusp anomalous dimension. The consistency relation leading to the structure of the $\mu$-anomalous dimension $\gamma_{S_B}^\kappa$ reads
\begin{align}
\gamma_{S_B}^{\kappa}(\ell_B,\eta_J,\mu,\nu) \, \delta(\ell_J) & = - \gamma_{H}^{\kappa}(\omega_i,\eta_J,\mu)\, \delta(\ell_J)\, \delta(\ell_B) -\sum_{i =a,b}\gamma_{B}^{\kappa_i}\Bigl(\ell_B,\mu,\frac{\nu}{\omega_i}\Bigr)\, \delta(\ell_J) \nn \\
& \quad - 2 p_T^J \,\gamma_{J}^{\kappa_J}(2 p_T^J \ell_J,\mu)\, \delta(\ell_B)-2 \gamma_{S_{R}}^{\kappa_J} \biggl(\frac{2\ell_J}{R},R\, \ell_B,\mu\biggr) 
\, . \end{align}
Inserting eqs.~(\ref{eq:anom_H}),~(\ref{eq:anom_B}),~(\ref{eq:anom_J}),~(\ref{eq:SR_app_factorize}) and~(\ref{eq:anom_Shemi}) then leads to 
\begin{align}
\gamma_{S_B}^{\kappa}(\ell_B,\eta_J,\mu,\nu) &=2\,\Gamma_{\rm cusp}[\al_s(\mu)] \biggl[ \,-\bfT_J^2 \, \frac{1}{\mu} \, \mathcal{L}_0\Bigl(\frac{\ell_B}{\mu}\Bigr) +\biggl(\bfT_a^2 \,\ln\frac{\mu \,e^{\eta_J}}{\nu} + \bfT_b^2 \,\ln\frac{\mu\, e^{-\eta_J}}{\nu} \biggr) \de(\ell_B)\biggr]
\nn \\ & \quad
- \Bigl\{\gamma_{H}^{\kappa}[\al_s(\mu)] + \gamma_{J}^{\kappa_J}[\al_s(\mu)] + \gamma_{B}^{\kappa_a}[\al_s(\mu)]
+ \gamma_{B}^{\kappa_b}[\al_s(\mu)]
\nn \\ & \quad\qquad
 + 2 \gamma_{{\rm hemi}}^{\kappa_J}[\al_s(\mu)] \Bigr\} \,\delta(\ell_B)
\, .\end{align} 
Using color conservation, i.e.~$\bfT_J = -\bfT_a-\bfT_b$, and noting that the noncusp anomalous dimensions cancel each other at one loop, it is straightforward to check that \eq{S_BT} is consistent with this relation.

\section{Calculation of the soft function $S_B$}
\label{app:softfct}	

Here we outline the main steps for the one-loop computation of the wide-angle soft function for narrow jets ($R \ll R_0$) with a general jet veto. $S_B$ was defined in \subsec{softB} and results for various jet vetoes were given in eqs.~(\ref{eq:S_BC}),~(\ref{eq:S_BB}) and~(\ref{eq:S_BT}). Due to the fact that the jet region is not resolved by the wide angle soft modes, the contribution from the beam-beam dipole with the color factor $\bfT_a \cdot \bfT_b$ is just given by the result without any jet which is known for common measurements like (beam) thrust, C-parameter or the transverse momentum for back-to-back configurations~\cite{Fleming:2007xt,Hoang:2014wka,Chiu:2012ir}. The computation for the real radiation correction from the jet-beam dipoles can be performed similarly to the corresponding contribution for an energy veto in ref.~\cite{Ellis:2010rwa} summarized in their appendix B.1. It is convenient to take advantage of their results and calculate only the difference correction between the employed jet veto and the energy veto explicitly, which both have common soft IR divergences. For definiteness we consider the correction with the color structure $\bfT_a \cdot \bfT_J$, which can be written as 
\begin{align}\label{eq:S_aJ}
S^{(1)}_{B,aJ} &= -2g^2 \biggl(\frac{\mu^2 e^{\gamma_E}}{4\pi}\biggr)^{\epsilon} \bfT_a \cdot \bfT_J \int \frac{\df^d k}{(2\pi)^d} \, \frac{n_a \cdot n_J}{(n_a \cdot k)(n_J \cdot k)} \,\biggl(\frac{\nu}{2k_0\left|\cos\theta\right|}\biggr)^\eta\,2\pi \delta(k^2) \, \Theta(k_0) 
\nn \\
& \quad \times \Bigl\{\delta(\ell_B-k^0)+\Bigl[\delta(\ell_B-k^0 \left|\sin \theta\right| \tilde{f}(\cos \theta))-\delta(\ell_B-k^0)\Bigr]\Bigr\}
\nn \\
& \equiv S_{E} + \Delta S_{\Tau_B}
\, ,\end{align}
where $\cos \theta$ denotes the angle between the gluon momentum and the beam direction $\vec{n}_a$, i.e.~$\cos \theta = \vec{n}_a \cdot \vec{k}/|\vec{k}| = \tanh \eta$, and $\tilde{f}(\cos \theta) = f_B( \eta)$ is defined in terms of the veto-dependent function $f_B$ in \eq{TauB}. 
When $f_B(\eta) \to 1$ for $\eta \to \pm \infty$ this leads to rapidity divergences, which is for example the case for the transverse energy veto discussed in \subsec{pt}. To regulate these divergences we employ a factor $\nu^\eta/|2\,\vec{n}_a\sdt \vec{k}|^\eta$ arising from a modified version of the $\eta$-regulator in refs.~\cite{Chiu:2011qc,Chiu:2012ir}.\footnote{The rapidity regularization factor for the soft function needs to satisfy $\nu^\eta/(\bar{n}_a\sdt k)^\eta$ when the momentum $k$ becomes collinear to the beam direction $n_a$ in order to use the common result for the beam function matching coefficients (where precisely this factor is used for regularization).} Furthermore, we rescaled $\mu^2 \to \mu^2 e^{\gamma_E}/4\pi$  in \eq{S_aJ} anticipating $\MS$ renormalization. 

The unrenormalized result for the correction with an energy veto $S_E$ can be read off from eq.~(5.12) in ref.~\cite{Ellis:2010rwa}, 
\begin{align}\label{eq:S_E}
S_E  & = \frac{\alpha_s}{4\pi} \, \bfT_a \cdot \bfT_J \, \biggl[-\frac{16}{\mu}\, \mathcal{L}_1\Bigl(\frac{\ell_B}{\mu}\Bigr)+\biggl(\frac{8}{\epsilon}-8\ln(2-2\tanh \eta_J)\biggr) \frac{1}{\mu}\,\mathcal{L}_0\Bigl(\frac{\ell_B}{\mu}\Bigr) 
\\
& \quad + \biggl(-\frac{4}{\eps^2}+\frac{4\ln(2-2\tanh \eta_J)}{\epsilon}+4\,\Li_2(-e^{2\eta_J})-8\ln 2 \ln(1-\tanh \eta_J)+ \pi^2\biggr)\delta(\ell_B) \biggr]
\nn \, .\end{align}
The remaining correction $\Delta S_{\Tau_B}$ implementing the difference to the actual jet veto can be written as an integral over the angle $\cos \theta$, 
\begin{align}\label{eq:Delta_S}
\Delta S_{\Tau_B} &= -\frac{2\alpha_s}{\pi} \, \,\bfT_a \cdot \bfT_J\,\frac{\mu^{2\epsilon} e^{\gamma_E \epsilon}}{\ell_B^{1+2\epsilon}}\int_{-1}^{1} \df \cos \theta \, F(\cos \theta, \epsilon) \biggl[G\Bigl(\cos \theta, \epsilon, \eta, \frac{\nu}{\ell_B}\Bigr) - 1\biggr]
\,,\end{align}
where the function $F$ denotes the integrand for an energy veto (given in eq.~(B.2) of \cite{Ellis:2010rwa}), which also contains implicitly the dependence on the angle between beam and jet with $\cos \theta_{aJ} \equiv 1- n_a \cdot n_J \equiv n = \tanh \eta_J$, and the function $G$ encodes the additional factor for the specifically applied jet veto,
\begin{align}
F(u,\epsilon) & = (1-u)^{-1-\epsilon} (1+u)^{-\epsilon} \frac{1-n}{1-u n} \, {}_2\tilde{F}_1\biggl(\frac{1}{2},1,1-\epsilon,\frac{(1-n^2)(1-u^2)}{(1-un)^2}\biggr)
\, ,\nn \\
G\Bigl(u, \epsilon, \eta, \frac{\nu}{\ell_B}\Bigr) &=
\Bigl(\sqrt{1-u^2} \,\tilde{f}(u)\Bigr)^{2\epsilon+\eta} \biggl(\frac{\nu}{2 |u| \ell_B}\biggr)^\eta
\, .\end{align}
For all jet vetoes discussed in this paper $G$ reads explicitly
\begin{align}
& G^C (\cos \theta,\epsilon) = \biggl(\frac{1-\cos^2 \theta}{2}\biggr)^{2\epsilon} \, , \quad G^{\tau} (\cos \theta,\epsilon) = (1-\cos \theta)^{2\epsilon} \, , 
\nn \\
& G^{E_T}\Bigl(\cos \theta, \epsilon, \eta, \frac{\nu}{\ell_B}\Bigr)=(1-\cos^2 \theta)^{\epsilon+\frac{\eta}{2}} \biggl(\frac{\nu}{2 |\cos \theta| \ell_B}\biggr)^\eta
\,,\end{align}
where we dropped the rapidity regulator for \SCETa-type measurements.

To compute the integral \eq{Delta_S} to $\mathcal{O}(\epsilon^0,\eta^0)$ it is convenient to split it into two integration regions $-1 \leq \cos \theta \leq 1- \delta$ and $1- \delta \leq \cos \theta \leq 1$ with $1-\delta> \cos \theta_{aJ}$,  such that collinear divergences appear either in the jet or beam direction. Otherwise the choice of the cutoff parameter $\delta$ is irrelevant. For simplicity we take $\delta \ll 1$ and also expand in this parameter. We start with the contribution from the first integration domain, $\Delta S_{\Tau_B,1}$. The integrand $F$ is decomposed into a product of two functions $F_{J}$ and $\tilde{F}_J$, where $F_J$ has a power-like behavior for $\theta \to \theta_{aJ}$ (i.e.~for radiation close to the jet) and $\tilde{F}_{J}$ encodes the finite remainder, as discussed in ref.~\cite{Ellis:2010rwa} above and below eq.~(B.6). We can then write\footnote{Here we can drop the $\eta$-regulator since rapidity divergences can only arise for $\cos \theta \to 1$.} for $\Delta S_{\Tau_B,1}$
\begin{align}
 & \int_{-1}^{1-\delta} \df \cos \theta \, F_J(\cos \theta, \epsilon) \, \tilde{F}_J(\cos \theta, \epsilon) \bigl[ G(\cos \theta, \epsilon)-1\bigr] 
\nn \\
& = \tilde{F}_J(\cos \theta_{aJ}, \epsilon) \bigl[ G(\cos \theta_{aJ}, \epsilon)-1\bigr]  \int_{-1}^{1-\delta} \df \cos \theta \, F_J(\cos \theta,\epsilon) 
\nn \\
& \quad + \int_{-1}^{1-\delta} \df \cos \theta \, F_J(\cos \theta,\epsilon) \bigl[ \tilde{F}_J(\cos \theta, \epsilon)- \tilde{F}_J(\cos \theta_{aJ}, \epsilon) \bigr]\bigl[ G(\cos \theta, \epsilon)-1\bigr]
\nn \\
& \quad + \int_{-1}^{1-\delta} \df \cos \theta \, F_J(\cos \theta,\epsilon) \,\tilde{F}_J(\cos \theta_{aJ}, \epsilon)\bigl[ G(\cos \theta, \epsilon)-G(\cos \theta_{aJ}, \epsilon)\bigr]
\, ,\end{align}
where the integrands in the last two lines can be expanded in $\epsilon$ before the integration and the other integral can easily be carried out in $d$ dimensions.

For the correction $\Delta S_{\Tau_B,2}$ from the integration domain $1- \delta \leq \cos \theta \leq 1$, we perform a similar decomposition for both of the functions $F$ and $G$, such that $F_a$ and $G_a$ contain the power behavior for $\cos \theta \to 1$ (i.e.~for radiation close to beam $a$) and $\tilde{F}_a$ and $\tilde{G}_a$ contain the remainder. This leads to the integral for $\Delta S_{\Tau_B,2}$,  
\begin{align}
 &  \int_{1-\delta}^{1} \df \cos \theta \, F_a(\cos \theta, \epsilon) \, \tilde{F}_a(1, \epsilon) \, G_a\Bigl(\cos \theta, \epsilon, \eta, \frac{\nu}{\ell_B}\Bigr) \, \tilde{G}_a(1, \epsilon, \eta) 
\nn \\
& \quad - \int_{1-\delta}^{1} \df \cos \theta \, F_a(\cos \theta, \epsilon) \, \tilde{F}_a(1, \epsilon) +\mathcal{O}(\delta)
\, ,\end{align}
where both integrals can be carried out analytically without any additional expansions in $\epsilon$ or $\eta$. 

Adding the contributions from the two integration regions, the dependence on $\delta$ drops out. After expanding in $\eta$ and $\epsilon$, we obtain for the C-parameter veto, 
\begin{align}
\Delta S^C_{\Tau_B} & = \frac{\alpha_s}{4\pi} \, \bfT_a \cdot \bfT_J \, \biggl[\frac{16}{\mu}\, \mathcal{L}_1\Bigl(\frac{\ell_B}{\mu}\Bigr)+\biggl(-\frac{8}{\epsilon}+8\ln(1-\tanh^2 \eta_J)\biggr) \frac{1}{\mu}\,\mathcal{L}_0\Bigl(\frac{\ell_B}{\mu}\Bigr) 
\\
& \quad + \biggl(\frac{4}{\eps^2}-\frac{4\ln(1-\tanh^2 \eta_J)}{\epsilon}+8\,\Li_2\Bigl(\frac{1+\tanh \eta_J}{2}\Bigr)-2\ln^2\biggl(\frac{1+\tanh\eta_J}{2}\biggr)
\nn \\
& \quad -4\ln\biggl(\frac{1+\tanh\eta_J}{2}\biggr)\ln\biggl(\frac{1-\tanh\eta_J}{2}\biggr)+2\ln^2(2-2\tanh \eta_J) -\frac{5 \pi^2}{3}\biggr)\delta(\ell_B) \biggr] \nn
\,,\end{align}
for the beam thrust veto
\begin{align}
\Delta S^{\tau}_{\Tau_B} & = \frac{\alpha_s}{4\pi} \, \bfT_a \cdot \bfT_J \, \biggl[\frac{16}{\mu}\, \mathcal{L}_1\Bigl(\frac{\ell_B}{\mu}\Bigr)+\biggl(-\frac{8}{\epsilon}+8\ln(2-2\tanh |\eta_J|)\biggr) \frac{1}{\mu}\,\mathcal{L}_0\Bigl(\frac{\ell_B}{\mu}\Bigr) 
\nn \\
& \quad + \biggl(\frac{4}{\eps^2}-\frac{4\ln(2-2\tanh |\eta_J|)}{\epsilon}- 4\,\Li_2(-e^{2\eta_J})-4\,\Li_2\bigl(e^{-2|\eta_J|}\bigr)-8 \eta_J^2 \, \theta(-\eta_J)
\nn \\
& \quad +8\ln 2 \ln(1-\tanh \eta_J) - \pi^2
\biggr)\delta(\ell_B) \biggr]
\,,\end{align}
and for the transverse energy veto\footnote{Although the full $\epsilon$-dependence in the expression proportional to $1/\eta$ should be in principle kept unexpanded, this is only relevant to ensure that the coefficient of the $1/\eta$ pole is explicitly $\mu$-independent, which is also true order by order in its $\epsilon$ expansion. Here we only display the terms up to $\mathcal{O}(\epsilon^0)$ for better readability since these contain all relevant information. } 
\begin{align}
\Delta S^{E_T}_{\Tau_B} & = \frac{\alpha_s}{4\pi} \, \bfT_a \cdot \bfT_J \, \biggl[\frac{16}{\mu}\, \mathcal{L}_1\Bigl(\frac{\ell_B}{\mu}\Bigr)+\biggl(-\frac{8}{\eta}-\frac{4}{\epsilon}-8 \ln\Bigl(\frac{\nu}{\mu}\Bigr)+4\ln(4-4\tanh^2 \eta_J)\biggr) \frac{1}{\mu}\,\mathcal{L}_0\Bigl(\frac{\ell_B}{\mu}\Bigr) 
\nn \\
& \quad + \biggl(\frac{4}{\eta \,\eps}+ \frac{1}{\eps} \Bigl[4 \ln \Bigl(\frac{\nu}{\mu}\Bigr)-2\ln(4-4\tanh^2 \eta_J)\Bigr]-4\,\Li_2\Bigl(\frac{1-\tanh \eta_J}{2}\Bigr)
\nn \\
& \quad -4\ln\Bigl( \frac{1-\tanh \eta_J}{2}\Bigr)\ln\Bigl( \frac{1+\tanh \eta_J}{2}\Bigr) +2\ln^2( 2-2\tanh \eta_J)\biggr)\delta(\ell_B) \biggr]
\,.\end{align}
Combining these with \eq{S_E}, removing the UV and rapidity divergences by renormalization, and simplifying the resulting expressions leads to the one-loop results given in eqs.~(\ref{eq:S_BC}),~(\ref{eq:S_BB}) and~(\ref{eq:S_BT}).

\section{Fixed-order expansion of the cross section}
\label{app:FO}

We now present the singular cross section for the cumulative measurement of the jet mass $m_J$ 
in $pp \to L +1$ jet, employing a veto on the transverse momentum of additional jets $p_T^{\rm cut} \ll p_T^J$. We assume $R \ll R_0$ and $m_J \ll p_T^J R$ (i.e.~regime 2), and obtain our expressions by expanding the factorization theorem in \eq{fact_pTjet} to a given order in $\alpha_s$. By including the resummation we can predict logarithmic terms in the higher-order cross sections. 

We decompose the cross section integrated over $\Tau_J \leq \Tau_J^{\rm cut}$ as
\begin{align}\label{eq:FO_decomp}
\sigma_2(\Tau_J^{\rm cut},p_T^{\rm cut},\Phi, \kappa) & = \hat{\sigma}_B
\sum_{k, l}\sum_{n \geq 0} \Bigl(\frac{\alpha_s}{4\pi}\Bigr)^n\!\!
\int\! \frac{\df x_a'}{x_a'}\,\frac{\df x_b'}{x_b'}\,
\hat{\sigma}_{kl}^{(n)}\Bigl(\frac{x_a}{x_a'},\frac{x_b}{x_b'},\mu_0\Bigr) f_{k}(x_a',\mu_0) \,f_l(x_b',\mu_0),
\end{align}
where $\mu_0 = p_T^J$, $\hat{\sigma}_B \equiv H^{(0)}_\kappa (\Phi,\mu_0)$ denotes the partonic cross section at the Born level, and $k,l$ sum over parton flavors. We further decompose $\hat\sigma_{kl}^{(n)}$ as
\begin{align}
\hat{\sigma}_{kl}^{(n)}(z_a,z_b) = \bigl[\hat{\sigma}_\text{jet}\, \hat\sigma_{\text{rest},k,l}(z_a,z_b) \bigr]^{(n)} = \sum_{m\geq 0} \hat{\sigma}^{(m)}_\text{jet}\, \hat\sigma^{(n-m)}_{\text{rest},k,l}(z_a,z_b)
\,,\end{align}
separating out the contribution containing the jet mass and jet radius logarithms 
\begin{align}
\hat{\sigma}_\text{jet} \equiv\int\! \df s_J\, J_{\kappa_J}(s_J,\mu_0) \int\! \df k_J \, S_{R,\kappa_J}(k_J,p_T^\text{cut} R,\Rv,\mu_0) 
\,\theta\Big(\frac{2 \Tau_J^{\rm cut}}{R}  - \frac{s_J}{p_T^J R} - k_J\Big)
\,.\end{align}
The rest contains corrections from the hard function $H_\kappa = H^{(0)}_\kappa\,h_\kappa$, the wide-angle soft function $S_{B,\kappa}$ and the beam function matching coefficients $I_{ij}$ in \eq{fact_pTjet}, i.e. 
\begin{align}
\hat\sigma^{(n)}_{\text{rest},k,l} & = \bigl[h_\kappa \, I_{\kappa_ak}(z_a) \, I_{\kappa_bl}(z_b) \, S_{B,\kappa}\bigr]^{(n)}
\,.\end{align}
At tree-level
\begin{align}
\hat\sigma^{(0)}_{\text{rest},k,l}(z_a,z_b)  = \de_{\kappa_ak}\, \de_{\kappa_bl}\,\de(1-z_a)\, \de(1-z_b)
\,.\end{align}
At one-loop level all ingredients are known analytically and we obtain 
\begin{align}\label{eq:Oneloop_FO}
\hat{\sigma}^{(1)}_\text{jet}
& =-\frac{\Gamma^{J}_0}{2}\, L_J^2+  L_J\biggl[2 \Gamma^J_0\, L_R-\frac{\gamma_{J\, 0}}{2}\biggr] -\Gamma^J_0 (2L_R^2 + 2L_R L_B  + L_B^2) + j_\kappa^{(1)}+2\bfT_J^2 \,s^{(1)}  
\,, \nn \\
\Delta \hat{\sigma}_{\text{rest},kl}^{(1)}
& = \delta_{\kappa_a k} \,\delta_{\kappa_b l}\biggl[h^{(1)}_\kappa + (\Gamma_0^J -\Gamma^a_0 - \Gamma^b_0) L_B^2 +2 \Gamma^a_0 L_B \ln\biggl(\frac{\omega_a e^{-\eta_J}}{p_T^J}\biggr) +2 \Gamma^b_0 L_B \ln\biggl(\frac{\omega_b e^{\eta_J}}{p_T^J}\biggr)  \nn \\
& \quad -(\bfT_a^2+ \bfT_b^2 +\bfT_J^2) s^{(1)}   \biggr] + \biggl[\biggl(\Gamma_0\, L_B\, \tilde{p}^{(0)}_{\kappa_ak}(x_a)+\tilde{I}^{(1)}_{\kappa_ak} (x_a)\biggr)\delta_{\kappa_bl} + (a,k \leftrightarrow b,l)\biggr] 
\,. \end{align}	
Here we have abbreviated the logarithms that occur in these expressions as
\begin{align}
L_J \equiv \ln\biggl(\frac{2\Tau_J^{\rm cut}}{p_T^J}\biggr) = \ln\biggl(\frac{(m_J^{\rm cut})^2}{(p_T^J)^2}\biggr)  \, , \quad  L_B \equiv \ln\biggl(\frac{p_T^{\rm cut}}{p_T^J}\biggr)\, , \quad L_R \equiv \ln R \, .
\end{align}
We have written \eq{Oneloop_FO} in terms of coefficients of the anomalous dimension defined through the expansion
\begin{align} \label{eq:betafunction}
\Gcusp(\alpha_s)
= \sum_{n=0}^\infty \Gamma_n \Bigl(\frac{\alpha_s}{4\pi}\Bigr)^{n+1}
,\quad 
\gamma_{J}^{\kappa}= \sum_{n=0}^\infty \gamma^\kappa_{J \, n} \Bigl(\frac{\alpha_s}{4\pi}\Bigr)^{n+1} , \quad \beta(\alpha_s)
= - 2 \alpha_s \sum_{n=0}^\infty \beta_n\Bigl(\frac{\alpha_s}{4\pi}\Bigr)^{n+1} 
,\end{align}
with $\Gamma_n^i = \bfT_i^2 \Gamma_n$. The one loop coefficients are
\begin{align}
\Gamma_0 = 4 \, , \quad \gamma^q_{J \, 0} = 6 C_F \, , \quad \gamma^g_{J\, 0} = 2 \beta_0 \, .
\end{align}
The remaining constants appearing in \eq{Oneloop_FO} are given by\footnote{We did not distinguish the constant term of the collinear-soft function in \eq{S_J} from the associated term in the wide-angle soft function in \eq{S_BT} and denoted both with the same symbol $s^{(1)}$ for simplicity.} 
\begin{align}
 j_q^{(1)} = (7-\pi^2)C_F \, , \quad \, j_g^{(1)} = \Bigl(\frac{4}{3}-\pi^2\Bigr)C_A +\frac{5}{3} \beta_0 \, , \quad  s^{(1)} =\frac{\pi^2}{6} \, .
\end{align}
The functions $\tilde{p}^{(0)}_{ij}$ are directly related to the splitting functions at $\mathcal{O}(\alpha_s)$  and given by 
\begin{align}
\tilde{p}^{(1)}_{qq}(z) & = C_F \Bigl[2\mathcal{L}_0(1-z)-\theta(1-z)(1+z)\Bigr] \, , \nn \\
 \tilde{p}^{(1)}_{qg}(z) & = T_F\, \theta(1-z)\, (1-2z+2z^2) \, ,  \nn \\
 \tilde{p}^{(1)}_{gg}(z) & = 2C_A \biggl[\mathcal{L}_0(1-z) + \theta(1-z)\Bigl( \frac{1-z}{z}+z(1-z) -1\Bigr)\biggr] \, , \nn \\
\tilde{p}^{(1)}_{gq}(z) &= C_F\, \theta(1-z)\, \frac{2-2z+z^2}{z}  \, .
\end{align}
The matching functions $\tilde{I}_{ij}$ encoding collinear initial state radiation effects are given by 
\begin{align}
&\tilde{I}_{qq}^{(1)}(z) = C_F \theta(1-z) 2 (1-z) \, , \quad  \tilde{I}_{qg}^{(1)}(z) = T_F\theta(1-z) 4z (1-z) \, , \nn \\
&\tilde{I}_{gg}^{(1)}(z) = 0 \, , \quad \tilde{I}_{gq}^{(1)}(z) = C_F\theta(1-z) 2z \, .
\end{align}
We also display the logarithmic dependence of the two loop result. Here we only show explicitly the terms associated with either jet mass or jet radius logarithms. These read for $\hat{\sigma}_\text{jet}$ 
\begin{align}
\hat{\sigma}^{(2)}_\text{jet}
& =\frac{\bigl(\Gamma^J_0\bigr)^2}{8} L_J^4+  \frac{\Gamma^J_0}{4} L_J^3 \biggl[- 4 \Gamma^J_0 L_R+ 2 \beta_0  +  \gamma_{J \,0} \biggr] + \frac{1}{2} L_J^2 \biggl[ \bigl(\Gamma^J_0\bigr)^2 (6 L_R^2 + 
 2 L_ R L_B +  L_B^2)\nn \\
& \qquad  - 2
  \Gamma^J_0 (2\beta_0+\gamma_{J \,0}) L_R - \Gamma^J_1-\Gamma^J_0\Bigl( j_\kappa^{(1)} +2s^{(1)}+\frac{\pi^2}{6} \Gamma^J_0 \Bigr) +\frac{\gamma_{J \,0}}{4}(2\beta_0 + \gamma_{J \,0}) \biggr] \nn \\
& \quad+ L_J \biggl[-2\bigl(\Gamma^J_0\bigr)^2 L_R(2L_R^2+ 2L_R L_B + L_B^2)+ \Gamma^J_0(2 \beta_0 + \gamma_{J \,0}) L_R^2 +  \frac{\Gamma^J_0\gamma_{J \,0}}{2} \,L_B (2L_R+L_B)\nn \\
& \qquad+ 2\biggl(\Gamma^J_1 +\Gamma^J_0\Bigl(j^{(1)}_\kappa + 2s^{(1)} +\frac{\pi^2}{6} \Gamma^J_0\Bigr)\biggr) L_R  -\frac{\gamma_{J \,1} }{2}-\gamma_{{\rm hemi}\,1} -\frac{\gamma_{J \,0}}{2}\Bigl(j^{(1)}_\kappa + 2s^{(1)} +\frac{\pi^2}{6} \Gamma^J_0\Bigr) \nn \\
& \qquad -\beta_0(j^{(1)}_\kappa + 2s^{(1)})  + \zeta_3(\Gamma^J_0)^2 \biggr] +  2\bigl(\Gamma^J_0\bigr)^2 L_R (L_R^3 + 2L_R^2 L_B + 2 L_R L_B^2  + L_B^3 \Bigr)\nn \\
& \quad   +2\Gamma^J_0\beta_0 L_R L_B (L_R+L_B) -\frac{\pi^2}{3} (\Gamma^J_0\bigr)^2 L_R^2-2L_R(L_R+L_B) \biggl(\Gamma^J_1+\Gamma^J_0\Bigl(j^{(1)}_\kappa + 2s^{(1)}\Bigr)\biggr) \nn \\
 & \quad + L_R \biggl(\Gamma_0^J \Bigl(\frac{\pi^2}{6}\gamma_{J \,0} -2\zeta_3 \Gamma_0^J\Bigr) - \Delta \gamma^{\rm alg}_{S_R\, 1}(\Rv)\biggr) +S_{\rm hemi}^{({\rm NG},2)}\Bigl(\frac{2\Tau_J^{\rm cut}}{p_T^{\rm cut} R^2}\Bigr)  \nn \\
 & \quad + \textrm{(terms involving only $L_B$ and $\ln \Rv$)}   \, .
\end{align}Here the term $ S_{\rm hemi}^{({\rm NG},2)}(x)$ encodes the nonglobal structures and can be directly read off from refs.~\cite{Hornig:2011iu,Kelley:2011ng},
\begin{align} \label{eq:ShemiNGL}
 S_{\rm hemi}^{({\rm NG},2)}(x) & = \bfT_J^2 \biggl\{-\frac{4 \pi^2}{3} C_A \ln^2 x +\biggl[C_A\Bigl(-\frac{4}{3}+\frac{44 \pi^2}{9}-8\zeta_3\Bigr)+T_F n_f\Bigl(\frac{8}{3}-\frac{16\pi^2}{9}\Bigr)\biggr] \left|\ln x\right|\biggr\} \nn \\
 & \quad + \textrm{(nonlogarithmic terms)} \, .
\end{align}
The nonlogarithmic terms in \eq{ShemiNGL} must be kept when including this term, since they are of the same size as the logarithms for the regions we consider.
The required anomalous dimension coefficients at two-loop order are given by
\begin{align}
\Gamma_1
&= \Bigl( \frac{268}{9} -\frac{4\pi^2}{3} \Bigr)C_A  - \frac{80}{9} \,T_F n_f  \, , \nn \\
\gamma^{q}_{J\,1} & =C_F\biggl[\Bigl(\frac{146}{9}-80\zeta_3\Bigr)C_A+(3-4\pi^2+48\zeta_3)C_F+\Bigl(\frac{121}{9}+\frac{2\pi^2}{3}\Bigr)\beta_0\biggr]  \, , \nn \\
\gamma^{g}_{J\,1} & =\Bigl(\frac{182}{9}-32\zeta_3\Bigr)C_A^2+\Bigl(\frac{94}{9}-\frac{2\pi^2}{3}\Bigr)C_A \beta_0+2\beta_1 \, , \nn \\
\gamma^{\kappa_J}_{{\rm hemi}\,1} & =\bfT_J^2 \biggl[\Bigl(-\frac{64}{9}+28\zeta_3\Bigr)C_A+\Bigl(-\frac{56}{9}+\frac{\pi^2}{3}\Bigr)\beta_0\biggr] \, .
\end{align}
At two loops, clustering corrections due to jet algorithm employed for the jet veto algorithm enter in the noncusp anomalous dimension of the csoft function,
\begin{align}
\gamma_{S_R \, 1}(\Rv) = 2\gamma_{{\rm hemi}\,1} + \Delta \gamma^{\rm alg}_{S_R\, 1}(\Rv) 
\,.\end{align}
The term $\Delta\gamma^{\rm alg}_{S_R\, 1}(\Rv)$ is currently not known.

\phantomsection
\addcontentsline{toc}{section}{References}
\bibliographystyle{../jhep}
\bibliography{../softfunc}

\end{document}